\documentclass[12pt,journal,onecolumn]{IEEEtran}
\usepackage{times,epsfig,amsbsy,amssymb,float,color}
\usepackage[cmex10]{amsmath}
\usepackage{mdwmath,setspace}
\usepackage{mdwtab}
\usepackage{amsfonts}
\usepackage{subfigure}
\usepackage{amsmath,amssymb}
\usepackage{cases}
\usepackage{booktabs}
\usepackage{algorithm}
\usepackage{stfloats}
\usepackage{color}
\usepackage[numbers,sort&compress]{natbib}
\usepackage{algorithmicx}
\usepackage{algpseudocode}
\usepackage{mathrsfs}
\algdef{SE}[DOWHILE]{Do}{doWhile}{\algorithmicdo}[1]{\algorithmicwhile\ #1}

\newtheorem{theorem}{Theorem}
\newtheorem{lemma}{Lemma}
\newtheorem{proposition}{Proposition}
\title{\huge A Multilevel Framework for Lattice Network Coding}
\author{
\IEEEauthorblockN{  Yi Wang, Alister Burr, Qinhui Huang and Mehdi Molu\\ \thanks{The work described in this paper was supported by the European Commission Framework 7 Programme under grant agreement 318177 (DIWINE).}}

}
\begin{document}
\maketitle\pagestyle{empty}
\maketitle\thispagestyle{empty}
\begin{abstract}
We present a general framework for studying the multilevel structure of  lattice network coding (LNC), which serves as the theoretical fundamental for solving the ring-based LNC problem in practice, with greatly reduced decoding complexity. Building on the framework developed, we propose a novel lattice-based network coding solution, termed layered integer forcing (LIF), which applies to any lattices having multilevel structure. The theoretic foundations of the developed multilevel framework lead to a new general lattice construction approach, the elementary divisor construction (EDC), which shows its strength in improving the overall rate over multiple access channels (MAC) with low computational cost. We prove that the EDC lattices subsume the traditional complex construction approaches. Then a soft detector is developed for lattice network relaying, based on the multilevel structure of EDC. This makes it possible to employ iterative decoding in lattice network coding, and simulation results show the large potential of using iterative multistage decoding to approach the capacity.  
\end{abstract} 


\section{Introduction}
\label{sec:introduction}
There has recently been a resurgence in research on lattice codes for wireless communications, as a result of two recent developments. The first is recent work \cite{Erez.Zamir, Forney.Coset.Codes} which has shown that lattice codes with lattice decoding are capable of approaching channel capacity. The second is their application to physical layer network coding (PNC) \cite{Zhang.PLNC.ACM.2006} for ultra-dense wireless multihop networks \cite{Yi.CLDLC.2015}. In particular Nazer and Gastpar have developed compute-and-forward (C\&F) \cite{Nazer.Gastpar.TIT.2011}, which applies structured nested lattice codes to PNC for multiuser relay networks. However it is difficult to increase the transmission rate using previous lattice constructions such as construction A lattices, since this requires linear channel codes over large finite fields, for which the decoding complexity is typically unaffordable.  This paper lays the foundations for a multilevel structure for lattice codes, and uses it to introduce a general lattice construction approach and two multistage decoding approaches which greatly simplify decoding, and which can exploit iterative techniques to approach capacity.

Feng \textit{et al.} formulated a general algebraic framework for lattice network coding (LNC) \cite{Feng.AlgeAppro.TIT.2013}, giving practical design guidelines for C\&F. Previous work, e.g. in \cite{Feng.AlgeAppro.TIT.2013, Qifu.Eisenstein, Nara.Eisenstein}, has given LNC design guidelines when quotient lattices are constructed from existing channel codes using complex construction A. In this paper, we consider a multilevel structure for lattice network coding, which provides a practical solution to the ring-based network coding problem. We also propose an efficient lattice construction approach (which we term the elementary divisor construction (EDC)) based on the theorems developed, which also subsumes the most important previous lattice constructions. The EDC lattice has a multilevel algebraic structure, and is well suited for multistage decoding. Note that the recently proposed product construction \cite{Multistage.Product.Construction.Nara} used in C\&F is a special case of EDC. The EDC approach is a straightforward result of the theoretic framework developed in section \ref{sec:Multilevel.Lattice.Network.Coding}. We give explicit representation of the generator matrix for the EDC lattice, propose  a new concept of the primary sublattice, and derive the nominal coding gain and kissing numbers for the EDC lattice in all forms.  The main contributions of the paper are summarised below:  

\begin{enumerate}
\item We develop a generic multilevel lattice network coding scheme based on some algebraic theorems. This approach keeps beneficial compatibility of the traditional LNC scheme, whereas enabling more flexible coding design techniques. Note that MLNC makes also no particular assumption about the structure of the underlying nested lattice code.     
\item We propose a novel lattice network decoding approach based on MLNC, termed layered integer forcing (LIF), which
	\begin{itemize}
	\item improves the overall throughput for network coding with greatly reduced decoding complexity.
	\item decodes lattices which are no longer a vector space.
	\item allows flexible linear labelling design for additional performance enhancement. 
	\end{itemize}
\item We develop a modified Viterbi algorithm which implements LIF. 
\item Building on the algebraic framework developed for MLNC, we present a novel lattice construction approach (EDC approach), show its good structure properties (e.g. the explicit form of the generator matrix) in reducing the decoding complexity, and derive its nominal coding gain and kissing numbers. Mathematically we also prove that EDC lattices subsume the most important complex lattice constructions. 
\item We propose a soft detector specifically designed for EDC lattices (as an alternative to LIF for decoding EDC lattices). We evaluate its non-binary extrinsic information transfer characteristics, and propose an iterative multistage decoding approach for EDC lattices, which shows a substantial improvement in decoding performance.
\item We show how  multistage detection, iteration-aided multistage detection, and LIF can be applied to MLNC. We also show, by simulation, that iterative decoding performs better than the Viterbi detection approach used in the traditional LNC.  This provides the basis for further work, and opens a new research area of iterative decoding for lattice network coding.
\end{enumerate}  

This paper lays the foundations for a new research area in multistage and iterative decoding for lattice network coding. We expect that it will provide the basis for extensive further work, both to explore the rich algebraic features of the new construction, and to exploit it in practical implementations of LNC in 5G wireless systems.  

The remainder of this paper is organised as follows. In section \ref{sec:Algebra.Preliminaries} we review some algebraic preliminaries which will be useful in setting up our multilevel framework. Section \ref{sec:Multilevel.Lattice.Network.Coding} studies the algebraic properties of MLNC and presents the practically feasible encoding and decoding solutions. Section \ref{sec:EDC.Overall} presents a new general lattice construction approach based on MLNC theorems developed and proves that it subsumes some important lattice constructions that have been widely known. Section \ref{sec:Iterative.Detection.EDC} presents the soft detector for MLNC and studies the iterative decoding and multistage decoding approaches designed for MLNC. Section \ref{sec:simulations} presents the simulation results based on different decoding modes. Section \ref{sec:conclusions} concludes the paper and presents the future work.

\subsection{Notations}
Notations used throughout this paper are defined as follows. $\mathbb{N}$, $\mathbb{Z}$ and $\mathbb{C}$ denote the fields of natural numbers, integers and complex numbers, respectively. $\mathbb{F}_q$, $q>1$, $q\in\mathbb{Z}$ denotes the finite field of size $q$. $\mathbb{F}_{\mathbf{q}}^n$ denotes an $n$-tuple finite field where the field size for the $i^{\mathrm{th}}$ dimension $i\in\{1,2,\cdots,n\}$ is determined by $q_i\in\mathbb{Z}$. We also use boldface lower-case to denote a vector, i.e. $\mathbf{a}=[a^1,a^2,\cdots,a^n]$. $\mathbf{V}^{\setminus i}\triangleq[V^1,V^2,\cdots,V^{i-1},V^{i+1},\cdots,V^n]$ represents a set including all elements except the $i^{\mathrm{th}}$ one. The upper-case letter (e.g. $V$) represents a random variable and its realisation is denoted by the lower-case $v$. The direct sum and direct product are denoted as $\oplus$ and $\times$, respectively.   


\section{Algebra Preliminaries}
\label{sec:Algebra.Preliminaries}
We present some definitions and theorems in abstract algebra, which can be found in relevant textbooks, e.g.  \cite{Lessons.Rings.Modules}. \vspace{-2em}

\subsection{Ideal and Principal Ideal Domain}
Let $R$ be a commutative ring with identity $1$, and $R^*=R\backslash 0$. A \textit{unit} $\mathcal{U}(R)$ in $R$ refers to any element $x$ in $R$ such that $xr=rx=1$ for some $r\in R$. Any root of unity in a ring $R$ is a unit. An element $x$ in $R$ is called a \textit{zero divisor} $\mathcal{Z}(R)$ if $xr=rx=0$ for some $r\in R^*$. An element $p\in R$, $p\notin\mathcal{Z}(R)$, $p\notin\mathcal{U}(R)$, is called a \textit{prime} in $R$ when $p~|~ab$ for some $a,b\in R^*$, then either $p~|~a$ or $p~|~b$.  

An \textit{ideal} $\mathcal{I}$ of $R$ is a non-empty subset of $R$ that is closed under subtraction (which implies that $\mathcal{I}$ is a group under addition), and is defined by:
\begin{enumerate}
\item $\forall a,b\in\mathcal{I}$, $a-b\in\mathcal{I}$.
\item $\forall a\in\mathcal{I}$, $\forall r \in R$, then $ar\in R$ and $ra\in R$.
\end{enumerate}

If $A=\{a_1,\cdots,a_m\}$ is a finite non-empty subset of $R$, we use $\langle a_1,\cdots,a_m\rangle$ to represent the ideal generated by $A$, i.e. $$\langle a_1,\cdots,a_m\rangle=\{a_1r_1+\cdots+a_mr_m:r_1,\cdots,r_m\in R  \}$$ Note that $R$ has at least two ideals $\{0\}$ and $\{R\}$. 

An ideal $\mathcal{I}$ of $R$ is said to be \textit{proper} if and only if $1\notin\mathcal{I}$. An ideal $\mathcal{I}_{\mathrm{max}}$ is said to be \textit{maximal} if $\mathcal{I}_{\mathrm{max}}$ is a proper ideal and the only ideals that include $\mathcal{I}_{\mathrm{max}}$ are $R$ and $\mathcal{I}_{\mathrm{max}}$ itself. We say that an equivalence relation $a\sim b$ on the  set $R$ is defined by $\mathcal{I}$ if and only if $a-b\in\mathcal{I}$. 

An ideal $\mathcal{I}$ of $R$ is \textit{principal} if $\mathcal{I}$ is generated by a single element $a\in\mathcal{I}$, written as $\mathcal{I} = \langle a \rangle$. A \textit{principal ideal ring} is a ring whose every ideal is principal. If $R$ is a principal ideal ring without zero divisors, then $R$ forms an ideal domain, and more precisely, a \textit{principal ideal domain} (PID). Examples of PIDs include the ring of integers, the ring of Gaussian integers $\mathbb{Z}[i]$ and the ring of Eisenstein integers $\mathbb{Z}[\omega]$.

%
\vspace{-1em}
\subsection{Modules over PID and Structure Theorem} \label{sec:Modules.Over.PID}
Again, let $S$ be a commutative ring with identity 1. An $S$-module $M$ over $S$ is an abelian group $(M,+)$ under a binary operation $+$, together with a function $\mathscr{F}:S\times M\longmapsto M$ which satisfies the same conditions as those for vector space. Note that modules over a field are the same as vector spaces. An $S$-submodule of $M$ is a subgroup $N$ of $M$ which is closed under the action of ring elements, and hence the submodule $N$ forms also an $S$-module under the restricted operations. 

An $S$-module is said to be finitely generated (\textit{f.g.}) if $M$ has a finite basis $\{m_1,\cdots,m_n\}$ such that $\sum_i Rm_i = M$.
 
The annihilator of an element $m\in M$ is the set of elements $s\in S$ such that $sm = 0$. The annihilator of $M$ is the elements $s\in S$ such that $\{sm=0 |\forall m\in M\}$, denoted by $\mathrm{Ann}_S(M) = \bigcap \{\mathrm{Ann}_S (m)|m\in M  \}$. If $M$ is a free $S$-module, then $\mathrm{Ann}_S(M)=\langle 0 \rangle$. 

If $M$ is annihilated by ideal $\mathcal{I}$ of $S$, we can make $M$ into a quotient $S$-module $M/N$ by defining an action on $M$ satisfying, $$ m(s+\mathcal{I})=ms,~~~~\forall m\in M$$ 

The torsion submodule $M_{\mathrm{Tor}}$ of $M$ is defined by:$$M_{\mathrm{Tor}}=\{m\in M : \mathrm{Ann}_{S}(m)\neq \{0\} \}$$ A torsion free module is trivial.

Let $M$ and $N$ be two $S$-modules. An $S$-module homomorphism is a map $\phi: M\longmapsto N$, which respects the $S$-module structures of $M$ and $N$, i.e.,
$$\phi(s_1m_1+s_2m_2) = s_1\phi(m_1) \odot s_2\phi(m_2)$$
$\forall s_1,s_2\in S$, $\forall m_1,m_2\in M$. An $S$-module homomorphism $\phi:  M\longmapsto N$ is called an $S$-module isomorphism if it is both injective and surjective, which is denoted by $M\cong N$. The kernel of $\phi$ denotes the elements in $M$ which makes the image of $\phi$ equal to zero.

\section{Multilevel Lattice Network Coding} \label{sec:Multilevel.Lattice.Network.Coding}
\subsection{Algebraic Approach for Multilevel Structure} \label{sec:Multilevel.Structure}
Briefly if there is a matrix $\mathbf{G}_{\Lambda}\in\mathbb{C}^{n^{\prime}\times n}$, $n^{\prime}\leq n$ such that all its $n^{\prime}$ row vectors $\mathbf{g}_{\Lambda,1}, \cdots, \mathbf{g}_{\Lambda,{n^{\prime}}}\in\mathbb{C}^{n}$ are linearly independent, the set of all $S$-linear combinations of $\mathbf{g}_{\Lambda,1}, \cdots, \mathbf{g}_{\Lambda,{n^{\prime}}}$ forms an $S$-lattice $\Lambda\in\mathbb{C}^n$, written by, $\Lambda=\{\mathbf{s}\mathbf{G}_{\Lambda}: \mathbf{s}\in S^{n^{\prime}}\}$, where $\mathbf{G}_{\Lambda}$ is called the lattice generator. 

Following the explanation in section \ref{sec:Modules.Over.PID}, an $n$-dimensional $S$-lattice is  precisely an $S$-module over PID, and similarly the sublattice $\Lambda^{\prime}$ in $\Lambda$ forms a $S$-submodule. The partition of the $S$-lattice, denoted by $\Lambda/\Lambda^{\prime}$ represents $|\Lambda:\Lambda^{\prime}|<\infty$ (the index of $\Lambda^{\prime}$) equivalence classes.  Assume $S$ is a PID, we have the following theorem. 

\begin{theorem}\label{Theorem.Lattice.Decomposition}
Let $\Lambda$ and $\Lambda^{\prime}$ be $S$-lattices and $S$-sublattices, $\Lambda^{\prime}\subseteq\Lambda$, $|\Lambda:\Lambda^{\prime}|<\infty$ such that $\Lambda/\Lambda^{\prime}$ has nonzero annihilators. Then $\Lambda/\Lambda^{\prime}$ is the direct sum of a finite number of quotient sublattices, 
\begin{align}
\Lambda/\Lambda^{\prime} &= \Lambda_{p_1}/\Lambda^{\prime}_{p_1}\oplus\Lambda_{p_2}/\Lambda^{\prime}_{p_2}\oplus\cdots\oplus\Lambda_{p_m}/\Lambda^{\prime}_{p_m} \label{equ:Layered.Sublattices}
\end{align}
where $\Lambda_{p_i}/\Lambda^{\prime}_{p_i}\triangleq\{\lambda\in \Lambda/\Lambda^{\prime}: p_i^{\gamma}\lambda = 0\}$ for some $\gamma\geqslant 1$, and every $p_i$, $i=1,2,\cdots,m$ is a distinct prime over $S$. 
\end{theorem}

\begin{IEEEproof} The quotient $S$-lattice $\Lambda/\Lambda^{\prime}$ has non-zero annihilators; this implies that $\Lambda/\Lambda^{\prime}$ forms a $f.g.$ torsion module. Let $\lambda\in\Lambda/\Lambda^{\prime}$ and suppose that $\mathrm{Ann}_S(\Lambda/\Lambda^{\prime})=Sa$, where $a\in S$ and $a\neq 0$ (the property of the torsion module). Since $S$ is also a unique factorization domain, so $a=p_1^{\gamma_1}p_2^{\gamma_2}\cdots,p_m^{\gamma_m}$. We now write $a_i = a/p_i^{\gamma_i}$ which is the product of irreducible factors that are relatively prime to $p_i$. There must exist $s_1,s_2,\cdots,s_m$ in $S$ such $\sum_{i=1}^m s_ia_i = 1$ since $\mathrm{gcd}(a_1,a_2,\cdots,a_m) = 1$; Now we have
\begin{align}
s_ip_i^{\gamma_i}a_i\lambda = 0
\end{align}   
since $a$ annihilates $\Lambda/\Lambda^{\prime}$. Theorem \ref{Theorem.Lattice.Decomposition} states that the sublattice $\Lambda_{p_i}/\Lambda^{\prime}_{p_i}$ must satisfy the condition  $p_i^{\gamma_i}\lambda = 0$ for some $\gamma_i$s. Hence if $\lambda$ is annihilated by some powers of $p_i$, then $s_ia_i\lambda\in\Lambda_{p_i}/\Lambda^{\prime}_{p_i}$. Based on the statements above, $\sum_{i=1}^ms_ia_i\lambda = \lambda$, this proves that the $S$-lattice 
\begin{align}
\lambda\in \Lambda_{p_1}/\Lambda^{\prime}_{p_1}+\Lambda_{p_2}/\Lambda^{\prime}_{p_2}+\cdots+\Lambda_{p_m}/\Lambda^{\prime}_{p_m}
\end{align}

We suppose $\lambda_i\in\Lambda_{p_i}/\Lambda^{\prime}_{p_i}$ and $\sum_{i=1}^m \lambda_i = 0$. Then 
\begin{align}
a_i\sum_{j=1}^m \lambda_j = a_i\lambda_i=0 \label{equ:prove1.3}
\end{align}
where ($\ref{equ:prove1.3}$) follows from the fact that $a_i\lambda_j = 0$ for $i\neq j$. Since $a_i$ is non-zero, $\lambda_i$ must be zero. Based on the same proof, we can conclude that $\{\lambda_i=0|\forall i=1,2,\cdots,m\}$ provided that $\sum_{i=1}^m \lambda_i = 0$. This suggests that every $\lambda\in\Lambda/\Lambda^{\prime}$ can be uniquely expressed as the summation of the primary sublattice $\lambda_i$, $i=1,2,\cdots,m$. It implies that there exists a map $\pi$:
\begin{align}
&\Lambda_{p_1}/\Lambda^{\prime}_{p_1}\oplus\Lambda_{p_2}/\Lambda^{\prime}_{p_2}\oplus\cdots\oplus\Lambda_{p_m}/\Lambda^{\prime}_{p_m} \notag \\
&\longmapsto \Lambda_{p_1}/\Lambda^{\prime}_{p_1}+\Lambda_{p_2}/\Lambda^{\prime}_{p_2}+\cdots+\Lambda_{p_m}/\Lambda^{\prime}_{p_m} \label{equ:prove2.3}
\end{align}
defined by 
\begin{align}
&\pi(\Lambda_{p_1}/\Lambda^{\prime}_{p_1}\oplus\Lambda_{p_2}/\Lambda^{\prime}_{p_2}\oplus\cdots\oplus\Lambda_{p_m}/\Lambda^{\prime}_{p_m}) \notag \\
=&\Lambda_{p_1}/\Lambda^{\prime}_{p_1}+\Lambda_{p_2}/\Lambda^{\prime}_{p_2}+\cdots+\Lambda_{p_m}/\Lambda^{\prime}_{p_m} \label{equ:prove3.3}
\end{align}
which is an $S$-module isomorphism, and also that
\begin{align}
 \Lambda_{p_i}/\Lambda^{\prime}_{p_i} \cap \sum_{j=1,j\neq i}^m \Lambda_{p_j}/\Lambda^{\prime}_{p_j} = \mathbf{0} \label{equ:prove4.3}
\end{align}
This proves that the sum $\sum_{j=1}^m \Lambda_{p_j}/\Lambda^{\prime}_{p_j} $ is direct, and hence the map $\pi$ is an identity map which belongs to automorphism. Theorem \ref{Theorem.Lattice.Decomposition} is thus proved. \end{IEEEproof}

Theorem \ref{Theorem.Lattice.Decomposition} proves that $\Lambda/\Lambda^{\prime}$ can be decomposed into the direct sum of $m$ sublattices $\Lambda_{p_i}/\Lambda_{p_i}^{\prime}$ (the primary sublattices) which itself forms a new lattice  system.  Hence   $\Lambda/\Lambda^{\prime}$ can be regarded as an $m$ layer quotient lattice. 
 
\begin{theorem}\label{Theorem.Primary.Sub.Iso}
Every primary sublattice $\Lambda_{p_i}/\Lambda^{\prime}_{p_i}$ is isomorphic to a direct sum of cyclic $p_i$-torsion modules:
\begin{align}
\Lambda_{p_i}/\Lambda^{\prime}_{p_i} \cong S/\langle p_i^{\theta_1}\rangle\oplus S/\langle p_i^{\theta_2}\rangle\oplus\cdots\oplus S/\langle p_i^{\theta_t}\rangle \label{equ:theorem3.1}
\end{align} 
for some integers $1\leq\theta_1\leq\theta_2\leq\cdots\leq\theta_t$ which are uniquely determined by $\Lambda_{p_i}/\Lambda^{\prime}_{p_i}$. 
\end{theorem}

\begin{IEEEproof} Theorem \ref{Theorem.Lattice.Decomposition} implies that $\Lambda_{p_i}/\Lambda^{\prime}_{p_i}$ is an $f.g.$ torsion module. Here we write $M_{p_i}= \Lambda_{p_i}/\Lambda^{\prime}_{p_i}$; let $x_1,\cdots,x_f$ be generators for $M_{p_i}$ where $f$ is minimal. This means that $M_{p_i}/Sx_1$ is generated by $f-1$ elements, and hence it is a direct sum of $\leq f-1$ cyclic torsion modules. Thus if $\lambda\in M_{p_i}$ satisfies $p_i^{\theta_t}\lambda = 0$ and $p_i^{\theta_t-1}\lambda\neq 0$, then $M_{p_i}=S\lambda\oplus N\cong (S/\langle p_i^{\theta_t}\rangle)^k \oplus N$, where $N$ is the submodule which is not annihilated by $p_i^{\theta_t}$ although it is annihilated by some other powers of $p_i$. Given this, we have $M_{p_i}/pM_{p_i} \cong (S/\langle p \rangle)^k\oplus N/pN$ (if $S/\langle p \rangle$ exists), the dimension of the second term, $\mathrm{dim} (N/pN) = \mathrm{dim} (M_{p_i}/pM_{p_i}) - k$. It is clear that the dimension is reduced when the power of $p_i$ increases, until the process ends. This proves that $\Lambda_{p_i}/\Lambda^{\prime}_{p_i}$ consists of a direct sum of cyclic torsion modules, and hence must be isomorphic to a direct sum of quotient rings over some powers of $p_i$. This proves the existence of (\ref{equ:theorem3.1}).

There exists a chain $0=p_i^{\theta_t}M_{p_i}\subset\cdots\subset p_i^2M_{p_i}\subset p_iM_{p_i}\subset M_{p_i}$. Consider that $p_i^{\theta-1}M_{p_i}\cong p_i^{\theta-1}S/p_i^{\theta_1}S\oplus\cdots\oplus p_i^{\theta-1}S/p_i^{\theta_t}S \cong S/\langle p_i^{\theta_t-\theta+1}\rangle\oplus\cdots\oplus S/\langle p_i^{\theta_s - \theta+1}\rangle$. This follows from the third isomorphism theorem and that for those $\theta\geq \theta_i$, $p_i^{\theta}(S/\langle p_i^{\theta_i} \rangle) = 0$. Hence $p_i^{\theta-1}M_{p_i}/p_i^{\theta}M_{p_i}\cong (S/\langle p_i \rangle)^k$ forms a vector space over $S/\langle p_i \rangle$ where $k$ is the number of elementary divisors $p_i^{\alpha}$ with $\alpha\geq\theta$. Thus, $\mathrm{dim}(p_i^{\theta-1}M_{p_i}/p_i^{\theta}M_{p_i})$ is the number of elements in (\ref{equ:theorem3.1}) whose $\theta_i\geq \theta$. This proves that the dimension of $p_i^{\theta-1}M_{p_i}/p_i^{\theta}M_{p_i}$ is invariant with $M_{p_i}$, and the number of summands in a particular form $S/\langle p_i^{\theta_i} \rangle$ is uniquely determined by $M_{p_i}$. This proves the uniqueness of (\ref{equ:theorem3.1}). 
\end{IEEEproof}

Theorem \ref{Theorem.Primary.Sub.Iso} implies that the quotient primary $S$-sublattice system $\Lambda_{p_i}/\Lambda_{p_i}^{\prime}$ is isomorphic to a cyclic $p_i$-torsion module. The right-hand side of (\ref{equ:theorem3.1}) can be viewed as the message space of $\Lambda_{p_i}/\Lambda_{p_i}^{\prime}$ which is detailed in Lemma \ref{Lemma.Map.Sub.Lattice}. 

\begin{lemma}\label{Lemma.Map.Sub.Lattice} There exists a map:
\begin{align}
\phi_i: \Lambda_{p_i} \longmapsto  \bigoplus_j S/\langle p_i^{\theta_{j}}\rangle   \label{equ:Lemma.1}
\end{align}
which is a surjective $S$-module homomorphism with kernel $\mathcal{K}(\phi_i)=\Lambda_{p_i}^{\prime}$. To ease the abstract representation, we consider $\Lambda_{p_i}^{\prime} = \Lambda^{\prime}$ in the sequel. Thus, $\mathcal{K}(\phi_i) = \Lambda^{\prime}$ for $i=1,2,\cdots,m$. If the message space is taken as the canonical decomposition of (\ref{equ:theorem3.1}), i.e. $\mathbf{w}^i=\bigoplus_j S/\langle	p_i^{\theta_{j}}\rangle$, there exists a surjective homomorphism $\phi$ and also an injective map $\tilde{\phi}: (\mathbf{w}^1,\cdots,\mathbf{w}^m)\longmapsto \Lambda$ such that 
\begin{align}
\phi(\tilde{\phi}(\mathbf{w}^1\oplus\cdots\oplus\mathbf{w}^m) ) = \mathbf{w}^1\oplus\cdots\oplus\mathbf{w}^m \label{equ:Lemma.2}
\end{align}
\end{lemma}

\begin{IEEEproof} The statement of (\ref{equ:Lemma.1}) follows immediately from Theorem \ref{Theorem.Primary.Sub.Iso} and the first isomorphism theorem. The statement of (\ref{equ:Lemma.2}) follows from Theorem \ref{Theorem.Lattice.Decomposition}, Theorem \ref{Theorem.Primary.Sub.Iso} and the first isomorphism theorem.  
\end{IEEEproof}

\begin{lemma}\label{Lemma.GenMatr.Sub.Lattice}
The generator matrix of the $S$-sublattice $\Lambda_{p_i}$ at the $i^{\mathrm{th}}$ layer can be expressed in the form of:
\begin{align}
\mathbf{G}_{\Lambda_{p_i}} = \begin{bmatrix}
\mathrm{Diag}(\underbrace{\mathbf{p}_1^{\theta_1}\cdots\mathbf{p}_{i-1}^{\theta_{i-1}},\mathbf{I}_{t},\mathbf{p}_{i+1}^{\theta_{i+1}}\cdots\mathbf{p}_m^{\theta_m}}_{k} ) & \mathbf{0} \\
\mathbf{0} & \mathbf{I}_{n-k}
\end{bmatrix} \mathbf{G}_{\Lambda}\label{equ:Lemma2.1}
\end{align} 
and
\begin{align}
\phi_i(\mathbf{w}\mathbf{G}_{\Lambda_{p_i}}) = \big(w^{i,1}+\langle p_i^{\theta_1} \rangle,\cdots,w^{i,t}+\langle p_i^{\theta_t} \rangle \big) \label{equ:Lemma2.2}
\end{align}  
where $w^{i,t}\in S/\langle p_i^{\theta_t} \rangle$ and $\mathbf{w}\in \mathbf{w}^1\oplus\cdots\oplus \mathbf{w}^m$. $\mathbf{G}_{\Lambda}$ is the generator matrix of the fine lattice $\Lambda$, $\mathbf{p}_j^{\theta_j}$, $j=1,2,\cdots,m$ is a vector, with all elements being the same elementary divisor $p_j^{\theta_j}$ over $S$, and $t = \mathrm{dim}(\Lambda_{p_i}/\Lambda^{\prime})$. 
\end{lemma}

\begin{IEEEproof} Every matrix over a PID must have a Smith normal form (SNF) with unique invariant factors up to multiplication by units. This complies with the structure theorem of modules over PID in invariant factor form. Hence there exists an \textit{equivalent} SNF matrix $\mathbf{M}_{\mathrm{SNF}}$ such that $\mathbf{M}_{\mathrm{SNF}}\mathbf{G}_{\Lambda_{p_i}}$ is the generator matrix of the lattice $\tilde{\Lambda}^{\prime}$ which is isomorphic to the kernel $\mathcal{K}(\phi_i) = \Lambda^{\prime}$. Based on the theorems mentioned above and the fact that the invariant factors are uniquely determined, the invariant factors in $\mathbf{M}_{\mathrm{SNF}}$ must be some powers of $p_i$ which naturally satisfies the divisibility relations, and we claim that, now $\tilde{\Lambda}^{\prime} = \Lambda^{\prime}$. The statement of (\ref{equ:Lemma2.2}) follows from Lemma 1. 
\end{IEEEproof}

Lemma \ref{Lemma.GenMatr.Sub.Lattice} shows a way to produce the quotient $S$-sublattice of each layer defined in Theorem \ref{Theorem.Lattice.Decomposition}. $\Lambda_{p_i}/\Lambda^{\prime}$ forms an independent lattice system, and the direct sum of all $\Lambda_{p_i}/\Lambda^{\prime}$, $i=1,2,\cdots,m$ is equal to $\Lambda/\Lambda^{\prime}$.

\subsection{Construction of multilevel lattice network coding} \label{sec:Construct.MLNC}
Based on the Theorems developed in section \ref{sec:Multilevel.Structure}, we show in this subsection the detailed description of the MLNC scheme, and a way of multilevel network decoding (named layered integer forcing), which provides an efficient way of decoding the linear combination of the multi-source messages with greatly reduced complexity. 
\begin{figure*}[t!]
  \centering
\begin{minipage}[t]{0.8\linewidth}
\centering
  \includegraphics[width=1\textwidth]{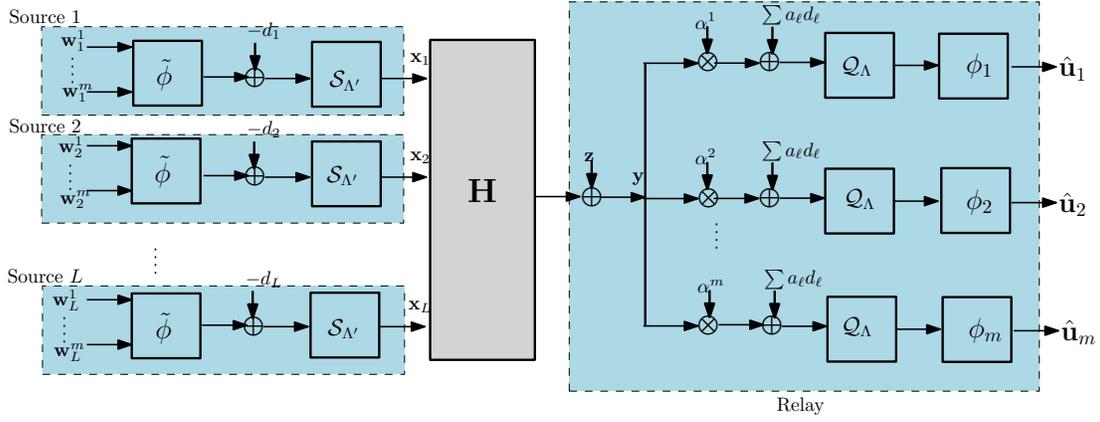}
\end{minipage}
\caption{\small System diagram of the multilevel lattice network coding and multistage decodeing. The righr-hand side of $\mathbf{H}$ represents the decoding for a single relay.} \label{fig:System}
\end{figure*}

\subsubsection*{Traditional Approach} Theorem \ref{Theorem.Lattice.Decomposition}, Theorem \ref{Theorem.Primary.Sub.Iso} and Lemma \ref{Lemma.Map.Sub.Lattice} imply that the message space with large cardinality may be expressed as a set of smaller message spaces over the hybrid finite field and finite chain ring. Fig. \ref{fig:System} depicts a multilevel lattice network coding architecture, with $L$ sources and a single relay. The encoder $\mathcal{E}_{\ell}$ at the $\ell^{\mathrm{th}}$ source maps the original message $\mathbf{w}_{\ell}=\mathbf{w}_{\ell}^{1}\oplus\cdots\oplus\mathbf{w}_{\ell}^{m}$ to a fine lattice point $\Lambda$ (assuming $n$-dimension) via the injective map $\tilde{\phi}$ defined in Lemma 1. Then we add a dither $\boldsymbol d_{\ell}\in\mathbb{C}^n$ which is uniformly  distributed over the fundamental Voronoi region $\mathcal{V}_{\Lambda^{\prime}}$ of $\Lambda^{\prime}$. The dithered lattices pass through a nested shaping operator in order to restrain the power consumption. This operation is performed via the sublattice quantization:
\begin{align}
\lambda_{\ell}^{\prime} = \mathcal{Q}_{\Lambda^{\prime}}(\tilde{\phi}(\mathbf{w}_{\ell}^{1}\oplus\cdots\oplus\mathbf{w}_{\ell}^{m})+\boldsymbol d_{\ell})
\end{align}   
where $\lambda_{\ell}^{\prime}\in\Lambda^{\prime}$, and $\mathcal{Q}_{\Lambda^{\prime}}(\cdot): \mathbb{C}^n \longmapsto \Lambda^{\prime}$ is a coarse lattice quantizer. The output of the $\ell^{\mathrm{th}}$ source is given by:
\begin{align}
\mathbf{x}_{\ell} &= \mathcal{E}_{\ell}(\mathbf{w}_{\ell}^{1}\oplus\cdots\oplus\mathbf{w}_{\ell}^{m}) \notag	\\
&= \tilde{\phi}(\mathbf{w}_{\ell}^{1}\oplus\cdots\oplus\mathbf{w}_{\ell}^{m}) + \boldsymbol d_{\ell} - \lambda_{\ell}^{\prime} \label{equ:xl}
\end{align} 

Note that $\mathbf{x}_{\ell}$ is uniformly distributed over $\mathcal{V}_{\Lambda^{\prime}}$ due to the effect of the dither. The average power of the transmitted signal $\mathbf{x}_{\ell}$ is given by:
\begin{align}
P = \frac{1}{n\mathrm{Vol}(\mathcal{V}_{\Lambda^{\prime}})}\int_{\mathcal{V}_{\Lambda^{\prime}}} \parallel\mathbf{x}_{\ell}\parallel^2\mathrm{d}\mathbf{x_{\ell}}
\end{align}
which is the second moment per dimension of $\mathbf{x}_{\ell}$ over $\mathcal{V}_{\Lambda^{\prime}}$. The message space at each source consists of a direct sum of $m$ small message spaces (assuming there are $m$ levels ) over different finite fields or chain rings. The encoder $\mathcal{E}_{\ell}$ constructs a one-to-one relation between the message space and the coset system $\Lambda/\Lambda^{\prime}$.  

At the relay, given the received signals $\mathbf{y}$ and an $S$-integer vector $\mathbf{\tilde{a}}=[\tilde{a}_{1},\tilde{a}_{2},\cdots,\tilde{a}_{L}]^T \in S^{L}$, the decoder aims at computing a new lattice point which is regarded as an $S$-linear combination of transmitted lattice points from all sources. The homomorphism designed for the coset system will be used for decoding the lattice point to a linear combination of the original messages. We assume in this paper that the fading coefficients $\mathbf{h}=[h_1,h_2,\cdots,h_{L}]$, and dithers are perfectly known at the relay. The decoder can be described, generally, by:
\begin{align}
\mathcal{D}: (\mathbb{C}^n, \mathbb{C}^L, S^L, \mathbb{C},\mathbb{C}^{n\times L}) \longmapsto W \label{equ:Decoder.Traditional}
\end{align}
Thus, the output of $\mathcal{D}(\mathbf{y}|\mathbf{h},\tilde{\mathbf{a}},\alpha, \mathbf{d})$ is the estimates of the linear combination of the original messages of each source. Here $\alpha$ is a scaling factor \cite{Nazer.Gastpar.TIT.2011} which maximises the computation rate. Note that the aforementioned decoder (\ref{equ:Decoder.Traditional}) may vary according to the specific problem. There may be additional information available to the decoder, and the decoder may also have extra outputs. However, basically the core idea for the decoding remains the same. Based on the quotient lattice $\Lambda/\Lambda^{\prime}$, we have:
\begin{align}
\hat{\mathbf{u}} &= \mathcal{D}(\mathbf{y}|\mathbf{h},\mathbf{a},\alpha, \mathbf{d}) \notag \\
&\overset{(a)}{=} \phi\bigg(\mathcal{Q}_{\Lambda}\Big(\alpha\mathbf{y} - \sum_{\ell=1}^L	\tilde{a}_{\ell}\boldsymbol d_{\ell}	\Big) \bigg) \label{equ:Decoder.Tradi.1}\\
&\overset{(b)}{=}  \phi\bigg(\mathcal{Q}_{\Lambda}\Big( \sum_{\ell=1}^{L} \tilde{a}_{\ell} \big(	\tilde{\phi}(\mathbf{w}_{\ell}) - \lambda_{\ell}^{\prime}					\big) +\mathbf{n}_{\mathrm{eff}}	\Big) \bigg) \label{equ:Decoder.Tradi.2}\\ 
&\overset{(c)}{=} \phi\bigg(\sum_{\ell=1}^L\tilde{a}_{\ell}\tilde{\phi}(\mathbf{w}_{\ell}) + \mathcal{Q}_{\Lambda}(\mathbf{n}_{\mathrm{eff}})				\bigg) \label{equ:Decoder.Tradi.3} \\
&\overset{(d)}{=} \bigoplus_{\ell=1}^{L} a_{\ell}\mathbf{w}_{\ell} \boxplus \phi\left(\mathcal{Q}_{\Lambda}(\mathbf{n}_{\mathrm{eff}}) \right) \label{equ:Decoder.Tradi.4}
\end{align}   
where (a) follows from the fact that we expect to quantize a set of scaled received signals which are subtracted from the corresponding dithers. (b) follows from the manipulation of:
\begin{align}
\alpha\mathbf{y} = \sum_{\ell=1}^L \tilde{a}_{\ell}\mathbf{x}_{\ell} + \sum_{\ell=1}^L	\tilde{a}_{\ell}\boldsymbol d_{\ell}+ \overbrace{\sum_{\ell=1}^{L} (\alpha h_{\ell} - \tilde{a}_{\ell} )\mathbf{x}_{\ell} + \alpha \mathbf{z}}^{\mathbf{n}_{\mathrm{eff}}} \label{equ:alpha.y}
\end{align} 
(c) follows from the definition of the lattice quantizer, and (d) follows from the properties of a surjective module homomorphism, and also Lemma 1. Note that here $\phi(\tilde{a}_{\ell}) = a_{\ell}\in \mathbf{w}^1\oplus\cdots\oplus \mathbf{w}^m$. 

Equations (\ref{equ:Decoder.Tradi.1}) - (\ref{equ:Decoder.Tradi.4}) reveal the decoding operations for the traditional lattice-based PNC. We are able to decode a linear combination of messages $\bigoplus_{\ell=1}^{L} a_{\ell}\mathbf{w}_{\ell}$ over all sources without errors provided that $\phi\left(\mathcal{Q}_{\Lambda}(\mathbf{n}_{\mathrm{eff}}) \right) = 0$. Thus, the successful decoding is guaranteed iff the effective noise is quantized to the kernel of $\phi$, $\mathcal{K}(\phi)$.  

The problems left unsolved are:  1. how to exploit rich ring features in order to make it practically applicable in lattice-based network coding. 2. when the cardinality (the coset representatives) of $\Lambda/\Lambda^{\prime}$ is large, the complexity of the lattice quantizer becomes unmanageable, which restricts the application of LNC. What is the practical lattice network decoding approach that could greatly relieves the decoding load in LNC. We study a new decoding solution which is specifically designed in terms of MLNC, and which relaxes the two problems mentioned.

\subsubsection*{Layered Integer Forcing} 
The breakthrough of MLNC (based on Theorems and Lemmas in section \ref{sec:Multilevel.Structure}) is that
\begin{itemize}
\item The original message space over $\Lambda/\Lambda^{\prime}$ can be decomposed into a direct sum of $m$ smaller message spaces in terms of $\Lambda_{p_i}/\Lambda^{\prime}$, $i=1,2,\cdots,m$.
\item The relay can decode each layer independently; thus the decoder tries to infer and forward a linear combination of messages of each layer \textit{separately} over the message subspace defined in Theorem \ref{Theorem.Primary.Sub.Iso}.  
\end{itemize}

Let us recall the traditional decoding operations explained in (\ref{equ:Decoder.Tradi.1}) - (\ref{equ:Decoder.Tradi.4}). If we are only concerned with the linear combination of a particular layer, the quantization of the effective noise need not necessarily be the kernel of $\phi$. There must exist other lattice points in $\Lambda/\Lambda^{\prime}$ such that the homomorphism of these points does not interfere with the linear combination of that layer following the aforementioned theorems. 

\begin{theorem}\label{Theorem.NewLattice.Decoding}
There exists a quotient $S$-lattice $\Lambda/\Lambda_i^{\prime}$ with generator matrices $\mathbf{G}_{\Lambda}$ for $\Lambda$, and $\mathbf{G}_{\Lambda_i^{\prime}}$ for $\Lambda_i^{\prime}$, which satisfies:
\begin{align}
\mathbf{G}_{\Lambda_i^{\prime}} = \begin{bmatrix}
\mathrm{Diag}(\underbrace{\mathbf{I},p_i^{\theta_1},\cdots, p_i^{\theta_t},\mathbf{I} }_{k} ) & \mathbf{0} \\
\mathbf{0} & \mathbf{I}_{n-k}
\end{bmatrix} \mathbf{G}_{\Lambda}\label{equ:Theorem5.1}
\end{align} 
and there is a surjective $S$-module homomorphism $\varphi_i$:
\begin{align}
\varphi_i: \Lambda \longmapsto S/\langle p_i^{\theta_1}\rangle\oplus S/\langle p_i^{\theta_2}\rangle\oplus\cdots\oplus S/\langle p_i^{\theta_t}\rangle \label{equ:Theorem5.2}
\end{align}
whose kernel $\mathcal{K}(\varphi_i)=\Lambda_i^{\prime}$. The quotient $S$-lattice $\Lambda/\Lambda_i^{\prime}$ is isomorphic to the direct sum of cyclic modules:
\begin{align}
\Lambda/\Lambda_i^{\prime} \cong S/\langle p_i^{\theta_1}\rangle\oplus S/\langle p_i^{\theta_2}\rangle\oplus\cdots\oplus S/\langle p_i^{\theta_t}\rangle \label{equ:Theorem5.3}
\end{align}
\end{theorem}

\begin{IEEEproof} By applying the equivalent SNF explained in the proof of Lemma \ref{Lemma.GenMatr.Sub.Lattice}, we can prove (\ref{equ:Theorem5.1}) in similar way. Now we begin the proof of (\ref{equ:Theorem5.2}). The sublattice $\Lambda_i^{\prime}$ can be written as:
\begin{align*}
\Lambda_i^{\prime} = \{\mathbf{w}\mathbf{G}_{\Lambda}: \mathbf{w}^i\in \langle p_i^{\theta_1} \rangle \oplus\cdots\oplus \langle p_i^{\theta_t} \rangle		\}
\end{align*}
in terms of the generator matrix for $\Lambda_{i}^{\prime}$ in (\ref{equ:Theorem5.1}). It is clear that:
\begin{align*}
\varphi_i(\mathbf{w}\mathbf{G}_{\Lambda}) = 0 ~~\mathbf{iff}~~ \mathbf{w}^i\in \langle p_i^{\theta_1} \rangle \oplus\cdots\oplus \langle p_i^{\theta_t} \rangle
\end{align*}  
and hence, the kernel of $\varphi_i$, $\mathcal{K}(\varphi_i)$ must be $\Lambda_i^{\prime}$. It is also obvious from (\ref{equ:Theorem5.2}) that $\varphi_i$ is indeed surjective and $S$-linear. The proof of (\ref{equ:Theorem5.3}) follows immediately from the first isomorphism theorem.
\end{IEEEproof}

Note that although both $\Lambda_{p_i}/\Lambda^{\prime}$ and $\Lambda/\Lambda_i^{\prime} $ are isomorphic to $S/\langle p_i^{\theta_1}\rangle\oplus S/\langle p_i^{\theta_2}\rangle\oplus\cdots\oplus S/\langle p_i^{\theta_t}\rangle$, \textit{they belong to different coset systems}. $\Lambda_{p_i}/\Lambda^{\prime}$ is related to the construction of lattices that have multilevel structure, whereas  $\Lambda/\Lambda_i^{\prime} $ is related to the decoding issues, i.e. LIF. 

Theorem \ref{Theorem.NewLattice.Decoding} defines a new sublattice $\Lambda_i^{\prime}$ which plays a key role in decoding MLNC, as it is the kernel of the quotient $S$-lattice that possesses a surjective homomorphism $\varphi_i$ for the $i^{\mathrm{th}}$ layer. Hence it is possible to decode an $S$-linear combination of fine lattice points to an $S$-linear combination of the original messages of the $i^{\mathrm{th}}$ layer. This is explained  in Lemma 3.    

\begin{lemma} \label{Lemma.NewLattice.Map} 
Given the embedding injective map $\tilde{\phi}:(\mathbf{w}^1,\cdots,\mathbf{w}^m)\longmapsto \Lambda$, there exists a surjective $S$-module homomorphism $\varphi_i$, $i=1,2,\cdots,m$, defined in (\ref{equ:Theorem5.2}), satisfying:
\begin{equation}\label{equ:Lemma3}
\varphi_i\big(	\tilde{\phi}(\mathbf{w}^1\oplus\cdots\oplus\mathbf{w}^m)	\big) \\
= 
\begin{cases}
\mathbf{w}^i 
,\mathbf{w}^i\notin \langle p_i^{\theta_1} \rangle \oplus\cdots\oplus \langle p_i^{\theta_t} \rangle \\ 
 0 
,~~ \mathbf{w}^i\in \langle p_i^{\theta_1} \rangle \oplus\cdots\oplus \langle p_i^{\theta_t} \rangle  
\end{cases} 
\end{equation}  
\end{lemma}

\begin{IEEEproof} The injective mapping $\tilde{\phi}$ is an inverse operation of the homomorphism $\phi$ defined in terms of the quotient $S$-lattice $\Lambda/\Lambda^{\prime}$, which maps the messages into a lattice point $\Lambda$, as explained in Lemma 1. Following the second statement of Theorem \ref{Theorem.NewLattice.Decoding}, the $S$-module homomorphism $\varphi_i$ of the $i^{\mathrm{th}}$ layer indeed maps the lattice point $\Lambda$  to the message subspace. When $\mathbf{w}^i\in \langle p_i^{\theta_1} \rangle \oplus\cdots\oplus \langle p_i^{\theta_t} \rangle$, $\tilde{\phi}(\mathbf{w}^1\oplus\cdots\oplus\mathbf{w}^m)$ maps to the lattice point $\Lambda_i^{\prime}$ which is the kernel of $\varphi_i$. Hence, according to (\ref{equ:Theorem5.1}), the linear labelling of the new coset system in $\Lambda/\Lambda_i^{\prime}$  coincides with the labelling of the $i^{\mathrm{th}}$ layer of $\tilde{\phi}$. This proves Lemma 3. 
\end{IEEEproof}   

Based on Lemma 3, it is now possible to decode the linear combination of the messages of each layer separately and independently. Assuming the messages at the $i^{\mathrm{th}}$ layer is of interest, the relay computes:
\begin{align}
\hat{\mathbf{u}}^i &= \mathcal{D}^{i}(\mathbf{y}|\mathbf{h},\mathbf{a}^i,\alpha^i, \mathbf{d}) \\
&= \varphi_i\bigg(\mathcal{Q}_{\Lambda}\Big(\alpha^i\mathbf{y} - \sum_{\ell=1}^L	\tilde{a}_{\ell}^i\boldsymbol d_{\ell}	\Big) \bigg)
\end{align} 
where 
\begin{align}
\mathcal{D}^i: (\mathbb{C}^n, \mathbb{C}^L, S^L, \mathbb{C},\mathbb{C}^{n\times L}) \longmapsto W^i
\end{align}
and $\alpha^i\in\mathbb{C}$ and $\mathbf{a}^i$ are scaling parameter and $S$-integer coefficients of the $i^{\mathrm{th}}$ layer, respectively, which are determined by some optimisation criterion in terms of the quotient $S$-lattice $\Lambda/\Lambda_i^{\prime}$.  

Theorem \ref{Theorem.NewLattice.Decoding} and Lemma \ref{Lemma.NewLattice.Map} lay the foundation of the layered integer forcing. The linear combination of $\hat{\mathbf{u}}^i$ can be recovered in terms of LIF by:
\begin{align}
\hat{\mathbf{u}}^i &\overset{(d)}{=}  \varphi_i\bigg(\mathcal{Q}_{\Lambda}\Big( \sum_{\ell=1}^{L} \tilde{a}_{\ell}^i \big(	\tilde{\phi}(\mathbf{w}^1_{\ell}\oplus\cdots\oplus\mathbf{w}^m_{\ell}) - \lambda_{\ell}^{\prime}					\big) +\mathbf{n}_{\mathrm{eff}}	\Big) \bigg)  \notag \\
&\overset{(e)}{=} \varphi_i\bigg(\sum_{\ell=1}^L\tilde{a}_{\ell}^i\tilde{\phi}(\mathbf{w}^1_{\ell}\oplus\cdots\oplus\mathbf{w}^m_{\ell}) - \lambda_{\ell}^{\prime} - \lambda_{i,\ell}^{\prime} + \mathcal{Q}_{\Lambda}(\mathbf{n}_{\mathrm{eff}})	\bigg) \notag \\
&\overset{(f)}{=} \varphi_i\bigg(\sum_{\ell=1}^L\tilde{a}_{\ell}^i\tilde{\phi}(\mathbf{w}^1_{\ell}\oplus\cdots\oplus\mathbf{w}^m_{\ell}) \bigg) \boxplus \varphi_i\bigg( \mathcal{Q}_{\Lambda}(\mathbf{n}_{\mathrm{eff}})	\bigg) \notag\\
&\overset{(g)}{=} \bigoplus_{\ell=1}^{L}a_{\ell}^i\mathbf{w}^i_{\ell} \boxplus \varphi_i\bigg(\mathcal{Q}_{\Lambda}(\mathbf{n}_{\mathrm{eff}}) \bigg) \label{equ:layer.operation}
\end{align} 
where (d) follows from (\ref{equ:xl}) and basic arithmetic manipulations; (e) follows from the definition of the lattice quantizer $\mathcal{Q}_{\Lambda}$, and also the $S$-linear combination of the lattice points is restricted in $\mathcal{V}_{\Lambda_i^{\prime}}$; (f) follows from the property of a surjective $S$-module homomorphism, and also the fact that $\lambda^{\prime} \subseteq\lambda_{i}^{\prime}$ and $\mathcal{K}(\varphi_i) = \lambda_{i}^{\prime}$. (g) follows from Lemma 3, and note that $\varphi_i(\tilde{a}_{\ell}^i) = a_{\ell}^i\in W^i$. 

\begin{lemma} \label{Lemma.Comb.Recover.Prob}
The linear combination of the messages at the $i^{\mathrm{th}}$ layer $\hat{\mathbf{u}}^i = \bigoplus_{\ell=1}^{L}a_{\ell}^i\mathbf{w}^i_{\ell}$ can be recovered iff $\mathcal{Q}_{\Lambda}(\mathbf{n}_{\mathrm{eff}})\in\Lambda_i^{\prime}$. Thus, $\mathrm{Pr}(\hat{\mathbf{u}}^i\neq\mathbf{u}^i) = \mathrm{Pr}(\mathcal{Q}_{\Lambda}(\mathbf{n}_{\mathrm{eff}})\notin \Lambda^{\prime}_i )$.
\end{lemma}

\begin{IEEEproof} Following (\ref{equ:layer.operation}), it is clear that $\hat{\mathbf{u}}^i = \bigoplus_{\ell=1}^{L}a_{\ell}^i\mathbf{w}^i_{\ell}$ can be decoded correctly iff $\varphi_i\Big(\mathcal{Q}_{\Lambda}(\mathbf{n}_{\mathrm{eff}}) \Big) = 0$. According to Theorem \ref{Theorem.NewLattice.Decoding}, the kernel of $\varphi_i$, $\mathcal{K}(\varphi_i) = \Lambda_i^{\prime}$; thus the quantization of the effective noise $\mathcal{Q}_{\Lambda}(\mathbf{n}_{\mathrm{eff}}) \in \Lambda_i^{\prime}$. This proves Lemma \ref{Lemma.Comb.Recover.Prob}. 
\end{IEEEproof}    

Lemma \ref{Lemma.Comb.Recover.Prob} reveals that the lattice $\Lambda_i^{\prime}$ defined in Theorem \ref{Theorem.NewLattice.Decoding} plays a key role in decoding the messages of the $i^{\mathrm{th}}$ layer. 

The message space of the traditional C\&F scheme is determined by the size of the lattice partition. Hence, to increase the network throughput, the sublattice $\Lambda^{\prime}$ needs to be more sparse in order to allow the messages to be over a larger field or commutative ring (LNC). In this case, the decoding complexity is normally unaffordable. 

One example is associated with a group of lattice codes directly designed in the Euclidean space, e.g. complex low density lattice codes (CLDLC). It has prohibitive computational complexity when the cardinality of the quotient lattice is too large, since the decoding metrics are continuous functions (a mixture of multiple probability density functions), and the periodic extension that occurs at the variable nodes \cite{Yi.CLDLC.2015} runs over a large $S$-integer set, which seriously increases the overall computational costs over the iterative parametric belief propagation decoding, even if the Gaussian mixture reduction algorithm is employed. 

The $S$-lattices can also be  constructed through the existing channel codes based on some lattice construction approaches (e.g. construction A, D). However, the decoding complexity of the channel codes over a large algebraic field increases rapidly, e.g. a small increase of the memory for convolutional codes gives rise to an exponential increase in the number of trellis states, making the codes eventually undecodable. When the cardinality of the quotient lattices become larger, the decoding complexity for convolutional codes with even small memory is  unmanageable, but the performance is still very poor.

MLNC together with LIF provides a realistic solution to this problem. Being supported by the Theorems and Lemmas in sections \ref{sec:Multilevel.Structure} and \ref{sec:Construct.MLNC},  the quotient $S$-lattice having large cardinality can be decomposed into  some primary quotient $S$-sublattices which have smaller cardinalities. Each primary quotient sublattice forms a layer, and determines the message subspace over this layer. With the aid of the lattices $\Lambda_i^{\prime}$, we can perform multilevel lattice decoding at the relay, where the linear combination of the messages of all sources at each layer can be independently recovered over the message subspace. In this case, the overall computational loads are greatly relaxed.

LNC \cite{Feng.AlgeAppro.TIT.2013} shows the possibility of ring-based linear network coding, extending the traditional linear network coding defined over the finite field to a more general notion. Furthermore, MLNC leads to a  practically feasible encoding and decoding design approach for lattice network coding over commutative rings, thus, with greatly reduced decoding complexity. MLNC inherently gives an appealing solution for this since now we are able to construct multiple layers based on the decomposition theory  mentioned above, with each layer operating over a finite field or chain ring in a new coset system. Note that the elements in a finite chain ring can be uniquely represented by $\nu+1$ elements over a fixed residue field where $\nu$ is the nilpotency index of this finite chain ring. We will introduce this in the subsequent sections.    

\subsection{Achievable Rates and Probability of Error} \label{sec:achievable.rates}
As discussed in section \ref{sec:Construct.MLNC}, the message of the $i^{\mathrm{th}}$ layer corresponds to the decomposed quotient $S$-sublattice $\Lambda_{p_i}/\Lambda^{\prime}$, which should be decoded separately at each layer, based on a new $S$-lattice partition $\Lambda/\Lambda_{i}^{\prime}$. Suppose that each layer is given an $S$-integer coefficient vector $\mathbf{a}^i\in S^L$, and $\mathbf{A}=[{\mathbf{a}^1}|{\mathbf{a}^2}|\cdots|{\mathbf{a}^m}]\in S^{L\times m}$, we can obtain the achievable rate  following Nazer and Gastpar's method, under the assumption of that $S$ is Gaussian integers $\mathbb{Z}[i]$

\begin{theorem}\label{Theorem.Comp.Rate}
Given channel fading vector $\mathbf{h}\in\mathbb{C}^L$, non-zeros $S$-integer coefficient matrix $\mathbf{A}\notin\{\mathbf{0}\}$, and the message subspace $W^i = \big(\mathbb{Z}[i]/\langle p_i\rangle\big)^k$, the probability of decoding error $\mathrm{Pr}(\hat{\mathbf{u}}^i\neq\mathbf{u}^i|\mathbf{h},\mathbf{A})$ can be arbitrarily small if the overall message rate $\mathcal{R}$ satisfies:
\begin{align}
\mathcal{R}<\mathcal{R}(\mathbf{h},\mathbf{A}) = \sum_i^m\log_2\Bigg( \bigg( \parallel \mathbf{a}^i\parallel^2 - \frac{P^i|\mathbf{h}^\dagger\mathbf{a}^i |^2}{1+P^i \parallel \mathbf{h}\parallel^2 } \bigg)^{-1}	\Bigg) \label{equ:comp.bound}
\end{align}
for sufficiently large lattice dimension $n$ and prime factor $p_i$. $P^{(i)}$ is defined by
\begin{align}
P^{(i)} = \frac{1}{n\mathrm{Vol}(\mathcal{V}_{\Lambda/\Lambda_i^{\prime}})}\int_{\mathcal{V}_{\Lambda/\Lambda_i^{\prime}}} \parallel\mathbf{x}_{\ell}\parallel^2\mathrm{d}\mathbf{x_{\ell}}
\end{align}
\end{theorem}

\begin{IEEEproof} Suppose there are $m$ layers, we can construct a quotient $\mathbb{Z}[i]$-lattice $\Lambda/\Lambda_i^{\prime}$ which is isomorphic to the message subspace $W^i$. The computation rate of each layer follows from Nazer and Gastpar's method in \cite{Nazer.Gastpar.TIT.2011}. Since each layer is decoded independently, the sum of computation rate of all layers is the overall achievable rate. 
\end{IEEEproof}

Recall Lemma 4, the error probability of decoding a linear combination $\mathbf{u}$ in terms of $\mathbf{a}^i$ for the $i^{\mathrm{th}}$ level is equal to the probability of $\mathrm{Pr}(\mathcal{Q}_{\Lambda}(\mathbf{n}_{\mathrm{eff}})\notin \Lambda^{\prime}_i )$. The union bound of the error probability for MLNC is given by:

\begin{theorem}\label{Theorem.Eroor.Prob}
Given $\mathbf{h}\in\mathbb{C}^L$, non-zeros $S$-integer coefficient matrix $\mathbf{A}\notin\{\mathbf{0}\}$, and the optimal scaling factor $\alpha_\mathrm{opt}$, the union bound of the error probability in decoding the linear combinations of all levels in MLNC is given by:
\begin{align}
&\mathrm{Pr}\bigg(\hat{\mathbf{u}}\neq\mathbf{u}|\mathbf{h},\mathbf{A},\alpha_{\mathrm{opt}}\bigg) \notag \\
= & \boldsymbol E_{p(\mathcal{Z})}\bigg[\mathrm{Pr}(\hat{\mathbf{u}}^i\neq\mathbf{u}^i|\mathbf{h},\mathbf{A},\alpha_{\mathrm{opt}})\bigg] \notag \\
\lessapprox & \boldsymbol E_{p(\mathcal{Z})}\Bigg[ \mathcal{N}(\Lambda/\Lambda_i^{\prime}) \exp{\Big( \frac{-d^2(\Lambda / \Lambda_i^{\prime})}{4(N_0|\alpha_{\mathrm{opt}}|^2+ P^i||\alpha_{\mathrm{opt}}\mathbf{h} - \mathbf{a}^i ||^2) }\Big)	}	\Bigg] \label{equ:error.prob}
\end{align} 
where $\mathcal{Z}$ is a random variable with its outcomes taking on $\{r=\frac{\mathrm{dim}(\mathbf{u}^i)}{\mathrm{dim}(\mathbf{u})}| i=1,2,\cdots,m \}$.
\end{theorem}

\begin{IEEEproof} At the $i^{\mathrm{th}}$ layer, the decoding operates over the lattice partition of $\Lambda/\Lambda_i^{\prime} = \{\Lambda\setminus\Lambda_i^{\prime}\}\cup \{\mathbf{0}\}$. $d(\Lambda/\Lambda_i^{\prime})$ is the minimum inter-coset distance of the lattice partition $\Lambda/\Lambda_i^{\prime}$ defined by
\begin{align*}
d(\Lambda/\Lambda_i^{\prime}) = \min ||\boldsymbol\vartheta_1 - \boldsymbol\vartheta_2||^2, \boldsymbol\vartheta_1,\boldsymbol\vartheta_2\in\Lambda/\Lambda_i^{\prime}, \boldsymbol\vartheta_1\neq\boldsymbol\vartheta_2
\end{align*}
We denote $\mathcal{N}(\Lambda/\Lambda_i^{\prime})$ as the number of $d(\Lambda/\Lambda_i^{\prime})$ in the $i^{\mathrm{th}}$ layer coset system. Following the steps in  \cite{Feng.AlgeAppro.TIT.2013}, we can prove that the probability error of effective noise quantization is bounded by the probability of effective noise which is not within the Voronoi region $\mathcal{V}_{\mathbf{0}}$:
\begin{align*}
\mathrm{Pr}\Big(\mathcal{Q}_{\Lambda}(\mathbf{n}_{\mathrm{eff}})\notin \Lambda^{\prime}_i|\mathbf{h},\mathbf{a}^i,\alpha_{\mathrm{opt}} \Big)\leq \mathrm{Pr}\Big(\mathbf{n}_{\mathrm{eff}}\notin \mathcal{V}_{\mathbf{0}}^i|\mathbf{h},\mathbf{a}^i,\alpha_{\mathrm{opt}} \Big)
\end{align*}
with 
\begin{align*}
\mathcal{V}_{\mathbf{0}}^i = \{\vartheta\in\mathbb{C}^n: ||\vartheta-\mathbf{0}||^2\leq||\vartheta-\lambda||^2,\forall\lambda\in	\Lambda\setminus\Lambda_i^{\prime}		\}
\end{align*}
The probability of $\mathrm{Pr}(\mathbf{n}_{\mathrm{eff}}\notin \mathcal{V}_{\mathbf{0}}^i \mathbf{h},\mathbf{a}^i,\alpha_{\mathrm{opt}})$ is upper bounded by the term within the bracket of (\ref{equ:error.prob}). The proof closely follows from the method given in \cite{Feng.AlgeAppro.TIT.2013}, based on the Chernoff inequality, the moment generating function of a complex Gaussian random vector, and hypercube Voronoi region $\Lambda_i^{\prime}$. We refer to \cite{Feng.AlgeAppro.TIT.2013} for the detailed proof, and also \cite{Qifu.Eisenstein} for the proof under Eisenstein integers. Since each layer decodes the linear combination independently, the average error probability is the expectation of $\mathrm{Pr}(\hat{\mathbf{u}}^i\neq\mathbf{u}^i|\mathbf{h},\mathbf{A},\alpha_{\mathrm{opt}})$ over the probability function $p(\mathcal{Z})$. According to Lemma 3, we know that the probability $\mathrm{Pr}(\hat{\mathbf{u}}^i\neq\mathbf{u}^i|\mathbf{h},\mathbf{A},\alpha_{\mathrm{opt}})\leq \mathrm{Pr}(\mathbf{n}_{\mathrm{eff}}\notin \mathcal{V}_{\mathbf{0}}^i|\mathbf{h},\mathbf{a}^i,\alpha_{\mathrm{opt}} )$; this gives (\ref{equ:error.prob}).
\end{IEEEproof}

One way to design the homomorphism of $\Lambda/\Lambda_i^{\prime}$ at the $i^{\mathrm{th}}$ layer is implied in Theorem \ref{Theorem.Eroor.Prob}. Thus, $\mathcal{N}(\Lambda/\Lambda_i^{\prime})$ should be minimised and $d(\Lambda/\Lambda_i^{\prime})$ is maximized such that the probability of error is as small as possible at the $i^{\mathrm{th}}$ layer. It is clear that MLNC has good flexibility in the design of the homomorphism, which determines the achievable rate at some levels.

\section{Elementary Divisor Construction}\label{sec:EDC.Overall}
In this section, we study a new lattice construction approach, based on the Theorems and Lemmas developed in section \ref{sec:Multilevel.Structure}. 

\begin{lemma} \label{Lemma.Sublattice.Decomp}
Let $\Lambda$ and $\Lambda^{\prime}$ be $S$-lattices and $S$-sublattices, $\Lambda^{\prime}\subseteq\Lambda$, $|\Lambda:\Lambda^{\prime}|<\infty$ such that $\Lambda/\Lambda^{\prime}$ has a nonzero annihilator $\varpi$ which can be uniquely factorised into distinct powers of primes in $S$, $\varpi=\mathcal{U}(S)p_1^{\gamma_1}p_2^{\gamma_2}\cdots p_m^{\gamma_m}$. Then $\Lambda/\Lambda^{\prime}$ is the direct sum of a finite number of quotient sublattices, $\Lambda_{p_i}/\Lambda^{\prime}=\{\lambda\in\Lambda/\Lambda^{\prime}:p_i^{\gamma_i}\lambda	=0\}$, $i=1,2,\cdots,m$, and given by,
\begin{align}
\Lambda/\Lambda^{\prime} &= \Lambda_{p_1}/\Lambda^{\prime}\oplus\Lambda_{p_2}/\Lambda^{\prime}\oplus\cdots\oplus\Lambda_{p_m}/\Lambda^{\prime} \label{equ:Layered.Sublattices.1} 
\end{align}
\end{lemma}

\begin{IEEEproof} Lemma 5 is a special case of Theorem \ref{Theorem.Lattice.Decomposition} where the annihilator of $\Lambda/\Lambda^{\prime}$ is a single $S$-integer. Therefore $\Lambda/\Lambda^{\prime}$  must be the direct sum of some new quotient $S$-lattices. The annihilator of the  $\Lambda_{p_i}/\Lambda^{\prime}$ is precisely $p_i^{\gamma_i}$. 
\end{IEEEproof}



\subsection{Elementary Divisor Construction} \label{sec:EDC}
We outline a possible lattice construction solution based on Lemma 5 and the statements in section \ref{sec:Multilevel.Structure}.

\textit{Elementary Divisor Construction (EDC):} Let $p_1,p_2,\cdots,p_m$ be some distinct primes in a PID $S$, and $\varpi=\mathcal{U}(S) p_1^{\gamma_1}p_2^{\gamma_2}\cdots p_m^{\gamma_m}$ is a unique factorisation, $\gamma_i\geq 1$.  Let $\mathcal{C}^1, \mathcal{C}^2, \cdots, \mathcal{C}^m$ be $m$ $[n,k_i]$ linear codes over $S/\langle p_1^{\gamma_1} \rangle, S/\langle p_m^{\gamma_m} \rangle, \cdots,S/\langle p_m^{\gamma_m} \rangle$, respectively. The elementary divisor construction lattice is defined by:
\begin{align}
\Lambda \triangleq \{	\lambda\in S^n : \tilde{\sigma}(\lambda)\in \mathcal{C}^1\oplus \mathcal{C}^2\oplus \cdots\oplus\mathcal{C}^m \} \label{equ:EDC.Definition}
\end{align}     
and the sublattice is:
$$\Lambda^{\prime} \triangleq \{\varpi\lambda : \lambda\in S^n	\}$$  
where $\tilde{\sigma}: S^n\longmapsto (S/\langle p_1^{\gamma_1}\rangle)^n\oplus (S/\langle p_2^{\gamma_2}\rangle)^n\oplus\cdots\oplus (S/\langle p_m^{\gamma_m}\rangle)^n$ is a natural map obtained by extending the ring homomorphism $\sigma: S \longmapsto S/\langle p_1^{\gamma_1}\rangle\times S/\langle p_2^{\gamma_2}\rangle)\times\cdots\times S/\langle p_m^{\gamma_m}\rangle$ to multiple dimensions. Apparently $\Lambda^{\prime}\subseteq \Lambda$. The message space under EDC is 
\begin{align}
W= (S/\langle p_1^{\gamma_1}\rangle)^{k_1}\oplus\cdots\oplus (S/\langle p_m^{\gamma_m}\rangle)^{k_m} \label{equ:Message.Space.EDC}
\end{align}
where $k_i$ is the message length of the $i^{\mathrm{th}}$ layer which sums up to $k=\sum_{j=1}^m k_j$.  

The elementary divisor construction is a straightforward extension of Lemma 5, which defines a class of lattices constructed by $m$ linear codes, with each operating over either a finite field or a finite chain ring. Hence the quotient $\Lambda/\Lambda^{\prime}$ must consist of $m$ primary sublattices $\Lambda_{p_i}/\Lambda^{\prime}$, with each constructed by the $i^{\mathrm{th}}$ linear code. The primary sublattices $\Lambda_{p_i}$ of the $i^{\mathrm{th}}$ layer is defined by: 
\begin{align}
\Lambda_{p_i} \triangleq \{\lambda_{p_i}\in \delta_i S: \tilde{\sigma}_i(\lambda_{p_i}) \in C^i )  \} \label{equ:EDC.Primary.Lattice.Definition}
\end{align}
where $\tilde{\sigma}_i$ is a natural map:
\begin{align}
\tilde{\sigma}_i: (\delta_iS)^n \longmapsto (\delta_iS/p_i^{\gamma_i}\delta_iS)^n \cong (S/\langle p_i^{\gamma_i}   \rangle)^{k_i} \label{equ:EDC.Primary.Map}
\end{align}
obtained by extending the ring homomorphism $\sigma_i: \delta_i S \longmapsto \delta_iS/ \langle p_i^{\gamma_i}\delta_iS\rangle$ to multiple dimensions. The scaling factor $\delta_i = \frac{\varpi}{p_i^{\gamma_i}}$  can be proved in terms of the proof in Theorem \ref{Theorem.Lattice.Decomposition}. 

We consider three scenarios based on different algebraic fields which the linear codes may belong to. 
\begin{figure*}[!t]
\centering
  \subfigure[Layer 1]{
    \label{subfig:F3F4.Label.1} 
    \begin{minipage}[b]{0.5\textwidth}
      \centering
      \includegraphics[width=1\textwidth]{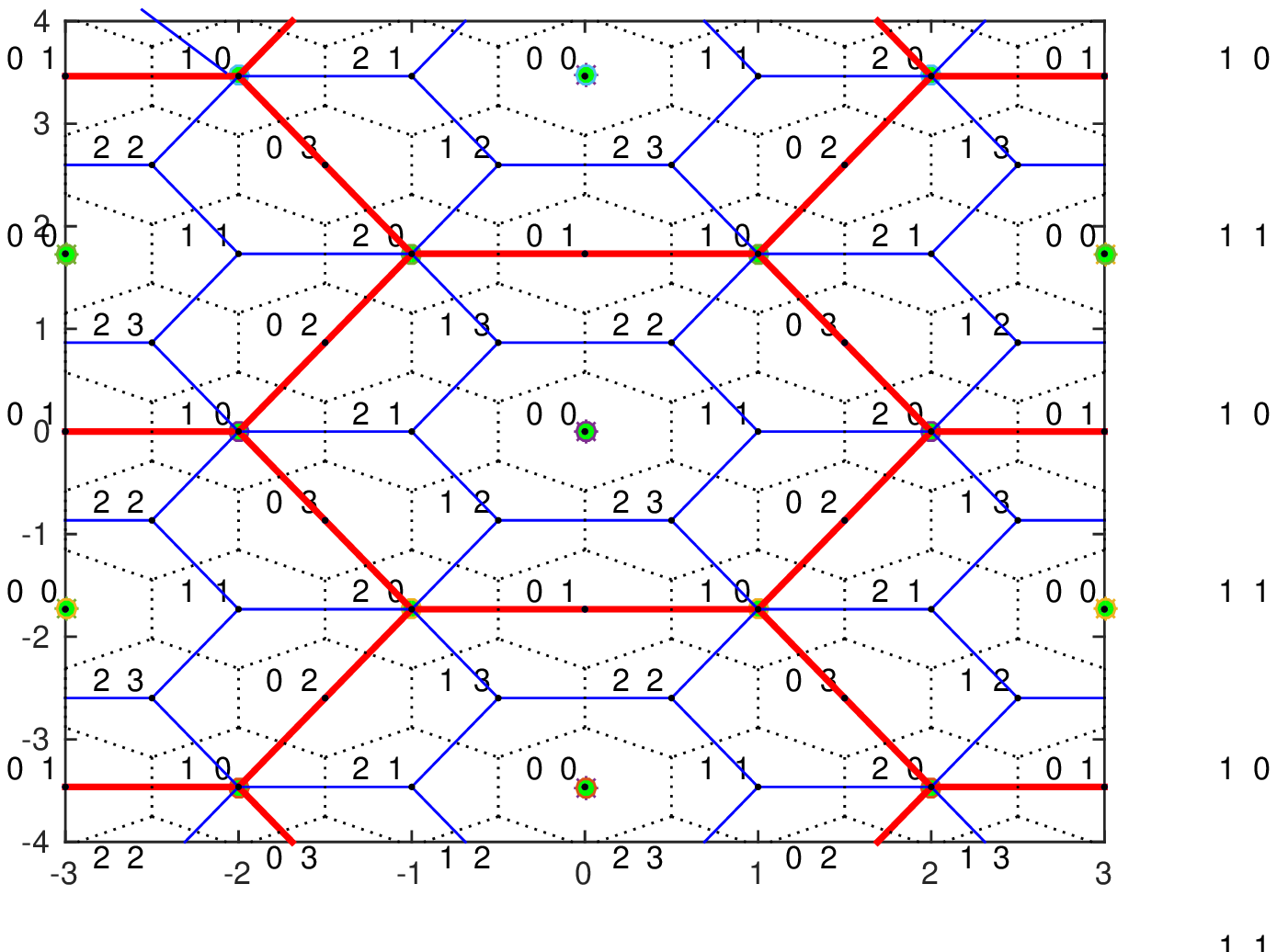}
    \end{minipage}}%
\centering
  \subfigure[Layer 2]{
    \label{subfig:F3F4.Label.2} 
    \begin{minipage}[b]{0.5\textwidth}
      \centering
      \includegraphics[width=1\textwidth]{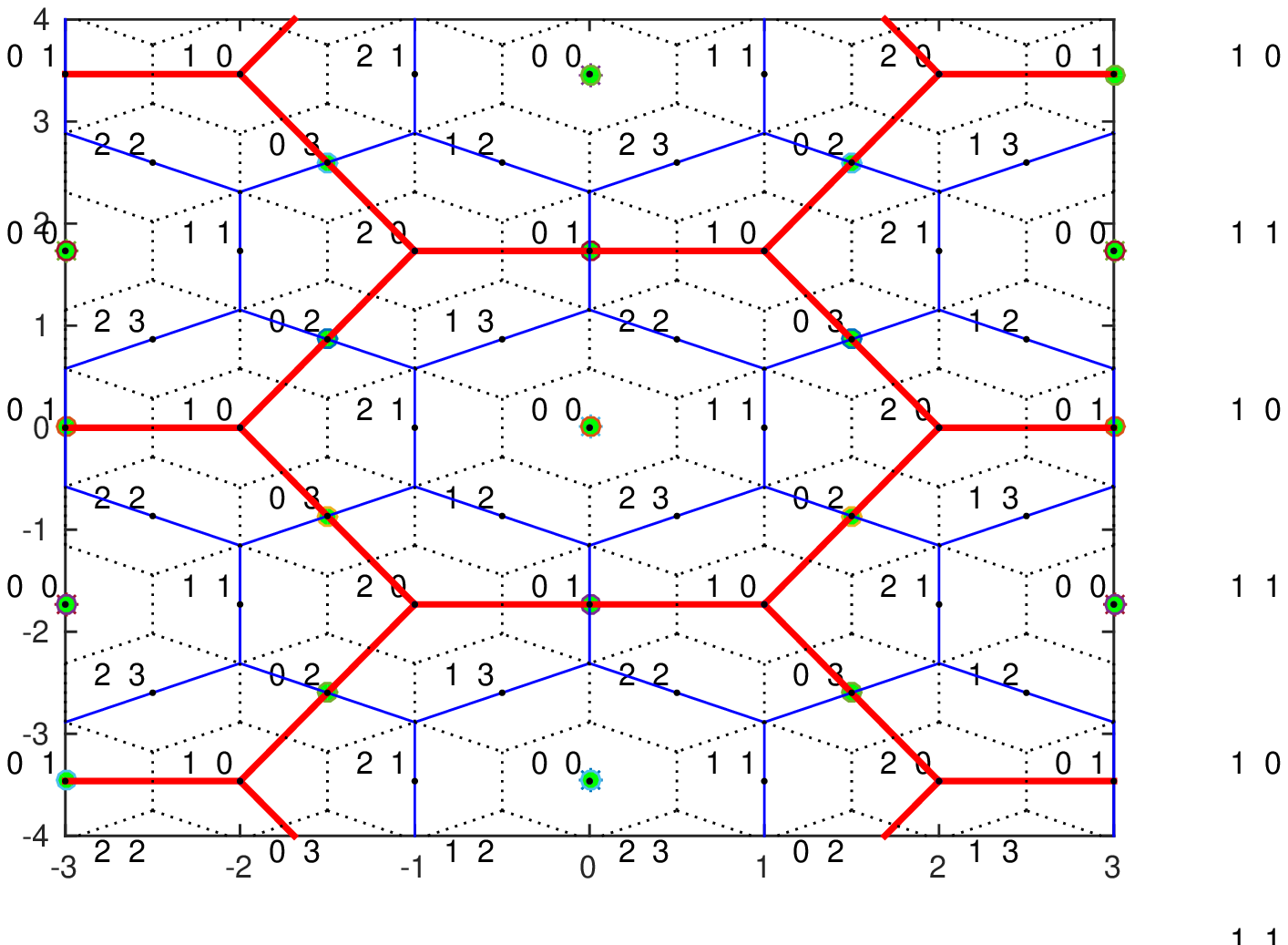}
    \end{minipage}}
  \caption{\small Layer structure of a 2-layer EDC lattice. The green points and blue lines represent the primary sublattices and Voronoi region of $\mathcal{V}_{\Lambda_i^{\prime}}$ for the corresponding layers, respectively. Dotted lines represent the Voronoi region of the fine lattice.}
  \label{fig:F3F4.Label} 
\end{figure*}

\subsubsection*{\textbf{Scenario 1}} Assume that the primary sublattice at each layer is constructed by a linear code over a finite field, thus, $\gamma_1=\gamma_2=\cdots=\gamma_m=1$. Then, $\mathcal{C}^i\in (\delta_iS/\langle p_i\delta_i\rangle)^{n}$. Since the coarse lattice $\Lambda^{\prime}$ is generated by a single element $\varpi$,  $\Lambda/\Lambda^{\prime}$ forms a cyclic torsion module which allows us to produce the generator matrix of the $i^{\mathrm{th}}$ layer lattice $\Lambda_{p_i}$. It will have a form described in Lemma \ref{Lemma.GenMatr.Sub.Lattice}, given by:
\begin{align}
\mathbf{G}_{\Lambda_{p_i}} = \begin{bmatrix} 
\mathrm{Diag}\left(\mathbf{p}_1^{(k_1)}\cdots \mathbf{p}_{i-1}^{(k_{i-1})}, \mathbf{I}_{k_i}, \mathbf{p}_{i+1}^{(k_{i+1})}\cdots \mathbf{p}_m^{(k_m)}\right) & \mathbf{0} \\
\mathbf{0} & \mathbf{I}_{mn-k}
\end{bmatrix} \mathbf{G}  \label{equ:sublattice.Generator}
\end{align}
where $\mathbf{p}_i^{(k_i)}$ is a length-$k_i$ vector with each element $p_i$. 
$\mathbf{G}_{\Lambda_{p_i}}$ in (\ref{equ:sublattice.Generator}) gives the generator matrix for the $i^{\mathrm{th}}$ layer lattices, when the message input 
\begin{align}
\mathbf{w}=[\mathbf{w}^1,\mathbf{w}^2,\cdots, \mathbf{w}^m, \underbrace{\tilde{d}_1\cdots\tilde{d}_m}_{mn-k}]
\end{align} 
where $\mathbf{w}^i\in (\delta_iS/\langle p_i\delta_i \rangle)^{k_i}$, $\tilde{d}_i\in S^{n-k_i}$.

Since EDC lattices are constructed by some linear codes, the matrix $\mathbf{G}$ must include the generator matrix of each linear code $\mathcal{C}^i$. Let $\tilde{\sigma}_i([\mathbf{I}_{k_i}~~ \mathbf{B}^i_{k_i \times (n-k_i)}])$ be a generator matrix for a linear code $\mathcal{C}^i$ (without loss of generality, we consider that the linear code is systematic in this case.), then $\mathbf{G}$ is an $n\times n$ matrix defined below, 
\begin{align}
\mathbf{G} = \begin{bmatrix}
\mathbf{I}_{k_1} &  \mathbf{B}_{k_1 \times (n-k_1)}^1  \\
\mathbf{I}_{k_2} &  \mathbf{B}_{k_2 \times (n-k_2)}^2  \\
\vdots	& \vdots\\
\mathbf{I}_{k_m} &  \mathbf{B}_{k_m \times (n-k_m)}^m  \\
\mathbf{0}  & \varpi\mathbf{I}_{n-k_1}  \\
\vdots	& \vdots\\
\mathbf{0}  & \varpi\mathbf{I}_{n-k_m}
\end{bmatrix} \label{equ:G.Generator}
\end{align} 

Equation (\ref{equ:G.Generator}) follows from Lemma 5 and part of the proof of Theorem \ref{Theorem.Lattice.Decomposition} (i.e. (\ref{equ:prove3.3})). The generator matrix of the coarse lattice $\Lambda^{\prime}$ is therefore given by,
\begin{align}
\mathbf{G}_{\Lambda^{\prime}} = \begin{bmatrix}
\mathrm{Diag}\left(\mathbf{I}_{\sum_{j=1}^{i-1}k_j}, \mathbf{p}_i^{(k_i)},\mathbf{I}_{\sum_{j=i+1}^{m}k_j}  \right) & \mathbf{0} \\
\mathbf{0} & \mathbf{I}_{mn-k}
\end{bmatrix} \mathbf{G}_{\Lambda_{p_i}}\label{equ:Coarse.Generator.Case1}
\end{align}  


It can be easily observed that these generator matrices are consistent with the Theorems and Lemmas proposed in section \ref{sec:Multilevel.Lattice.Network.Coding}. Note that the generator matrix for linear code $\mathcal{C}^i$ is $\tilde{\sigma}_i([\mathbf{I}_{k_i}~~ \mathbf{B}^i_{k_i \times (n-k_i)}])$ where $\tilde{\sigma}_i$ is defined in (\ref{equ:EDC.Primary.Map}). Theorem \ref{Theorem.NewLattice.Decoding} establishes the theoretic fundamental for low-complexity lattice decoding (i.e. LIF) of MLNC, and states that     there exists a surjective $S$-module homomorphism $\varphi_i$ which satisfies Lemma \ref{Lemma.NewLattice.Map}, with kernel  $\mathcal{K}(\varphi_i)=\Lambda_i^{\prime}$, which plays a key role in decoding the $i^{\mathrm{th}}$ layer linearly combined messages. Its generator matrix has a form:
\begin{align}
\mathbf{G}_{\Lambda_i^{\prime}} = \begin{bmatrix}
\mathrm{Diag}(\underbrace{\mathbf{I}, \mathbf{p}_i^{(k_i)} ,\mathbf{I} }_{k} ) & \mathbf{0} \\
\mathbf{0} & \mathbf{I}_{mn-k}
\end{bmatrix} \mathbf{G}\label{equ:attice.Generator}
\end{align} 

We can easily verify $\Lambda/\Lambda_i^{\prime}\cong (S/\langle p_i\rangle)^{k_i}$ in terms of these generator matrices.  
 
\subsubsection*{\textbf{Scenario 2}} When $\forall i =1,2,\cdots,m$, $\gamma_i\neq 1$, the primary sublattice $\Lambda_{p_i}$ at each layer is constructed by a linear code over a finite chain ring $T = \delta_iS/\langle p_i^{\gamma_i}\delta_i\rangle$ \cite{Feng.FCR.TIT}. A finite chain ring is a finite local principal ideal ring, and the most remarkable characteristic of a finite chain ring is that its every ideal (including $\langle 0 \rangle$) is generated by the maximal ideal, which can be linearly ordered by inclusion, and hence, forms a chain. The finite chain ring $T$ has a unique maximal ideal and hence the resultant residue field is $Q=\delta_iS/\langle p_i\delta_i\rangle$ with size $q=|\delta_iS/\langle p_i\delta_i\rangle|$. The chain length of the ideals is indeed the nil-potency index of $p_i$ which is, in this case $\gamma_i$. We refer to $T$ a $(q,\gamma_i)$ chain ring. 

At the $i^{\mathrm{th}}$ layer, the generator matrix $\mathbf{G}_{\mathrm{FCR}}^i$ of a linear code over $T$ has a standard form given in (\ref{equ:FCR.Generator}), where $\mathbf{I}_{k_{i,t}^{\prime}}$ denotes an identity matrix with dimension ${k_{i,t}^{\prime}}$,\footnote{Here, the index $i$ used in ${k_{i,t}^{\prime}}$ is the indicator of layer.} $i=1,2,\cdots,m$ and $t=0,1,\cdots,\gamma_i-1$. Hence $\mathbf{G}_{\mathrm{FCR}}^i$ has a dimension $k^{\prime}_i\times n$ where $k^{\prime}_i = \sum_{t=0}^{\gamma_i-1}k_{i,t}^{\prime}$. Here $Z_{t,l}$, $l=t+1,2,\cdots,\gamma_i$, denotes a ${k_{i,t}^{\prime}}\times {k_{i,t+1}^{\prime}}$ ($k_{i,\gamma_i}^{\prime}=n-k_i^{\prime}$) matrix which is unique modulo $p_i^{\gamma_i-t}$ \cite{Finite.Rings.Identity}. In (\ref{equ:FCR.Generator}), $\boldsymbol{\mathcal{I}}^{*}_{p_i^{\gamma_i}}$ is an upper triangular matrix with dimension $k_i^{\prime}\times k_i^{\prime}$, and $\boldsymbol{\mathcal{B}}_{k_i^{\prime},n-k_i^{\prime}}$ has a dimension of $k_i^{\prime}\times (n-k_i^{\prime})$.  Note that the codeword is row spanned by $\mathbf{G}_{\mathrm{FCR}}^i$ and all rows of $\mathbf{G}_{\mathrm{FCR}}^i$ are linearly independent.   

\begin{figure*}[!t]
\begin{equation}
\mathbf{G}_{\mathrm{FCR}}^i = \left[ \begin{array}{ccccc|c}
\mathbf{I}_{k_{i,0}^{\prime}} & Z_{0,1}^i                         &Z_{0,2}^i    &\cdots  & Z_{0,\gamma_{i-1}}^i     & Z_{0,\gamma_i}^i   \\
0                             & p_i\mathbf{I}_{k_{i,1}^{\prime}}  &p_iZ_{1,2}^i &\cdots  &p_iZ_{1,\gamma_{i-1}}^i &p_iZ_{1,\gamma_i}^i    \\
0		 &		0		               &p_i^2\mathbf{I}_{k_{i,2}^{\prime}}   &\cdots  &p_i^2Z_{2,\gamma_{i-1}}^i &p_i^2Z_{2,\gamma_i}^i  \\
\vdots    &	    \vdots	               &\vdots			                        &\cdots  &\vdots	 &\vdots	   \\
0	     &		0	                   &0		&\cdots  &p_i^{\gamma_i-1}\mathbf{I}_{k_{i,\gamma_i-1}^{\prime}}    &p_i^{\gamma_i-1}Z_{\gamma_i-1,\gamma_i}^i
\end{array} \right] = \left[ \begin{array}{c|c}
\boldsymbol{\mathcal{I}}^{*}_{p_i^{\gamma_i}} & \boldsymbol{\mathcal{B}}_{k_i^{\prime},n-k_i^{\prime}}
\end{array} \right]  \label{equ:FCR.Generator}
\end{equation} 
\hrulefill
\end{figure*} 

To study the message space of the linear codes over the finite chain ring, we first examine the kernel of the generator matrix $\mathbf{G}_{\mathrm{FCR}}^i$.    This is equivalent to finding the null space for the encoder $\mathcal{E}^i: \mathbf{w}^i \longmapsto \mathcal{C}^i$, where $\mathcal{E}^i(\mathbf{w}^i) \triangleq \mathbf{w}^i\mathbf{G}_{\mathrm{FCH}}^i$ and $\mathbf{w}^i = [\mathbf{w}_{k_{i,0}^{\prime}},\mathbf{w}_{k_{i,1}^{\prime}},\cdots,\mathbf{w}_{k_{i,\gamma_i-1}^{\prime}}]$. Here $\mathbf{w}^i$ is grouped into blocks of size  $\mathbf{w}_{k_{i,t}^{\prime}}$ which corresponds  to the row blocks defined in (\ref{equ:FCR.Generator}). In order to obtain the all-zero codeword $\mathcal{C}^i=\mathbf{0}$, we solve the homogeneous system $\mathbf{w}^i\mathbf{G}_{\mathrm{FCH}}^i = \mathbf{0}$, which gives $\mathbf{w}_{k_{i,t}^{\prime}} \in p_i^{\gamma_i - t}T^{k_{i,t}^{\prime}}$ ,  $t=0,1,\cdots,\gamma_i-1$. This result is based on the fact that if $d\in T^n$, then $p_i^{t}d=0 \Longrightarrow d \in p_i^{\gamma_i-t}T^n$. The null space of the encoder $\mathcal{E}^i$ is therefore:


\begin{align}
\mathbf{w}^{\prime} = [p_i^{\gamma_i}T^{k_{i,0}^{\prime}}, \cdots,p_iT^{k_{i,\gamma_i-1}^{\prime}}  ] \label{equ:Null.Space}
\end{align}

According to the first isomorphism theorem, the codeword $\mathcal{C}^i$  is isomorphic to a direct summation:

\begin{align}
\mathcal{C}^i &\cong (T/ p_i^{\gamma_i} T)^{k_{i,0}^{\prime}}\oplus (T/ p_i^{\gamma_i-1} T)^{k_{i,1}^{\prime}}\oplus\cdots\oplus (T/ p_i T)^{k_{i,\gamma_i-1}^{\prime}} \notag \\
&\cong (\delta_iS/\langle p_i^{\gamma_i}\delta_i \rangle)^{k_{i,0}^{\prime}}\oplus (\delta_iS/\langle p_i^{\gamma_i-1} \delta_i\rangle)^{k_{i,1}^{\prime}}\oplus\cdots\oplus (\delta_iS/\langle p_i\delta_i \rangle)^{k_{i,\gamma_i-1}^{\prime}}
 \label{equ:FCR.Code.Message.Space}
\end{align}   

The right-hand side of (\ref{equ:FCR.Code.Message.Space}) denotes the message space $\boldsymbol{W}^i$ of the linear code over the finite chain ring $T$ in terms of the generator matrix $\mathbf{G}_{\mathrm{FCR}}^i$. Note that each component in the direct sum of (\ref{equ:FCR.Code.Message.Space}) forms another module or vector space, and the size of the $t^{\mathrm{th}}$ component is $q^{(\gamma_i-t)k_{i,t}^{\prime}}$. This leads to the overall message size $|C|= q^{\sum_{t=0}^{\gamma_i-1}(\gamma_i-t)k_{i,t}^{\prime}}$. Of course, we can obtain this result directly from the kernel of $\mathbf{G}_{\mathrm{FCR}}^i$, thus, $|C| = \prod_{t=0}^{\gamma_i-1}(p_i^{t}T)^{k_{i,t}^{\prime}}$ which gives the same result. 

Let $\tilde{\boldsymbol{p}}_i^{\gamma_i}$ be a length-$k_i^{\prime}$ vector:
\begin{align}
\tilde{\boldsymbol{p}}_i^{\gamma_i}\triangleq [\mathbf{p}_{i,(k_{i,0}^{\prime})}^{\gamma_i},\mathbf{p}_{i,(k_{i,1}^{\prime})}^{\gamma_i-1},\cdots, \mathbf{p}_{i,(k_{i,\gamma_i-1}^{\prime})}] \notag 
\end{align}
where $\mathbf{p}_{i,(k_{i,0}^{\prime})}^{\gamma_i}$ denotes a length-$k_{i,0}^{\prime}$ vector, with each component being $p_i^{\gamma_i}$. Note that $\tilde{\boldsymbol{p}}_i^{\gamma_i}$ is closely related to (\ref{equ:Null.Space}). Following Lemma \ref{Lemma.GenMatr.Sub.Lattice}, the generator matrix of the primary sublattice $\Lambda_{p_i}$ of the $i^{\mathrm{th}}$ layer in this scenario has a form:
\begin{align}
\mathbf{G}_{\Lambda_{p_i}} = \begin{bmatrix} 
\mathrm{Diag}\left(\tilde{\boldsymbol{p}}_1^{\gamma_1}\cdots \tilde{\boldsymbol{p}}_{i-1}^{\gamma_{i-1}}, \mathbf{I}_{k_i^{\prime}}, \tilde{\boldsymbol{p}}_{i+1}^{\gamma_{i+1}}\cdots \tilde{\boldsymbol{p}}_m^{\gamma_m}\right) & \mathbf{0} \\
\mathbf{0} & \mathbf{I}_{mn-k^{\prime}}
\end{bmatrix} \mathbf{G}  \label{equ:sublattice.Generator.FCH}
\end{align}
where $k^{\prime} = \sum_{i =1}^m k_i^{\prime}$. The EDC lattices in this scenario are constructed by some linear codes over different finite chain rings, and the matrix $\mathbf{G}$ must be associated with the generator matrix of each linear code $\mathcal{C}^i$ over the finite chain ring.  Let $\tilde{\sigma}_i(\mathbf{d}\cdot[\tilde{\boldsymbol{\mathcal{I}}}^{*}_{p_i^{\gamma_i}}~~ \tilde{\boldsymbol{\mathcal{B}}}_{k_i^{\prime},n-k_i^{\prime}}])$ be the codeword of $\mathcal{C}^i = \mathbf{w}^i\mathbf{G}_{\mathrm{FCR}}^i$ over the finite chain ring $T$, $\mathbf{d}\in \delta_iS^{k_i^{\prime}}$. Then, $\mathbf{G}$ in (\ref{equ:sublattice.Generator.FCH}) is an $mn\times n$ matrix defined below:
\begin{align}
\mathbf{G} = \begin{bmatrix}
\tilde{\boldsymbol{\mathcal{I}}}^{*}_{p_1^{\gamma_1}} & \tilde{\boldsymbol{\mathcal{B}}}_{k_1^{\prime},n-k_1^{\prime}}  \\
\tilde{\boldsymbol{\mathcal{I}}}^{*}_{p_2^{\gamma_2}} & \tilde{\boldsymbol{\mathcal{B}}}_{k_2^{\prime},n-k_2^{\prime}}   \\
\vdots	& \vdots\\
\tilde{\boldsymbol{\mathcal{I}}}^{*}_{p_m^{\gamma_m}} & \tilde{\boldsymbol{\mathcal{B}}}_{k_m^{\prime},n-k_m^{\prime}}   \\
\mathbf{0}  & \varpi\mathbf{I}_{n-k_1^{\prime}}  \\
\vdots	& \vdots\\
\mathbf{0}  & \varpi\mathbf{I}_{n-k_m^{\prime}}
\end{bmatrix} \label{equ:G.Generator.Case2}
\end{align} 

Hence, we are able to construct $\Lambda_{p_i}$ and hence the EDC lattice $\Lambda$ for this scenario based on the generator matrices presented above. Note that message space of each layer follows from (\ref{equ:FCR.Code.Message.Space}), and $k_{i,t}^{\prime}$ should be selected such that 
\begin{align}
\gamma_ik_i = \sum_{t=0}^{\gamma_i-1}(\gamma_i-t)k_{i,t}^{\prime} \label{equ:FCR.message.length.condition}
\end{align} 
in order to guarantee the consistency to the message size of the $i^{\mathrm{th}}$ layer EDC lattices defined in (\ref{equ:Message.Space.EDC}). It is easy to prove that there exists $k_{i,t}^{\prime}\in\mathbb{Z}^+$, $\forall t=0,1,\cdots,\gamma_i-1$, satisfying (\ref{equ:FCR.message.length.condition}).

The generator matrix of the coarse lattice $\Lambda^{\prime}$ is given by,
\begin{align}
\mathbf{G}_{\Lambda^{\prime}} = \begin{bmatrix}
\mathrm{Diag}\left(\mathbf{I}_{\sum_{j=1}^{i-1}k_j^{\prime}}, \tilde{\boldsymbol{p}}_i^{\gamma_i},\mathbf{I}_{\sum_{j=i+1}^{m}k_j^{\prime}}  \right) & \mathbf{0} \\
\mathbf{0} & \mathbf{I}_{mn-k}
\end{bmatrix} \mathbf{G}_{\Lambda_{p_i}}\label{equ:Coarse.Generator.Case2}
\end{align}

Following (\ref{equ:Coarse.Generator.Case2}), it is obvious that $\Lambda/\Lambda^{\prime}\cong\boldsymbol{W}^1\oplus\cdots\oplus\boldsymbol{W}^m$.   The generator matrix for $\Lambda_i^{\prime}$ has a form:
\begin{align}
\mathbf{G}_{\Lambda_i^{\prime}} = \begin{bmatrix}
\mathrm{Diag}\left(\mathbf{I}_{\sum_{j=1}^{i-1}k_j^{\prime}}, \tilde{\boldsymbol{p}}_i^{\gamma_i},\mathbf{I}_{\sum_{j=i+1}^{m}k_j^{\prime}}  \right) & \mathbf{0} \\
\mathbf{0} & \mathbf{I}_{mn-k}
\end{bmatrix}  \mathbf{G}\label{equ:lattice.Generator.case2}
\end{align}  
which will be used for LIF detection. 

Every ideal of $T$ is generated by the maximal ideal, which forms a chain with chain length $\gamma_i$. Hence the residue field $Q$ plays an important role in producing the linear codes over $T$. We now consider a matrix in the form of:
\begin{align}
\mathbf{G}_{\mathrm{D}}^i = \mathrm{Diag}\left(\mathbf{p}_{i,(k_{i,0}^{\prime})}^{0},\cdots, \mathbf{p}_{i,(k_{i,\gamma_i-1}^{\prime})}^{\gamma_i-1} \right)\begin{bmatrix}
\mathbf{g}^i_{k_{i,0}^{\prime}}\\
\mathbf{g}^i_{k_{i,1}^{\prime}}\\
\vdots \\
\mathbf{g}^i_{k_{i,\gamma_i-1}^{\prime}}\label{equ:construct.D.generator}
\end{bmatrix}
\end{align}
where $\mathbf{g}^i_{k_{i,t}^{\prime}}\in Q_{k_{i,t}^{\prime}\times n}^{*}$, and $Q_{k_{i,t}^{\prime}\times n}^{*}$ is a $k_{i,t}^{\prime}\times n$ matrix with each entry over the coset representative of the residue field $Q = \delta_iS/\langle p_i\delta_i\rangle$. Each row of $\mathbf{G}_{\mathrm{D}}^i$ must satisfy the condition that none of its rows are linear combinations of the other rows. The message space of $\mathbf{G}_{\mathrm{D}}^{i}$ could be partitioned into $\gamma_i-1$ levels. We first define the vector $\boldsymbol{\beta}_{k_{i,t}^{\prime}}^{(j)} = [\beta_{1}^{(j)},\beta_2^{(j)},\cdots,\beta_{k_{i,t}^{\prime}}^{(j)}]$, when $t=0$, where $j=0,1,\cdots\gamma_i-1$, is the level indicator, and $\boldsymbol{\beta}_{k_{i,t}^{\prime}}^{(j)} = [\beta_{k_{i,t-1}^{\prime}+1}^{(j)},\beta_{k_{i,t-1}^{\prime}+2}^{(j)},\cdots,\beta_{k_{i,t}^{\prime}}^{(j)}]$ when $t=1,2,\cdots,\gamma_i-1$. Accurately $\boldsymbol{\beta}_{k_{i,t}^{\prime}}^{(j)}$ represents a length-$k_{i,t}^{\prime}$ segment of the $j^{\mathrm{th}}$ level message over the vector space $Q^{k_{i,t}^{\prime}}$. The full message space of the $j^{\mathrm{th}}$ level is given by,
\begin{align}
\boldsymbol{\beta}^{(j)} = [p_i^j\boldsymbol{\beta}_{k_{i,0}^{\prime}}^{(j)}, p_i^{j-1}\boldsymbol{\beta}_{k_{i,1}^{\prime}}^{(j)}, p_i^{j-2}\boldsymbol{\beta}_{k_{i,2}^{\prime}}^{(j)}, \underbrace{\mathbf{0}\cdots\mathbf{0}}_{k_{i}^{\prime} - \sum_{t=0}^{j} k_{i,t}^{\prime} }  ] \label{equ:FCR.Beta}
\end{align}
where the powers of $p_i$ can not be negative integers. Hence the message space of  $\mathbf{G}_{\mathrm{D}}^{i}$ is $W^i = \boldsymbol{\beta}^{(0)}+\boldsymbol{\beta}^{(1)}+\cdots+\boldsymbol{\beta}^{(\gamma_i-1)}$. The codewords $\mathcal{C}^i$ can be produced by
\begin{align}
\mathcal{C}^i = W^i\mathbf{G}_{\mathrm{D}}^{i}&=\left(\boldsymbol{\beta}^{(0)}+\boldsymbol{\beta}^{(1)}+\cdots+\boldsymbol{\beta}^{(\gamma_i-1)}\right)\mathbf{G}_{\mathrm{D}}^{i} \notag \\ 
&= \mathbf{c}^i_0 + \mathbf{c}^i_1p_i +\cdots+\mathbf{c}_{\gamma_i-1}^ip_i^{\gamma_i-1} \label{equ:FCR.Const.D.Codeword}
\end{align}

Since none of the rows of $\mathbf{G}_{\mathrm{D}}^{i}$ are linear combinations of the other rows, $\mathbf{c}_{t}^i$ is therefore row spanned by 
\begin{align}
\mathbf{g}_{\mathbf{c}_t^i} = \begin{bmatrix}
\mathbf{g}^i_{k_{i,0}^{\prime}} ;&\mathbf{g}^i_{k_{i,1}^{\prime}}; & \cdots ;&
\mathbf{g}^i_{k_{i,t}^{\prime}}
\end{bmatrix}
\end{align}

It is obvious that $\mathbf{c}_t^i$, $t=0,1,\cdots,\gamma_i-1$ forms a set of nested codes $\mathbf{c}_0^i\subseteq\mathbf{c}_1^i\subseteq\cdots\subseteq \mathbf{c}_{\gamma_i-1}^i$ over $Q^{*}$. Following the $Q$-adic decompostion theorem of  finite chain ring \cite{Finite.Rings.Identity}\cite{Feng.FCR.TIT}, we assert that the codeword $\mathcal{C}^i$ in (\ref{equ:FCR.Const.D.Codeword}) generated by $\mathbf{G}_{\mathrm{D}}^{i}$ is indeed  over $T$. 

%

In terms of (\ref{equ:construct.D.generator}) and (\ref{equ:FCR.Beta}), the message space corresponding to $\mathbf{g}^i_{k_{i,t}^{\prime}}$ should be written as:
\begin{align}
W_t^i = \sum_{j=t}^{\gamma_i-1} p_i^{j-t} \boldsymbol{\beta}_{k_{i,t}^{\prime}}^{(j)}
\end{align} 
this complies with the $Q$-adic decomposition and leads to the result that the message space corresponding to  $\mathbf{g}^i_{k_{i,t}^{\prime}}$ is $(T/\langle p_i^{\gamma_i-t}\rangle)^{k_{i,t}^{\prime}}$. This implies that the right-hand side of (\ref{equ:FCR.Code.Message.Space}) is precisely the message space of $\mathbf{G}_{\mathrm{D}}^{i}$. Mathematically the primary sublattices $\Lambda_{p_i}$ can also be represented  in the form below:
\begin{align}
\Lambda_{p_i} = \bigcup\left\{ \underbrace{\sum_{j=0}^{\gamma_i-1}\sum_{\ell=1}^{\mathscr{K}_{j}^i } p_i^{j}\beta_{\ell}^{(j)}\mathbf{g}_{\ell}^i }_{(52)}+p_i^{\gamma_i}S^n|\mathbf{g}_{\ell}^i\in Q_{1\times n}			 \right\} \label{equ:FCR.Construction.D}
\end{align}
where $\mathscr{K}_{j}^i=k_{i,0}^{\prime}+\cdots+k_{i,j}^{\prime}$. It is interesting to see that (\ref{equ:FCR.Construction.D}) has the same structure as complex construction D. Now we conclude that the primary $S$-sublattices constructed by a linear code over a finite chain ring subsumes construction D.

Based on this result, we may now construct EDC lattices for this scenario using a set of nested linear codes over a finite field. Let $\mathbf{g}_{(n-k_i^{\prime})}^i\in Q^{*}_{n-k_i^{\prime}\times n}$ be an $(n-k_i^{\prime})\times n$ matrix,   then the $\mathbf{G}$ matrix is:
\begin{align}
\mathbf{G} = \begin{bmatrix}
{\mathbf{G}_{\mathrm{D}}^1}^T, &\cdots,&
{\mathbf{G}_{\mathrm{D}}^m}^T, &
{\varpi\mathbf{g}_{(n-k_i^{\prime})}^1}^T, & \cdots, &
{\varpi\mathbf{g}_{(n-k_i^{\prime})}^m}^T 
\end{bmatrix}^T
\end{align}

\subsubsection*{\textbf{Scenario 3}} This corresponds to a hybrid case of scenario 1 and 2, and we give the following summaries:
\begin{enumerate}
\item $m=1$, $\gamma_1=1$, then the EDC lattice in (\ref{equ:EDC.Definition}) is a complex construction A lattice which is indecomposable.
\item $m=1$, $\gamma_1>1$, $\gamma_1\in\mathbb{Z}^+$ then the EDC lattice in (\ref{equ:EDC.Definition}) is a complex construction D lattice which is indecomposable. 
\item $m>1$, $m,\gamma_i\in\mathbb{Z}^+$, $i=1,2,\cdots,m$, then the EDC lattice in (\ref{equ:EDC.Definition}) is decomposable, and consists of some sublattices constructed by either construction A or D. 
\end{enumerate}

Note that in 3), a new class of lattices over $S$ is generated by a number of linear codes over either finite field or chain ring, which generalises the scenario 1 and 2. Scenario 3 suggests that the design of EDC lattices is very flexible, and we also give more detailed discussion about why EDC lattices are good at low-complexity decoding and throughput improvement for PNC in the next sections. 

\subsection{Nominal coding gain and Kissing number} \label{sec:Kissing.Number}
In this section, we study the nominal coding gain and kissing number of the EDC lattices for all three scenarios. The definition such as the minimum-norm coset leaders and minimum Euclidean weight of the codeword follows from \cite{Feng.AlgeAppro.TIT.2013}.  
\subsubsection*{\textbf{Scenario 1}} We first study the nominal coding gain and kissing number of the $i^{\mathrm{th}}$ layer primary sublattices in this scenario. Following (\ref{equ:EDC.Primary.Lattice.Definition}) and (\ref{equ:EDC.Primary.Map}), we know that  $\mathcal{C}^i$ is a linear code of length $n$ over $\delta_iS/p_i\delta_iS$. Thus, $\mathbf{c}^i = (c_1^i+\langle \varpi\rangle,\cdots,c_n^i+\langle\varpi\rangle)\in \mathcal{C}^i$. We denote $\omega^{(i)}(\mathbf{c}^i)$ the Euclidean weight of a codeword $\mathbf{c}^i$ in $\mathcal{C}^i$, and $\omega_{\mathrm{min}}^{(i)}(\mathcal{C}^i)$ the minimum Euclidean weight of non-zero codewords in $\mathcal{C}^i$. Let $\vartheta$ be a scaling factor depending on which PID is used, and $ N(\omega_{\mathrm{min}}^{(i)}(\mathcal{C}^i))$ be the number of codewords in $\mathcal{C}^i$ with the minimum Euclidean weight $\omega_{\mathrm{min}}^{(i)}(\mathcal{C}^i)$. 
\begin{proposition} \label{prop:Coding.Gain.A.1}
Let $\mathcal{C}^i$ be a linear code over $\delta_iS/p_i\delta_iS$,  and  $\Lambda_{p_i}/\Lambda^{\prime}$ the primary quotient lattice system of the $i^{\mathrm{th}}$ layer constructed by $\mathcal{C}^i$, $\Lambda_{p_i}\supseteq\Lambda^{\prime}$, then the nominal coding gain is given by:
\begin{align}
\varrho(\Lambda_{p_i}/\Lambda^{\prime}) =  \frac{\omega_{\mathrm{min}}^{(i)}(\mathcal{C}^i)}{\vartheta |p_i|^{2(1-\frac{k_i}{n})}|\delta_i|^{2} } \label{NomGain.A.Primary.Text}
\end{align}
and the kissing number is:
\begin{equation}
K(\Lambda_{p_i}/\Lambda^{\prime}) = \left\{
             \begin{array}{ll}
              N(\omega_{\mathrm{min}}^{(i)}(\mathcal{C}^i))\left(\frac{\mathcal{N}_{\mathcal{U}(S)}}{|p_i|^2-1}\right)^{\frac{\omega_{\mathrm{min}}^{(i)}(\mathcal{C}^i)}{|\delta_i|^2}}, & |p_i|^2-1 \leq \mathcal{N}_{\mathcal{U}(S) }\\
              N(\omega_{\mathrm{min}}^{(i)}(\mathcal{C}^i)), &  \mathrm{Otherwise}
             \end{array}  
        \right.
\end{equation}
\end{proposition}
\begin{IEEEproof}
See Appendix \ref{APD:4}.
\end{IEEEproof}
Here $\mathcal{N}_{\mathcal{U}(S)}$ represents the number of units in $S$.

It is of interest to study the nominal coding gain and kissing number of $\Lambda/\Lambda^{\prime}$ in terms of the $m$ linear codes $\mathcal{C}^i$. Following the proof of Theorem \ref{Theorem.Lattice.Decomposition}, and the descriptions in section \ref{sec:EDC}, $\tilde{\mathbf{c}}=  \mathbf{c}^1 + \mathbf{c}^2 + \cdots + \mathbf{c}^m$, $\tilde{\mathbf{c}}\in \tilde{\mathcal{C}}$ and $\tilde{\mathcal{C}}\in (S/\langle \varpi\rangle)^n$. Thus, the nominal coding gain of EDC lattices is determined by the $m$ linear codes $\mathcal{C}^i$ over $\delta_iS/p_i\delta_iS$, $i=1,2,\cdots,m$.

\begin{proposition} \label{prop:Coding.Gain.A.2}
Let $\mathcal{C}^1,\cdots, \mathcal{C}^m$ be $m$ linear codes over $\delta_iS/p_i\delta_iS$, $i=1,2,\cdots,m$, respectively. Let $\tilde{\mathbf{c}}=  \mathbf{c}^1 + \mathbf{c}^2 + \cdots + \mathbf{c}^m$, $\tilde{\mathbf{c}}\in \tilde{\mathcal{C}}$ and  $\mathbf{c}^i\in\mathcal{C}^i$. The nominal coding gain of the EDC lattices $\Lambda/\Lambda^{\prime}$ in scenario 1 is given by 
\begin{align}
\varrho(\Lambda/\Lambda^{\prime}) = \frac{\omega_{\mathrm{min}}(\tilde{\mathcal{C}})\prod_{\ell=2}^m|p_j|^{\frac{2(k_{\ell} - k_1)}{n}}}{\vartheta |p_1|^{2(1-\frac{k_1}{n})}|\delta_1|^{2}}  
\label{NomGain.A.Fine.Text}
\end{align}
where $k_1\leq k_2 \leq \cdots \leq k_m$. 
\end{proposition}
\begin{IEEEproof}
See Appendix \ref{APD:4}.
\end{IEEEproof}

\subsubsection*{\textbf{Scenario 2}} This corresponds to the case where $\gamma_i >1$, $\gamma_i\in \mathbb{Z}$ for $i = 1,2,\cdots,m$. The primary sublattice of the $i^{\mathrm{th}}$ layer can be constructed by a linear code $\mathcal{C}^i$ over a finite chain ring $\delta_iS/\langle \varpi\rangle$, where $\delta_i = \frac{\varpi}{p_i^{\gamma_i}}$. This follows immediately from (\ref{equ:EDC.Primary.Lattice.Definition}) and (\ref{equ:EDC.Primary.Map}).  Here, we are more concerned with the nominal coding gain and kissing number when the $i^{\mathrm{th}}$ primary sublattice is constructed by a set of nested linear codes over the residue field $Q$, since the linear code over a finite field is easier to generate.  Let $\mathcal{C}^{i,0}\subseteq\cdots\subseteq\mathcal{C}^{i,\gamma_i-1}$ be nested linear codes of length-$n$ over $Q$, where $\mathcal{C}^{i,t}$ is an $[n,\sum_{\ell=0}^{t}k_{i,\ell}^{\prime}]$ linear code for the $t^{\mathrm{th}}$ nested code at the $i^{\mathrm{th}}$ layer, and we denote $\omega_{\mathrm{min}}^{(i,t)}(\mathcal{C}^{i,t})$ the minimum Euclidean weight of non-zero codewords in $\mathcal{C}^{i,t}$. We have:

\begin{proposition} \label{prop:Coding.Gain.D.1}
Let $\mathcal{C}^{i,0}\subseteq\cdots\subseteq\mathcal{C}^{i,\gamma_i-1}$ be $\gamma_i$ nested linear codes of length-$n$ over $Q$, and $\Lambda_{p_i}/\Lambda^{\prime}$ be the primary quotient lattice of the $i^{\mathrm{th}}$ layer constructed from $\mathcal{C}^{i,t}$, $t=0,1,\cdots,\gamma_i-1$, then the nominal coding gain of the $i^{\mathrm{th}}$ layer is lower bounded by
\begin{align}
\varrho(\Lambda_{p_i}/\Lambda^{\prime}) &\geq \frac{|p_i|^{\frac{2}{n}\sum_{t=0}^{\gamma_i-1}(\gamma_i-t)k_{i,t}^{\prime} }\min_{0\leq t \leq \gamma_i-1}\{|p_i|^{2t}\omega_{\mathrm{min}}^{(i,t)}(\mathcal{C}^{i,t}) \}} { \vartheta|\varpi|^2 }  \label{NomGain.D.Primary.Text}
\end{align}
and the kissing number is upper bounded by:
\begin{equation}
K(\Lambda_{p_i}/\Lambda^{\prime}) \leq \left\{
             \begin{array}{ll}
              \sum_{t=0}^{\gamma_i-1} N_t(\omega_{\mathrm{min}}^{(i,t)}(\mathcal{C}^{i,t}))\left(\frac{\mathcal{N}_{\mathcal{U}(S)}}{|p_i|^2-1}\right)^{\frac{\omega_{\mathrm{min}}^{(i,t)}(\mathcal{C}^{i,t})}{|\delta_i|^2}}, & |p_i|^2-1 \leq \mathcal{N}_{\mathcal{U}(S) }\\
              \sum_{t=0}^{\gamma_i-1} N_t(\omega_{\mathrm{min}}^{(i,t)}(\mathcal{C}^{i,t})), &  \mathrm{Otherwise}
             \end{array}  
        \right. \label{KissN.D.Primary.Text}
\end{equation}
\end{proposition}         
\begin{IEEEproof}
See Appendix \ref{APD:5}.
\end{IEEEproof}  

It is of interest to study the nominal coding gain of $\Lambda/\Lambda^{\prime}$ in this scenario. If each primary sublattice is constructed via a set of nested linear codes over a finite field $Q = \delta_iS/\langle p_i\delta_i\rangle$ for the $i^{\mathrm{th}}$ layer, the nominal coding gain $\varrho(\Lambda/\Lambda^{\prime})$ will be related to overall $\sum_{i=1}^m\gamma_{i}$ linear codes since there are $\gamma_i$ nested linear codes for each $i$. Let $\tilde{\mathcal{C}}$ be a composite code such that $\tilde{\mathbf{c}} = \mathbf{c}^1 + \cdots+\mathbf{c}^m$ where $ \mathbf{c}^i = \mathbf{c}^{i,0}+p_i\mathbf{c}^{i,1}+\cdots+p_i^{\gamma_i-1}\mathbf{c}^{i,\gamma_i-1}$. Hence $\mathcal{C}^i\in \delta_iS/\langle\varpi\rangle$ and $\tilde{\mathcal{C}}\in S/\langle\varpi\rangle$. We denote $\omega_{\mathrm{min}}(\tilde{\mathcal{C}})$ the minimum Euclidean weight of non-zero codewords in $\tilde{\mathcal{C}}$, then: 

\begin{proposition} \label{prop:Coding.Gain.D.2}
Let $\mathcal{C}^{i,0}\subseteq\cdots\subseteq\mathcal{C}^{i,\gamma_i-1}$ be $\gamma_i$ nested linear codes of length-$n$ over $Q$, and let  $\tilde{\mathcal{C}}$ be a composite code such that $\tilde{\mathbf{c}} = \mathbf{c}^1 + \cdots+\mathbf{c}^m$ where $ \mathbf{c}^i = \mathbf{c}^{i,0}+p_i\mathbf{c}^{i,1}+\cdots+p_i^{\gamma_i-1}\mathbf{c}^{i,\gamma_i-1}$. The nominal coding gain for $\Lambda/\Lambda^{\prime}$ in scenario 2 is given by:     
\begin{align}
\varrho(\Lambda/\Lambda^{\prime}) &= \frac{\omega_{\mathrm{min}} (\tilde{\mathcal{C}})}{\left(V(\mathcal{V}(\Lambda))\right)^{\frac{1}{n}}} \notag \\
&= \frac{\omega_{\mathrm{min}}(\tilde{\mathcal{C}})\prod_{i=1}^m|p_{i}|^{2\sum_{t=0}^{\gamma_{i}-1}(\gamma_{i}-t)\frac{k_{i,t}^{\prime}}{n} }}{\vartheta|\varpi|^2} \label{NomGain.D.Fine.Text}
\end{align}
\end{proposition}
\begin{IEEEproof}
See Appendix \ref{APD:5}.
\end{IEEEproof}  

\subsubsection*{\textbf{Scenario 3}} As explained in the preceding section,  in this case, $\gamma_i\geq 1$, $\gamma_i\in\mathbb{Z}$, and hence the EDC lattice consists of a number of primary sublattices which can be constructed by linear codes over either finite field or finite chain ring.  The nominal coding gain and kissing number of the primary sublattices in each case have been derived in Proposition \ref{prop:Coding.Gain.A.1} and \ref{prop:Coding.Gain.D.1}. We are more interested in the nominal coding gain of $\Lambda/\Lambda^{\prime}$ in this scenario. Again, we consider the primary sublattices of scenario 2 is constructed over a set of nested linear codes. Let $\tilde{\mathcal{C}}$ be a composite code such that $\tilde{\mathbf{c}} = \mathbf{c}^1 + \cdots+\mathbf{c}^m$ where 
\begin{equation}
\mathbf{c}^i = \left\{
             \begin{array}{ll}
               \mathbf{c}^i, ~~\mathcal{C}^i\in \delta_iS/p_i\delta_iS, & \gamma_i=1\\
             	  \mathbf{c}^{i,0}+p_i\mathbf{c}^{i,1}+\cdots+p_i^{\gamma_i-1}\mathbf{c}^{i,\gamma_i-1}; ~~\mathcal{C}^{i,t}\in Q, &\gamma_i>1
             \end{array}  
        \right.  \notag
\end{equation}
We can easily prove that $\varrho(\Lambda/\Lambda^{\prime})$ has similar form as (\ref{NomGain.D.Fine.Text}) if we set $k_{i,0}^{\prime} = k_i$ for $\gamma_i = 1$.


\section{Iterative Detection of EDC and the EXIT chart analysis}\label{sec:Iterative.Detection.EDC}
In this section we present an iteration-aided multistage decoding approach specifically designed for EDC, which provides a feasible way of improving the performance of decoding the linear combinations, and also of increasing the overall rate with low decoding-complexity. In the remainder of the paper, we consider $S$ to be a ring of Eisenstein integers $\mathbb{Z}[\omega]$. However, the results can be readily extended to other PIDs. 

Section \ref{sec:EDC.Overall}  clearly reveals the possible encoding structure for EDC.  Recalling the definition for EDC, we know that the map $\tilde{\sigma}:  S^n\longmapsto (S/\langle p_1^{\gamma_1}\rangle)^n\oplus (S/\langle p_2^{\gamma_2}\rangle)^n\oplus\cdots\oplus (S/\langle p_m^{\gamma_m}\rangle)^n$ is a natural projection of a surjective ring homomorphism $\sigma: S \longmapsto S/\langle p_1^{\gamma_1}\rangle\times S/\langle p_2^{\gamma_2}\rangle\times\cdots\times S/\langle p_m^{\gamma_m}\rangle \longleftrightarrow  \mathbb{F}_{\tilde{p}_1}\times\cdots\times\mathbb{F}_{\tilde{p}_m}$ by applying it element-wise \cite{Conway.Sloane} ($\gamma_i=1$, $\forall i=1,2,\cdots,m$). Note that in this case, $\sigma$ is actually an \textit{f.g.} abelian group homomorphism. It is easy to see that each level $S/\langle p_i\rangle$ is coded by an $[n,k_i]$ linear code $\mathcal{C}^{i}$ over $\mathbb{F}_{\tilde{p}_i}$ (a finite field or finite chain ring determined by $\tilde{p}_i$).  

The Type-1 Eisenstein primes are those primes $p\in\mathbb{Z}$ which either have a form $6j+5$, $j\in\mathbb{Z}$, or $p=2$. Their associates are also categorised as Type-1. The Type-2 Eisenstein primes have the form $\tau = a+b\omega$, $a,b\neq 0$ where the norm $\mathcal{N}(\tau)$ of $\tau$ is a prime $p\in\mathbb{Z}$ satisfying $p\equiv 1~\mathrm{mod}~ 6$. Note that if $\tau = a+b\omega$ is a prime in $\mathbb{Z}[\omega]$, $\tau^{\prime} = b+a\omega$ is also a prime in $\mathbb{Z}[\omega]$. Hence $\tau$ and $\tau^{\prime}$ are distinct primes categorised as Type-2. Together with the Type-3 Eisenstein primes, $\varpi\in\mathbb{Z}[\omega]$ can be uniquely decomposed into:
\begin{align}
\varpi = \mathcal{U}(\mathbb{Z}[\omega])2^\varrho \prod_{i=1}^{\kappa_1}\tau_i^{\mu_i} \prod_{j=1}^{\kappa_2}{\tau_j^{\prime}}^{\eta_j} \prod_{k=1}^{\kappa_3} p_k^{\beta_k} \cdot (1+2\omega)^{\varsigma} \label{equ:Eisenstein.Factorisation}
\end{align} 
 Accordingly,  $\mathbb{Z}[\omega]/\langle 2\rangle\cong\mathbb{F}_{2^2}$, $\mathbb{Z}[\omega]/\langle \tau\rangle\cong\mathbb{F}_{\mathcal{N}(\tau)}$,   $\mathbb{Z}[\omega]/\langle p\rangle\cong\mathbb{F}_{p^2}$,  $\mathbb{Z}[\omega]/\langle 1+2\omega\rangle\cong\mathbb{F}_{3}$.  

\subsection{Soft Detector for EDC} \label{sec:Soft.Detector}
Section \ref{sec:Multilevel.Lattice.Network.Coding} gives a general decoding method LIF for MLNC, based on the optimised scaling factor $\alpha$, $S$-integer coefficient vectors $\mathbf{\tilde{a}}_i$, and a good EDC lattice quantizer, e.g. a Viterbi decoder with modified metrics (see Appendix \ref{Appendix.3}). Thus,  when EDC is employed in MLNC, LIF is also feasible.   In this section, we explore another detection approach designed specifically for the EDC-based MLNC ( which follows from the structure of the EDC lattices).  Especially an iterative detector is developed, which exploits the multilevel structure gain of EDC by using multistage decoding. 

First, we consider the non-iterative multistage decoding. The detector tries to decode the linear function of each level stage-by-stage, with the aid of the \textit{a priori} information from the preceding layers. The detection structure is similar to the point-to-point multilevel codes, e.g. \cite{Huber.Multilevel.Codes,AGB.TIT} whereas here the \textit{a priori} information is the soft estimation. We develop a layered soft detector (LSD) which calculates the posteriori L-vector (a vector of Log-likelihood ratio) for each layer with the aid of the multiple \textit{a priori} L-vectors. The detailed derivation is given in Appendix \ref{APD:1}.   

The LSD decodes the linear function of each layer over the corresponding non-binary finite field, and hence the \textit{a priori} information of each layer is no longer a scalar value. We define the \textit{a priori} information $\mathbf{A}^i$ to be a vector-based random variable with realization:
\begin{align}
\mathbf{a}^i = \left[\log\left( \frac{\mathrm{Pr}(\xi|V^i=v^i_1) } {\mathrm{Pr}(\xi|V^i=0)}		\right)\cdots \log\left( \frac{\mathrm{Pr}(\xi|V^i=v^i_{{\tilde{p}_i}-1}) } {\mathrm{Pr}(\xi|V^i=0)}		\right) \right] \label{equ:a.priori.vector}
\end{align}  
where $V^i$ denotes the possible linear combinations at the $i^\mathrm{th}$ level, which is a uniformly distributed random variable whose $k^{\mathrm{th}}$ realization is $v^i_k\in\mathbb{F}_{\tilde{p}_i}$, $k=1,2\cdots {\tilde{p}_i}-1$. $\mathrm{Pr}(\xi|V^i=v_k^i)$ is the probability of the \textit{a priori} channel outputs $\Xi=\xi$ given the event $V^i=v_k^i$. Assume that $w_j^i\in\mathbb{F}_{\tilde{p}_i}$, $i=1,2,\cdots,m$, $j=1,2,\cdots,L$ to be the message of the $i^\mathrm{th}$ level and the $j^{\mathrm{th}}$ source, the linear function is defined by $f^{i}(w_1^i,\cdots,w_L^i) = \bigoplus_{\ell=1}^{L}a_{\ell}^i w_{\ell}^{i}$ over $\mathbb{F}_{\tilde{p}_i}$. Note that the integer coefficient $a_{\ell}^i$ can be determined either by the lattice reduction approach as introduced in \cite{Feng.AlgeAppro.TIT.2013, CLLL.Cong} over the $i^{\mathrm{th}}$ quotient lattice $\Lambda/\Lambda^{\prime}_i $ as defined in Theorem  \ref{Theorem.NewLattice.Decoding}, or by the maximum mutual information criterion as described later.  

In the multistage iterative decoding, the proposed LSD outputs the extrinsic L-vector $\mathbf{e}^i$ for the $i^{\mathrm{th}}$ level, based on the \textit{a priori} L-vector $\mathbf{a}^j, j\in\{1,\cdots,m\},j\neq i$. Assume that there is a two level EDC and  the decoding proceeds from layer 1 (which is regarded as the $1^{\mathrm{st}}$ stage decoding) to layer 2 (the $2^{\mathrm{nd}}$ stage decoding). The extrinsic ouputs of layer 1 feed into layer 2 to assist the $2^{\mathrm{nd}}$ stage decoding. With the aid of the \textit{a priori} L-value, layer 2 estimates and forwards the extrinsic information (which serves as the \textit{a priori} information of layer 1) to layer 1. The process is repeated and all layers are activated in turn for the second and subsequent iterations. We refer to this approach as the iterative MSD (IMSD) scheme for MLNC. The detection process is similar to iterative decoding of multilevel codes, e.g. \cite{Yi.Multilevel.TCOM} whereas the nature of the detection is different.  As the iteration proceeds, each layer will produce more reliable extrinsic L-vector $\mathbf{e}^i$ which also serves as the \textit{a priori} information of the soft-in soft-out non-binary decoder for the corresponding $\mathcal{C}^i$.  

\begin{figure*}[!t]
\centering
  \subfigure[]{
    \label{fig:EXIT_F3F4} 
    \begin{minipage}[b]{0.5\textwidth}
      \centering
      \includegraphics[width=1\textwidth]{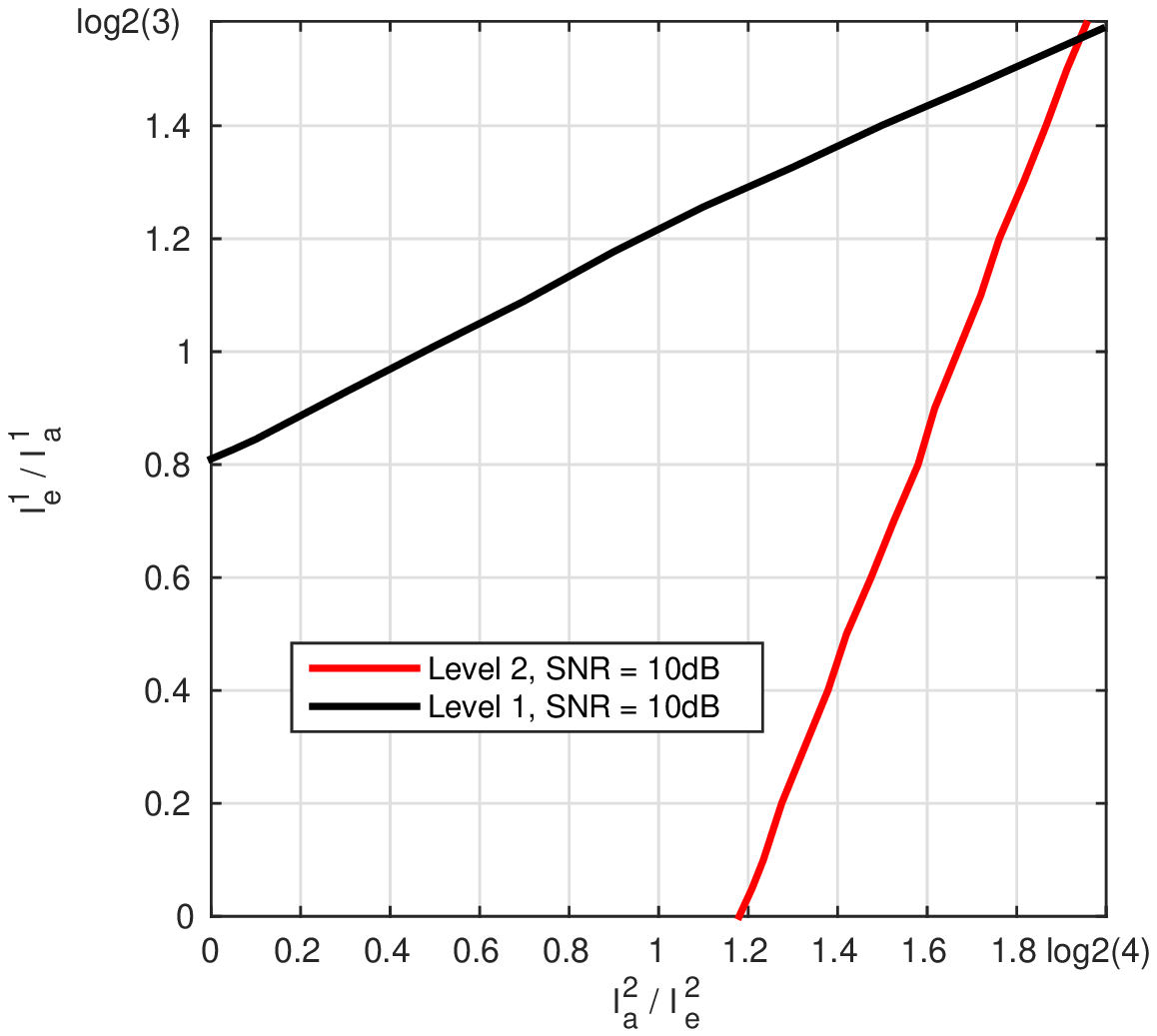}
    \end{minipage}}%
\centering
  \subfigure[]{
    \label{fig:Achievable.Rates} 
    \begin{minipage}[b]{0.5\textwidth}
      \centering
      \includegraphics[width=1\textwidth]{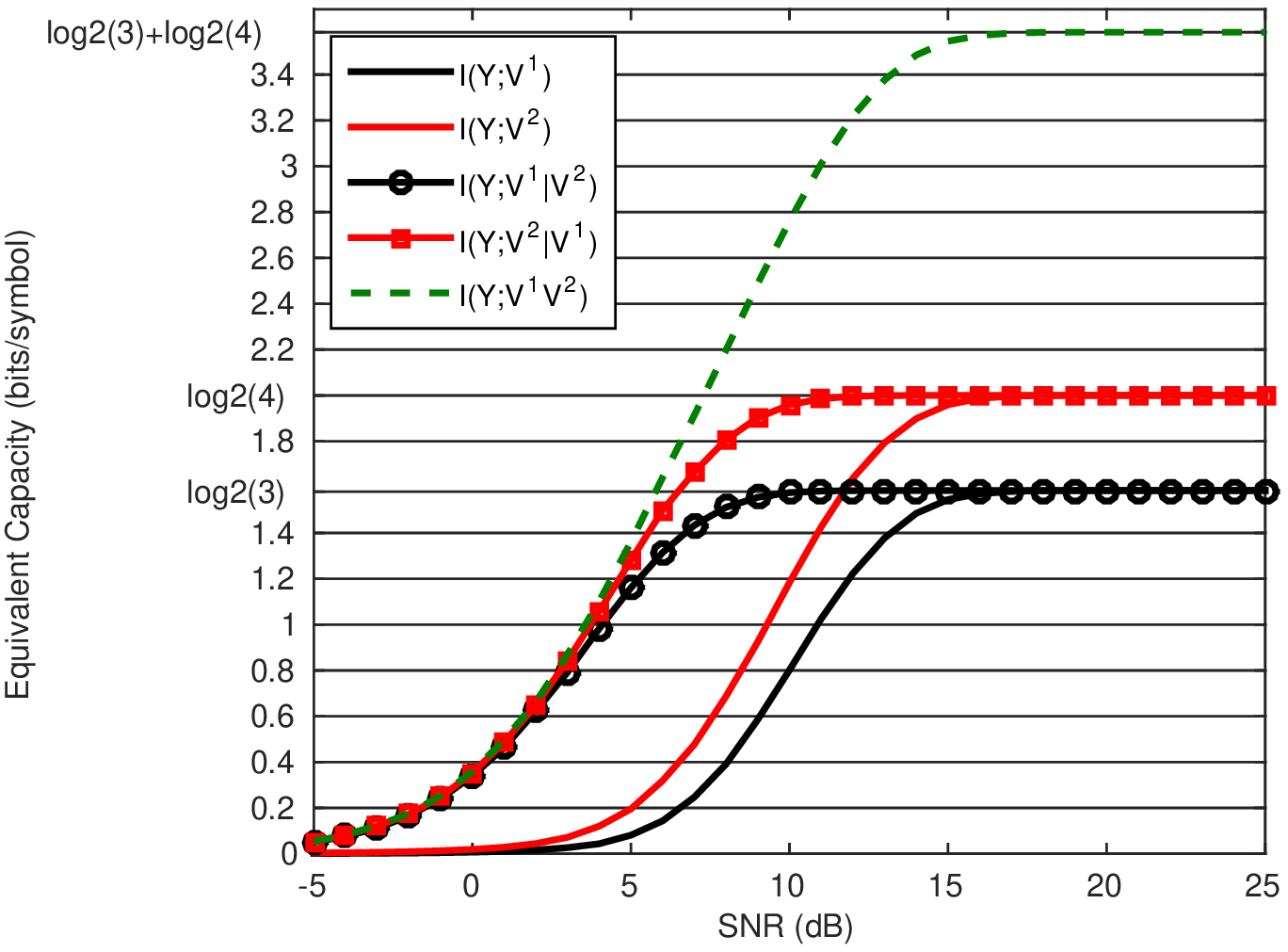}
    \end{minipage}}
\centering
  \subfigure[]{
    \label{fig:Achievable.Rates.Fix} 
    \begin{minipage}[b]{0.5\textwidth}
      \centering
      \includegraphics[width=1\textwidth]{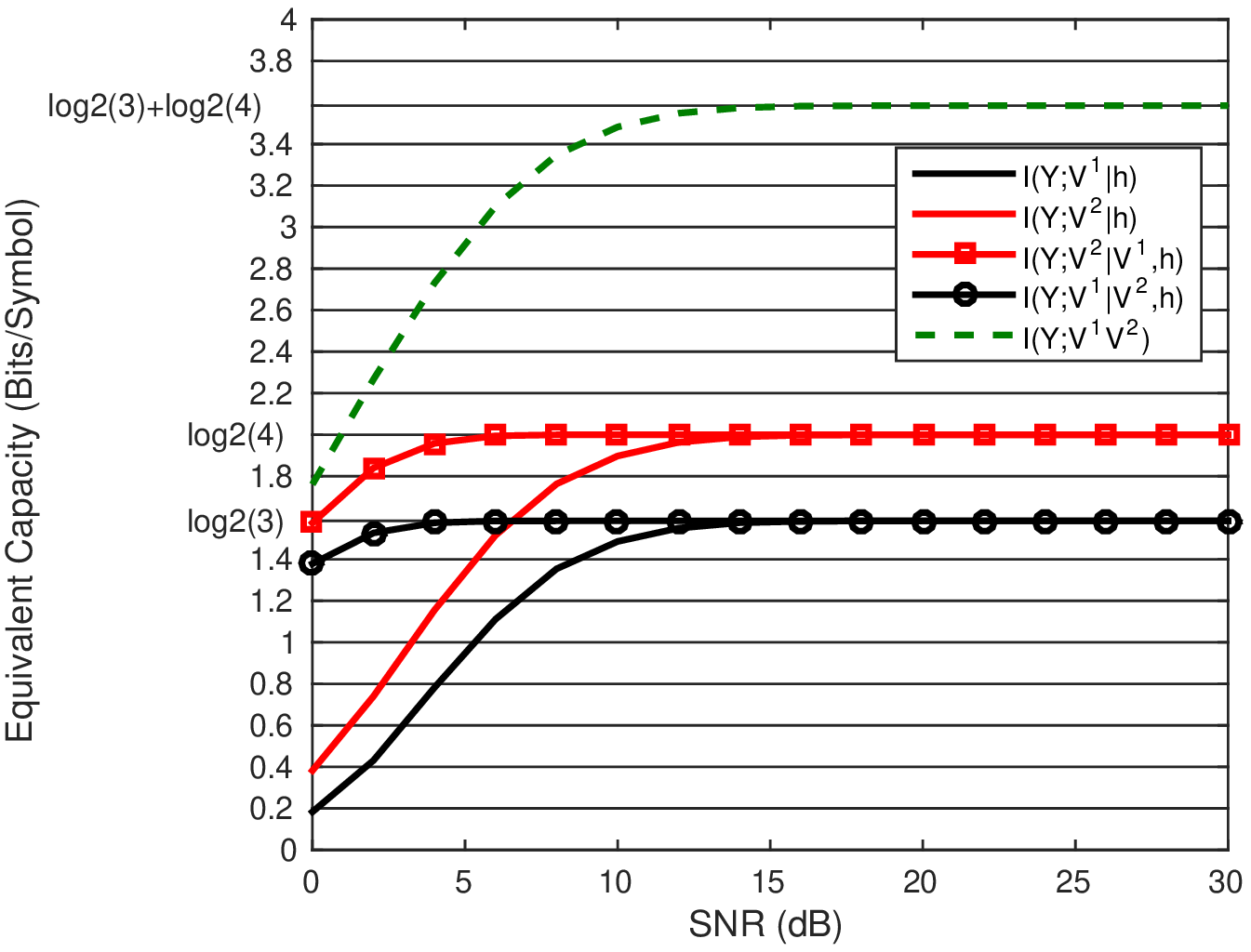}
    \end{minipage}}%
\centering
  \subfigure[]{
    \label{fig:Achievable.Rates.Ray} 
    \begin{minipage}[b]{0.5\textwidth}
      \centering
      \includegraphics[width=1\textwidth]{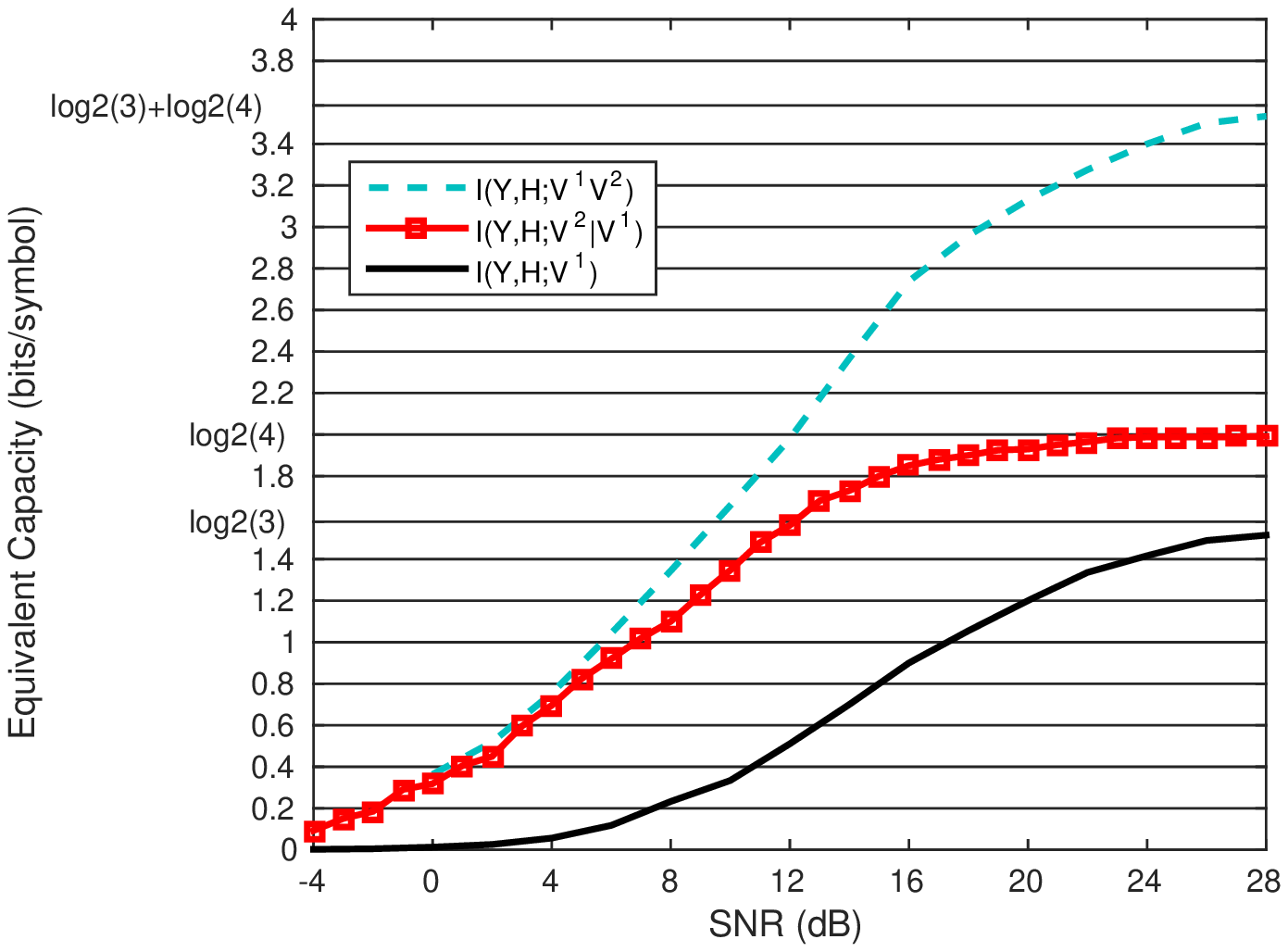}
    \end{minipage}}
  \caption{\small (a) The EXIT function for each layer. (b) Achievable information rates of the linear combinations at each layer; $h_1 = h_2 = 1$. (c) Achievable rates of the linear combinations at each layer with fixed fading coefficients $h_1=-1.17+2.15*1i$ and $h_2=1.25-1.63*1i$. (d) Achievable rates of the linear combinations at each layer with Rayleigh fading.}
\end{figure*}

\subsection{Non-Binary EXIT Chart Analysis}

We now evaluate the extrinsic information transfer characteristics of the soft detector developed in section \ref{sec:Soft.Detector}, based on the non-binary case. Alexei \textit{et al.} \cite{ten.Brink.Erasure.Channel} has proved, based on the binary iterative system, that the extrinsic information $E_k$ (the $k^{\mathrm{th}}$ time instant) of an \textit{a posteriori} probability (APP) decoder contains the same amount of information as the physical channel outputs $\mathbf{Y}$ and the outputs of the \textit{a priori} channel $\mathbf{Z}_{\setminus k}$. We can prove that when the extrinsic outputs are non-binary-based, this theorem also holds. In this case, $E_k$ becomes the vector-based random variable $\mathbf{E}_k$, and can be expressed as:
$$I(V_k|\mathbf{E}_k) = I(V_k|\mathbf{Y},\mathbf{Z}_{\setminus k})$$
 The proof \cite{Efficient.Computation.Soon} is based on the fact that $\mathrm{Pr}(V_k=v|\mathbf{e}_k) = \mathrm{Pr}(V_k=v|\mathbf{y},\mathbf{z}_{\setminus k})$. The average extrinsic information $I_{\mathrm{E}} = \frac{1}{N}\sum_{k=1}^NI(V_k;\mathbf{E}_k)$ can be obtained by:
\begin{figure*}
\begin{align}
I_{\mathrm{E}} = H(V) +  \mathbb{E}\Big[\frac{1}{N}\sum_{k=1}^N \sum_{\forall v}&\mathrm{Pr}(V_k=v|\mathbf{y},\mathbf{z}_{\setminus k})
\cdot\log(\mathrm{Pr}(V_k=v|\mathbf{y},\mathbf{z}_{\setminus k})) \Big]
\end{align}  
\hrulefill
\end{figure*}

Fig. \ref{fig:EXIT_F3F4} illustrates the extrinsic transfer characteristics for a two-level EDC lattice over the TWRC, where $\varpi=2+4\omega=2(1+2\omega)$. Based on the definition of EDC and (\ref{equ:Eisenstein.Factorisation}), the linear codes $\mathcal{C}^1$ and $\mathcal{C}^2$ are over $\mathbb{F}_3$ and a binary extension field $\mathbb{F}_{2^2}$, respectively. The extrinsic information $I_e^1$ for the linear combinations of the $1^{\mathrm{st}}$ level depends only on the \textit{a priori} information $I_a^2$ from the $2^{\mathrm{nd}}$ level, and similarly for $I_e^2$. It can be observed  in Fig. \ref{fig:EXIT_F3F4} that there is an increase of the average extrinsic information $I_e^1$ around $0.8$ at $10\mathrm{dB}$ when the soft detector has the ideal \textit{a priori} information at the $2^{\mathrm{nd}}$ level, compared to the non-iterative case. Hence the iteration-aided multistage detection implies a large potential to improve the reliability of decoding the linear combinations at each level. Due to space limitation for this paper, we show here results only for $h_1=h_2=1$. However the results can be easily extended to the faded MAC. Note that the optimal linear functions $f^1,\cdots,f^m$ should be selected in terms of:
\begin{align}
{f^1\cdots f^m} = \operatorname*{arg\,max}_{f^1\cdots f^m} I(Y;V^1V^2\cdots V^m) \label{equ:coefficient.selection.MI}
\end{align}         
which maximizes the achievable rate. Note that $V^i$ is a random variable with its outcomes from the linear function $f^i$. Hence the conditional probability density $\mathrm{Pr}(Y|V^i)$ is a function of the messages $w^i_j,j=1,2,\cdots,L$. Fig. \ref{fig:Achievable.Rates} gives the numerical integration results for the achievable rates at each level. It can be observed that the mutual information chain rule is satisfied, which gives   theoretical support for  mutistage iterative decoding. Fig. \ref{fig:Achievable.Rates} also well matches the EXIT chart results in Fig. \ref{fig:EXIT_F3F4}, e.g. the extrinsic information of the linear combinations for the first level is around $I(Y;V^1)=0.8$ and $I(Y;V^1|V^2)=1$ at $10\mathrm{dB}$ which precisely match the black line in Fig. \ref{fig:EXIT_F3F4}. Fig. \ref{fig:Achievable.Rates.Fix} and  Fig. \ref{fig:Achievable.Rates.Ray} give the achievable information rates of the linear combinations at each level based on the fixed fading  and Rayleigh fading, respectively. The detailed calculation of the these are described in Appendix \ref{APD:2}. 

It is seen that the maximum achievable rates for the network coded linear combinations are $\mathcal{R}^{(1)}=\log_23$ and $\mathcal{R}^{(2)}=\log_24$ for level 1 and 2. The allowable rate at a certain level is higher when  the \textit{a priori} information from another layer is available. We assume two memory 3, 1/2-rate convolutional codes are used at both levels (over $\mathbb{F}_3$ and $\mathbb{F}_{2^2}$ respectively). EDC lattices achieve overall rate $\frac{1}{2}\log_2(12)$, with the number of trellis states 27 and 64 at the corresponding levels. However, a single convolutional code over ring $R_{12}$ needs 1728 trellis states. The complexity reduction is obvious.      



\section{Simulation Results} \label{sec:simulations}
In this section, we evaluate the performance of the MLNC scheme, based on the detection approaches proposed. These results give strong support for the Theorems and Lemmas developed in previous sections. In this paper, we focus mainly on the applications of EDC lattices in MLNC. However, it is not necessarily limited to EDC lattices since MLNC design applies in principle to the general case. For example, high coding gain lattice codes (e.g. complex low density lattice codes \cite{Yi.CLDLC.2015}\cite{CLDLC.Origin} and signal codes \cite{Feder.ConvSignalCodes.TIT.2011}) which are directly designed in the geometric space can be used in the MLNC framework. This is interesting and will be investigated in our later work.    

We are mainly concerned (in this paper) with the performance of the multiple access channel (MAC) of the TWRC, which can be viewed as the building block for  more complicated network topologies. All simulations are based on a two-layer EDC lattice which has the same configuration. Thus, the two layers are constructed via linear codes $\mathcal{C}^1\in\mathbb{F}_3$ and $\mathcal{C}^2\in\mathbb{F}_{2^2}$. The linear codes at both layers are non-binary convolutional codes, with their  generator polynomials defined in Table \ref{tab:1}. Note that the decoder of the non-binary convolutional codes is based on the maximum a posterior (MAP) probability criteria and  modified BCJR algorithm, where the soft output of the component symbols is produced. We do  not give detailed explanation of the decoding in this paper since it is not our main concern, but we will provide the algorithm when requested. Unless otherwise stated, the convolutional decoder employs the same algorithm in the sequel. 
\begin{table}[h!]
\centering \caption{\small Code type and code rate assigned for each level.}
\begin{tabular}{c | c }
\toprule\hline  $i$ &  $\mathbf{g}(D)$   \rule{0pt}{2.6ex}
\rule[-1.2ex]{0pt}{0pt}\\
\hline \rule{0pt}{2.6ex}\rule[-1.2ex]{0pt}{0pt}
1  	& $[-2\omega^2+2\omega^2 D^3, 2\omega^2+(-2\omega^2)D+ 2\omega^2D^3]$    \\
2      & $\begin{bmatrix}-2+(1-\omega)D^2+(-2)D^3\\ -2 +(-2)D +(-2)D^2 +(1-\omega^2)D^3 \end{bmatrix}$
    \\
\hline\bottomrule
\end{tabular}
\label{tab:1}
\end{table}

\begin{figure}[t]
  \centering
\begin{minipage}[t]{1\linewidth}
\centering
  \includegraphics[width=0.7\textwidth]{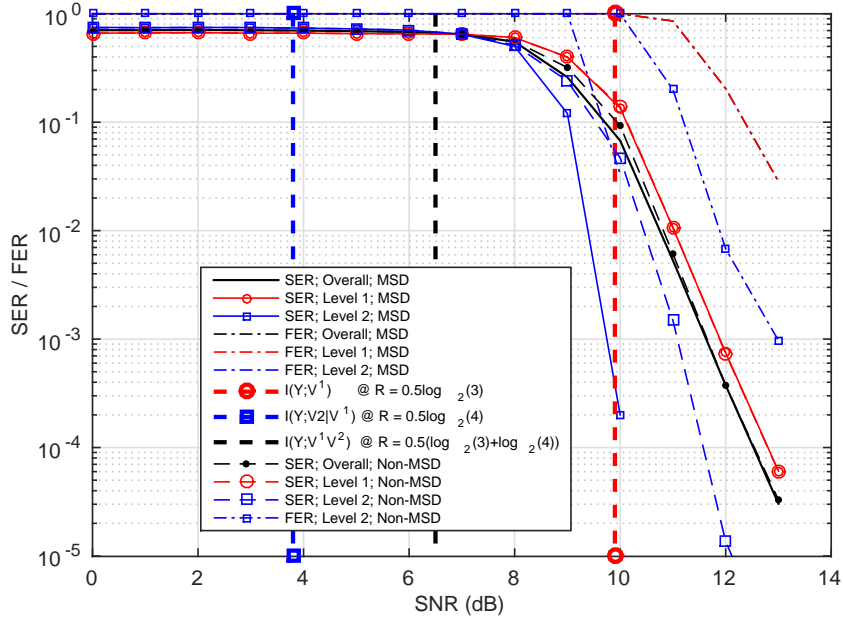}
\end{minipage}
\caption{\small SER and FER performance for an MLNC constructed from a two layer EDC Lattices; Soft detection; Multistage decoding/Non-Multistage decoding;  $\mathcal{R}_{\mathrm{mes}}^{(1)}=\frac{1}{2}\log_2(3)$; $\mathcal{R}_{\mathrm{mes}}^{(2)}=\frac{1}{2}\log_2(4)$; $h_1=h_2=1$.} \label{fig:BER.Case1}
\end{figure}

Fig. \ref{fig:BER.Case1} depicts the symbol-error rates and frame-error rates for EDC-based MLNC  as a function of SNR, where the soft detection approach is used. We examine the performance with and without multistage decoding when iterations and fading are not considered. The convolutional code  at the $i^{\mathrm{th}}$ level is defined as a $[2(\iota^i+\nu^i),\iota^i]$ linear block code, where $\iota^i$ and $\nu^i$ denote the data and memory length, respectively. Therefore the overall message rate is given by:
\begin{align*}
\mathcal{R}_{\mathrm{mes}} \approx \frac{1}{2}(\log_23+\log_24) ~\mathrm{bits/symbol}
\end{align*}
Note that we use the approximation sign here since the actual coding rate is smaller than $\frac{1}{2}$ due to the tail effect of memory. When the block length is sufficient large, this effect can be ignored. Without multistage decoding, it is observed from Fig. \ref{fig:BER.Case1} that the SER gap between layer 1 and 2 is around $0.8$dB at BER=$10^{-4}$, and layer 1 is $8$dB from the capacity of layer 1. When multistage decoding is performed from layer 1 to layer 2, we expect that the SER performance of layer 2 can be improved as a result of the additional \textit{a priori} soft information from layer 1. Note that layer 2 operates over $\mathbb{F}_{2^2}$ whereas its \textit{a priori} soft information is over $\mathbb{F}_3$. The simulation results confirm this anticipation in that the SER of layer 2 has $2$dB gain over non-MSD at $10^{-5}$. However this leads to only slightly better overall performance. When multistage processing starts from layer 1, it is obvious that MSD and non-MSD should give approximately the same performance at layer 1. The overall performance is dominated by the layer which has the worst SER performace over all layers, and in this case, it is layer 1. This explains the reason why the performance improvement of layer 2 gives small contribution in the overall SER. 

\begin{figure}[!t]
  \centering
\begin{minipage}[t]{1\linewidth}
\centering
  \includegraphics[width=0.7\textwidth]{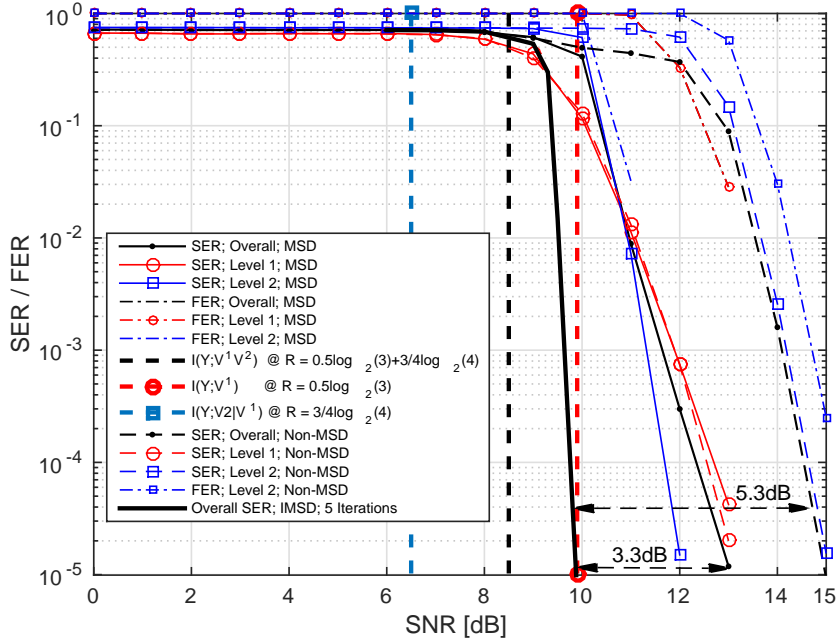}
\end{minipage}
\caption{\small  SER and FER performance for an MLNC constructed from a two layer EDC Lattices; Soft detection; Multistage decoding / Non-Multistage decoding / IMSD; Asymmetric coding rate;  $\mathcal{R}_{\mathrm{mes}}^{(1)}=\frac{1}{2}\log_2(3)$; $\mathcal{R}_{\mathrm{mes}}^{(2)}=\frac{3}{4}\log_2(4)$; $h_1=h_2=1$.} \label{fig:BER.Case2}
\end{figure}

To further increase network throughputs, and examine the performance of MSD based on the asymmetric coding rates over each level, the rate of layer 2 is set to $\mathcal{R}^{(2)}=\frac{3}{4}$. Thus, the sublattice $\Lambda_{p_2}$ is constructed via a higher rate linear code. The overall message rate is given by 
\begin{align*}
\mathcal{R}_{\mathrm{mes}} \approx \frac{1}{2}\log_23 + \frac{3}{4}\log_24 ~\mathrm{bits/symbol}
\end{align*} 
Note that the SER curve of level 1 (red dashed circle) without MSD should closely match that with MSD (red  solid circle)  when multistage decoding is used in layer 1. Simulations in Fig. \ref{fig:BER.Case2} confirm this. Based on the increased coding rate, we are more concerned with the SER performance of layer 2. It is observed from Fig. \ref{fig:BER.Case2} that the SER performance of layer 2 is greatly degraded if MSD is not employed, with approximately $3$dB loss at $10^{-5}$ compared to the half-rate code used at this level. However, when MSD is used, the SER (blue solid square) of layer 2 has more than $3$dB gain over the non-MSD case (blue dashed square) as a result of the reliable \textit{a priori} feedback from layer 1. The overall performance of MSD-based detection is determined mainly by layer 1, whereas  for non-MSD-based detection, the overall performance is dominated by layer 2. That is the reason why the overall SER of the MSD-based scheme performs better than the non-MSD scenario, with $2$dB gain obtained at $10^{-5}$. It is interesting to note that when the decoding of the $\Lambda_{p_i}/\Lambda^{\prime}$ which is constructed from a higher rate linear code occurs at a  later stage of MSD, the overall SER performance of MSD over non-MSD performs better. Hence, MSD is particularly suitable for the detection of EDC lattices in terms of MLNC design, since each layer of EDC operates over an asymmetric finite field or finite chain ring. Now the overall SER is $4.5$dB from the capacity. Note that the measure of SER is based on the correct recovery of the linear combinations of original messages at each source over the respective algebraic field. 

\textit{Iterative Multistage Decoding}: we believe that there is room to improve SER and FER performance further. Based on the soft detector developed in section \ref{sec:Iterative.Detection.EDC}, and also the soft decoder developed for the non-binary convolutional codes, we propose to apply the iterative technique to EDC-lattice-based MLNC.         

Fig. \ref{fig:BER.Case2} depicts the result when IMSD is used. It is observed that with $5$ iterations, the SER curve (black solid thick line) has a sharp turbo cliff reaching $\mathrm{SER}=10^{-5}$ at $10$dB, which is only $1.4$dB from the capacity. Thus, iterative decoding gives $3.3$dB gain over the traditional MSD decoding, and $5.3$dB gain over non-MSD decoding, as shown in the figure. Note that the simulation result is well consistent with the EXIT functions in Fig. \ref{fig:EXIT_F3F4}. When sufficient iterations are given, the L-value outputs from the soft detector at both layers are sufficiently reliable that the decoder can make the estimation with small probability of error. The simulation result also validates the soft detector algorithm specifically developed for EDC-based MLNC, and implies that there is large potential in employing iterative decoding in the multilevel lattice network coding.               

\textit{Layered Integer Forcing}: we have presented a general framework for the multilevel lattice network coding in section \ref{sec:Multilevel.Lattice.Network.Coding}. The work implies that any lattices with multilevel structure can be used in MLNC, and the essence of MLNC is to decode each layer separately such that the lattice decoder at each layer operates over smaller finite field or chain ring.  The layered integer forcing is a network decoding technique developed in terms of the algebraic structure of MLNC and hence, is generally applicable to any MLNC design. Thus, LIF is in principle capable of decoding EDC-lattice-based MLNC. According to Theorem \ref{Theorem.NewLattice.Decoding}, each layer forms a new quotient $S$-lattice $\Lambda/\Lambda_{i}^{\prime}$, and there exists a surjective $S$-module homomorphism $\varphi_i$ for the $i^{\mathrm{th}}$ layer such that $\mathcal{K}(\varphi_i)=\Lambda_i^{\prime}$. The general form of the generator matrix for $\Lambda_{i}^{\prime}$ based on the EDC lattice is given in (\ref{equ:attice.Generator}).  Note that $\Lambda_i^{\prime}$ is the coarse lattice for the new coset system $\Lambda/\Lambda_{i}^{\prime}$. 

In order to implement the LIF decoding for EDC lattices, we develop a modified Viterbi detector (see Appendix \ref{APD:3}) which can be viewed as a lattice quantizer based on the quotient $S$-lattice $\Lambda/\Lambda_{i}^{\prime}$ for the $i^{\mathrm{th}}$ layer, $i=1,2,\cdots,m$. 

\begin{figure}[!t]
  \centering
\begin{minipage}[t]{1\linewidth}
\centering
  \includegraphics[width=0.7\textwidth]{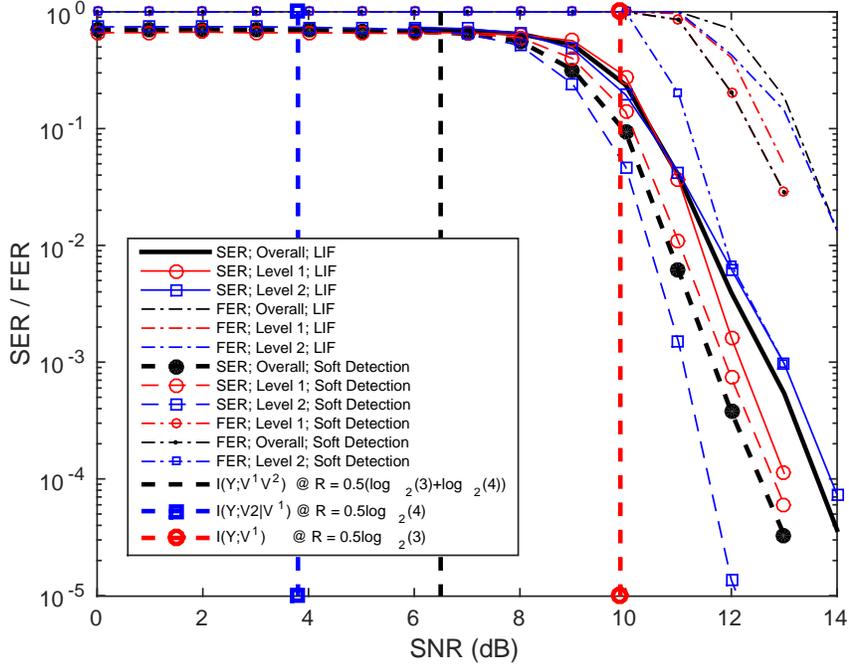}
\end{minipage}
\caption{\small SER and FER performance for an MLNC constructed from a two layer EDC Lattices; Soft detection; LIF;  $\mathcal{R}_{\mathrm{mes}}^{(1)}=\frac{1}{2}\log_2(3)$; $\mathcal{R}_{\mathrm{mes}}^{(2)}=\frac{1}{2}\log_2(4)$; $h_1=h_2=1$.} \label{fig:BER.Case3}
\end{figure}

Fig. \ref{fig:BER.Case3} illustrates the SER and FER performance based on LIF. It is observed (black solid line) that the overall SER has a good slope which validates the correctness of LIF and the modified Viterbi quantizer designed for the EDC lattice. The SER performance of LIF has approximately $0.8$dB loss at $10^{-5}$ in comparison to the soft detection approach. This is what we anticipate. First, the soft detection approach employs the BCJR algorithm for the convolutional decoding, which typically slightly outperforms  Viterbi detection. Then, the soft detector developed in section \ref{sec:Soft.Detector} and (\ref{equ:posteriori.1}) - (\ref{equ:posteriori.6}) outputs the soft information that the BCJR decoder uses to produce more reliable estimation than that for the Viterbi decoder. 

The soft detection approach is designed specifically for EDC lattices, and it is not strange to see that it gives better performance than LIF. Despite of this, we emphasise that LIF is universally applicable to any lattices having multilevel structure as detailed in section \ref{sec:Multilevel.Lattice.Network.Coding}, rather than just EDC lattices. For example, LIF is capable of solving MLNC problem when the lattices are directly designed in the Euclidean space (e.g. LDLC and signal codes). In summary, the application of the soft detection approach is more restrictive (which applies only to EDC lattices) and has relatively large complexity, but gives the best performance compared to LIF with Viterbi detection. However, LIF provides a solution for any kind of MLNC problem. Which method is preferable depends on the trade-off of  factors relevant to a particular scenario.    

\begin{figure}[!t]
  \centering
\begin{minipage}[t]{1\linewidth}
\centering
  \includegraphics[width=0.7\textwidth]{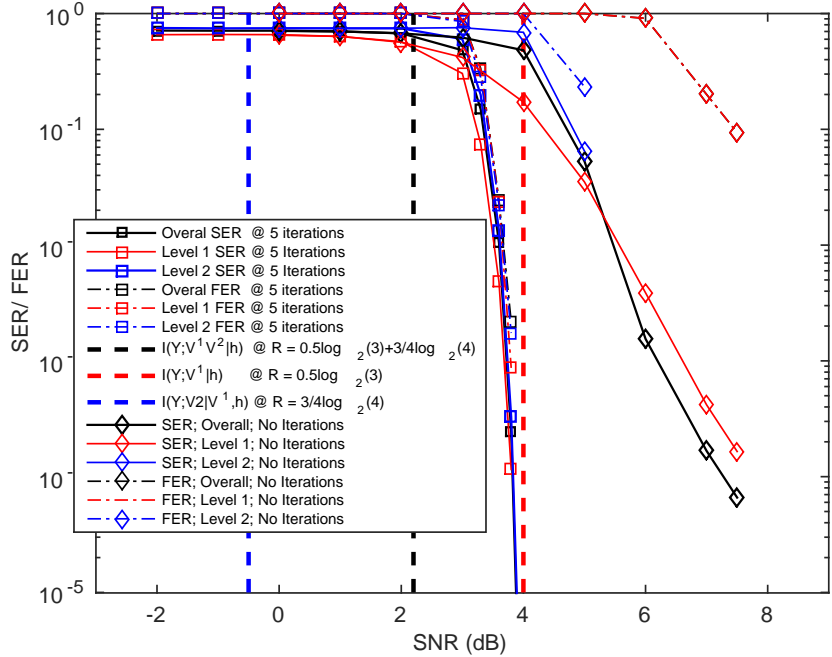}
\end{minipage}
\caption{\small SER and FER performance for an MLNC constructed from a two layer EDC Lattices; Soft detection; LIF;  $\mathcal{R}_{\mathrm{mes}}^{(1)}=\frac{1}{2}\log_2(3)$; $\mathcal{R}_{\mathrm{mes}}^{(2)}=\frac{1}{2}\log_2(4)$; $\mathbf{h} = [-1.17+2.15i, 1.25-1.63i]$.} \label{fig:BER.Case4}
\end{figure}

In Fig. \ref{fig:BER.Case4}, we also show the performance of the LSD when the fixed fading is considered. The channel fading vector is set to $\mathbf{h}=[-1.17+2.15i, 1.25-1.63i]$, which is the same as the fading vector used in scenario 1 of \cite{Feder.ConvSignalCodes.TIT.2011}. We employ a half-rate code for layer 1, and $\frac{3}{4}$-rate code for layer 2. The optimal $S$-integer vector for the two layers are selected in terms of (\ref{equ:coefficient.selection.MI}). We employ multistage decoding with 5 iterations between the two layers. A sharp turbo cliff occurs, which reaches $\mathrm{SER}=10^{-5}$ at $3.9$dB, approximately $1.7$dB from the capacity. When no iteration is employed, there is more than $5$dB loss. This implies that small number of additional iterations to generate more reliable values is worthwhile in improving the overall SER performance.  The iterative multistage soft detection for EDC lattices achieves the overall rate of $\mathcal{R}_{\mathrm{mes}}\approx 2.29$ bits/symbol at $3.9$dB. This demonstrates the potential of iterative decoding in improving the performance of physical layer network coding.

\section{Conclusions} \label{sec:conclusions}
The paper has laid the foundations for a new research area in multilevel lattices for LNC, and built on the theoretic work for MLNC which inherently allows practically feasible decoding design for network coding, and correspondingly we have developed a layered integer forcing approach which plays such a role. We have proposed a general lattice construction, i.e. EDC, based on MLNC theorems, given the generator matrix forms and shown its merits, especially for complexity reduction and code design flexibility. We have considered three possible EDC lattice structures, and mathematically proved that EDC subsumes the most important previous complex constructions, e.g. A and D. We have laid the foundations for another new research area in iteration-aided multistage decoding for EDC-based MLNC, which is based on the layered soft detector developed in section \ref{sec:Iterative.Detection.EDC}, and have explored its extrinsic information transfer characteristics. The results well support our viewpoint that LSD works well with multistage iterative decoding in MLNC, and provides better performance than the traditional non-iterative system. We have developed a modified Viterbi decoder based on LIF for EDC-based MLNC, and made performance comparison between iterative decoding, multistage decoding and LIF.    

%

We expect that all of these will provide the basis for extensive further work, both to explore the rich algebraic features of the new construction and the homomorphism design, and to exploit it in practical implementations of LNC in 5G wireless systems.   

\section{Acknowledgements}
The authors would like to thank Chen Feng from University of Toronto for useful discussions on lattice network coding.

\bibliographystyle{IEEEtran}

\bibliography{MLNC_Bib}

\begin{thebibliography}{10}
\providecommand{\url}[1]{#1}
\csname url@samestyle\endcsname
\providecommand{\newblock}{\relax}
\providecommand{\bibinfo}[2]{#2}
\providecommand{\BIBentrySTDinterwordspacing}{\spaceskip=0pt\relax}
\providecommand{\BIBentryALTinterwordstretchfactor}{4}
\providecommand{\BIBentryALTinterwordspacing}{\spaceskip=\fontdimen2\font plus
\BIBentryALTinterwordstretchfactor\fontdimen3\font minus
  \fontdimen4\font\relax}
\providecommand{\BIBforeignlanguage}[2]{{%
\expandafter\ifx\csname l@#1\endcsname\relax
\typeout{** WARNING: IEEEtran.bst: No hyphenation pattern has been}%
\typeout{** loaded for the language `#1'. Using the pattern for}%
\typeout{** the default language instead.}%
\else
\language=\csname l@#1\endcsname
\fi
#2}}
\providecommand{\BIBdecl}{\relax}
\BIBdecl

\bibitem{Erez.Zamir}
U.~Erez and R.~Zamir, ``Achieving 1/2 log (1+snr) on the awgn channel with
  lattice encoding and decoding,'' \emph{IEEE Trans. on Inf. Theory}, vol.~50,
  no.~10, pp. 2293--2314, Oct 2004.

\bibitem{Forney.Coset.Codes}
G.~Forney, M.~Trott, and S.-Y. Chung, ``Sphere-bound-achieving coset codes and
  multilevel coset codes,'' \emph{IEEE Trans. on Inf. Theory}, vol.~46, no.~3,
  pp. 820--850, May 2000.

\bibitem{Zhang.PLNC.ACM.2006}
S.~Zhang, ``Hot topic: physical-layer network coding,'' in \emph{in Proc. of
  ACM Mobicom}, 2006, pp. 358--365.

\bibitem{Yi.CLDLC.2015}
Y.~Wang and A.~Burr, ``Complex low density lattice codes to lattice network
  coding,'' in \emph{IEEE International Communications Conference (ICC)}, June
  2015.

\bibitem{Nazer.Gastpar.TIT.2011}
B.~Nazer and M.~Gastpar, ``Compute-and-forward: Harnessing interference through
  structured codes,'' \emph{IEEE Trans. Inf. Theory}, vol.~57, no.~10, pp.
  6463--6486, Oct 2011.

\bibitem{Feng.AlgeAppro.TIT.2013}
C.~Feng, D.~Silva, and F.~Kschischang, ``An algebraic approach to
  physical-layer network coding,'' \emph{IEEE Trans. Inf. Theory}, vol.~59,
  no.~11, pp. 7576--7596, Nov 2013.

\bibitem{Qifu.Eisenstein}
Q.~Sun, J.~Yuan, T.~Huang, and K.~Shum, ``Lattice network codes based on
  eisenstein integers,'' \emph{IEEE Trans. Commun.}, vol.~61, no.~7, pp.
  2713--2725, July 2013.

\bibitem{Nara.Eisenstein}
N.~Tunali, K.~Narayanan, J.~Boutros, and Y.-C. Huang, ``Lattices over
  eisenstein integers for compute-and-forward,'' in \emph{Communication,
  Control, and Computing (Allerton), 2012 50th Annual Allerton Conference on},
  Oct 2012, pp. 33--40.

\bibitem{Multistage.Product.Construction.Nara}
Y.-C. Huang and K.~Narayanan, ``Multistage compute-and-forward with multilevel
  lattice codes based on product constructions,'' in \emph{Information Theory
  (ISIT), 2014 IEEE International Symposium on}, June 2014, pp. 2112--2116.

\bibitem{Lessons.Rings.Modules}
D.~G. Northcott, \emph{Lessons on Rings, Modules and Multiplicities}, 1968.

\bibitem{Feng.FCR.TIT}
C.~Feng, R.~Nobrega, F.~Kschischang, and D.~Silva, ``Communication over
  finite-chain-ring matrix channels,'' \emph{IEEE Trans. Inf. Theory}, vol.~60,
  no.~10, pp. 5899--5917, Oct 2014.

\bibitem{Finite.Rings.Identity}
B.~R. McDonald, \emph{Finite Rings with Identity}, 1974.

\bibitem{Conway.Sloane}
J.~H. Conway and N.~J.~A. Sloane, \emph{Sphere Packings, Lattices and Groups},
  1998.

\bibitem{Huber.Multilevel.Codes}
U.~Wachsmann, R.~Fischer, and J.~Huber, ``Multilevel codes: theoretical
  concepts and practical design rules,'' \emph{IEEE Trans. Inf. Theory},
  vol.~45, no.~5, pp. 1361--1391, Jul 1999.

\bibitem{AGB.TIT}
A.~Burr and T.~Lunn, ``Block-coded modulation optimized for finite error rate
  on the white {G}aussian noise channel,'' \emph{IEEE Trans. Inf. Theory},
  vol.~43, no.~1, pp. 373--385, Jan. 1997.

\bibitem{CLLL.Cong}
Y.~H. Gan, C.~Ling, and W.~H. Mow, ``Complex lattice reduction algorithm for
  low-complexity full-diversity mimo detection,'' \emph{IEEE Trans. Sig.
  Processing}, vol.~57, no.~7, pp. 2701--2710, July 2009.

\bibitem{Yi.Multilevel.TCOM}
Y.~Wang and A.~Burr, ``Code design for iterative decoding of multilevel
  codes,'' \emph{IEEE Trans. Commun.}, vol.~63, no.~7, pp. 2404--2419, July
  2015.

\bibitem{ten.Brink.Erasure.Channel}
A.~Ashikhmin, G.~Kramer, and S.~ten Brink, ``Extrinsic information transfer
  functions: model and erasure channel properties,'' \emph{IEEE Trans. Inf.
  Theory}, vol.~50, pp. 2657--2673, Nov 2004.

\bibitem{Efficient.Computation.Soon}
J.~Kliewer, S.~X. Ng, and L.~Hanzo, ``Efficient computation of exit functions
  for nonbinary iterative decoding,'' \emph{IEEE Trans. Inf. Commun.}, vol.~54,
  no.~12, pp. 2133--2136, Dec 2006.

\bibitem{CLDLC.Origin}
Y.~Yona and M.~Feder, ``Complex low density lattice codes,'' in
  \emph{Information Theory Proceedings (ISIT), 2010 IEEE International
  Symposium on}, Jun. 2010, pp. 1027--1031.

\bibitem{Feder.ConvSignalCodes.TIT.2011}
O.~Shalvi, N.~Sommer, and M.~Feder, ``Signal codes: Convolutional lattice
  codes,'' \emph{IEEE Trans. Inf. Theory}, vol.~57, no.~8, pp. 5203--5226, Aug
  2011.

\end{thebibliography}

\appendices

\section{Layered Soft detector for EDC-based MLNC}\label{APD:1}
We show here the calculation of mutual information/ conditional mutual information  between the received superimposed signals (faded and noisy) and the network coded symbols at the $j^{\mathrm{th}}$ level. We denote by $\mathbf{w}_{j}=[w_j^1,\cdots,w_j^m]$, $j=1,2,\cdots,L$ the realizations of a vector-based random variable $\mathbf{W}_j$ representing the messages of all $m$ levels at the $j^{\mathrm{th}}$ source, and $\mathbf{w}^{i}=[w_1^i,\cdots,w_L^i]$, $j=1,2,\cdots,m$ the realizations of a vector-based random variable $W^i$ representing the messages of all $L$ sources for the $i^{\mathrm{th}}$ level.  We refer to $\mathbf{w}=[\mathbf{w}_1,\cdots,\mathbf{w}_L]$ as the realizations of another vector-based random variable $\mathbf{W}$.  Note that $w_j^i\in\mathbb{F}_{\tilde{p}_i}$ is the message of the $i^\mathrm{th}$ level and the $j^{\mathrm{th}}$ source, which is uniformly distributed over $\mathbb{F}_{\tilde{p}_i}$. $V^i$ is a random variable which takes on a set of possible values of $f^i(w_1^i,\cdots,w_L^i)\in\mathbb{F}_{\tilde{p}_i}$, and $v^i$ is the corresponding realization of $V^i$ for the $i^{\mathrm{th}}$ layer.


The \textit{a posteriori} probability of the event $V^i=v^i$ at the $i^{\mathrm{th}}$ level conditioned on the MAC outputs $Y=y$ and the \textit{a priori} channel outputs $\mathbf{\Xi} = \boldsymbol \xi = [\xi^1,\cdots,\xi^m]$, is given by
\begin{align}
&\mathrm{Pr}(V^{i} = v^i | y, \boldsymbol\xi ) \label{equ:posteriori.1} \\
 =& \sum_{V^i = v^i}\sum_{\mathbf{V}^{\setminus i}\in\mathbb{F}_{\mathbf{q}_{\setminus i}}^{m-1}} \frac{\mathrm{Pr}(Y|V^1\cdots V^m, \boldsymbol\xi)\mathrm{Pr}(\boldsymbol\xi |V^1\cdots V^m) \mathrm{Pr}(V^1\cdots V^m) } {\mathrm{Pr}(Y,\boldsymbol\xi)} \label{equ:posteriori.2} \\
 =& \mathrm{Pr}(\xi^i | V^i = v) \sum_{V^i = v}\cdot\sum_{\mathbf{V}^{\setminus i}\in\mathbb{F}_{\mathbf{q}_{\setminus i}}^{m-1}} \frac{\mathrm{Pr}(Y|V^1\cdots V^m)\mathrm{Pr}(\boldsymbol{\xi}^{\setminus i} |\boldsymbol{V}^{\setminus i}) \mathrm{Pr}(V^1\cdots V^m) } {\mathrm{Pr}(Y,\boldsymbol\xi)}  \label{equ:posteriori.3} \\
=&\frac{ \mathrm{Pr}(\xi^i | V^i = v)}{\mathrm{Pr}(Y,\boldsymbol\xi)}
\sum_{\substack{\forall(w_1^i,\cdots,w_L^i):\\ f^i(w_1^i,\cdots,w_L^i)=v}} \cdot \sum_{\forall\boldsymbol{V}^{\setminus i}\in\mathbb{F}_{\mathbf{q}_{\setminus i}} }\cdot 
\sum_{\substack{\forall(\mathbf{w}_{1}^{\setminus i},\cdots,\mathbf{w}_{L}^{\setminus i}):\\ f^{1}(\mathbf{w}^1)= v^{1},\cdots, f^{m}(\mathbf{w}^m)=v^{m} }} 
\mathrm{Pr}(Y|\mathbf{W}=\mathbf{w})\mathrm{Pr}(\boldsymbol{\xi}^{\setminus i} |\boldsymbol{V}^{\setminus i})\mathrm{Pr}(\mathbf{W}=\mathbf{w}) \label{equ:posteriori.4} 
\end{align}
where $\mathbb{F}_{\mathbf{p}_{\setminus i}}^{m-1}\triangleq[\mathbb{F}_{\tilde{p}_1}\cdots\mathbb{F}_{\tilde{p}_{i-1}}  ,\mathbb{F}_{\tilde{p}_{i+1}}\cdots\mathbb{F}_{\tilde{p}_{m}}]$ consists of  a set of finite field and finite chain ring. $\mathbb{F}_{\tilde{p}_i}$ is defined in section \ref{sec:EDC}. Note that if $\tilde{p}_i$ is not a prime number,  $\mathbb{F}_{\tilde{p}_i}$ can be decomposed furthermore in terms of the $p$-adic decomposition theorem \cite{Feng.FCR.TIT}, and small modifications of (\ref{equ:posteriori.4}) are required accordingly.  The conditional probability density function is given by,
\begin{align}
&\mathrm{Pr}(Y|\mathbf{W}=\mathbf{w}) \notag \\
=&\mathrm{Pr}(Y|\mathbf{w}_1,\cdots,\mathbf{w}_L) \notag \\
=& \frac{1}{\sqrt{\pi N_0}} e^{-\frac{\left|h_1\sigma^{-1}(\mathbf{w}_1^1\times\cdots\times\mathbf{w}_1^m)+\cdots+h_L\sigma^{-1}(\mathbf{w}_L^1,\times\cdots\times\mathbf{w}_L^m) - y   \right|^2}{N_0}} \label{equ:APD:Gaussin.Dist}
\end{align}
The \textit{a posteriori} L-value $d^i_k$ for the event $V^i=v^i_k$ is defined in (\ref{equ:posteriori.5}) which can be further separated into two terms in (\ref{equ:posteriori.6}), where $a_k^i$ is an element in (\ref{equ:a.priori.vector}) which serves as the \textit{a priori} L-value. Note that $v_k^i$ is the $k^{\mathrm{th}}$ realization of the random variable $V^i$.  Following (\ref{equ:a.priori.vector}), we have
$$\mathrm{Pr}(\xi|V=v_k^i) = \frac{e^{a_k^i}}{1+\sum_{k:\forall v_k^i\in\mathbb{F}_{\tilde{p}_i}, v_k^i\neq 0}e^{a_k^i}}$$
The second term of (\ref{equ:posteriori.6}) serves as the extrinsic L-value $e_k^i$ for the $i^{\mathrm{th}}$ level and the $k^{\mathrm{th}}$ realization of the vector-based random variable $\mathbf{E}^i$ which is the extrinsic information of the $i^\mathrm{th}$ level. 

\begin{align}
d_k^i &= \log\left( \frac{\mathrm{Pr}(V^{i} = v_k^i | y, \boldsymbol\xi )}{\mathrm{Pr}(V^{i} = 0| y, \boldsymbol\xi )} 	\right) \label{equ:posteriori.5} \\
&= a_k^i + \log \left(\frac{\sum_{\substack{\forall(w_1^i,\cdots,w_L^i):\\ f^i(w_1^i,\cdots,w_L^i)=v_k^i}} \cdot \sum_{\forall\boldsymbol{V}^{\setminus i}\in\mathbb{F}_{\mathbf{q}_{\setminus i}} }\cdot 
\sum_{\substack{\forall(\mathbf{w}_{1}^{\setminus i},\cdots,\mathbf{w}_{L}^{\setminus i}):\\ f^{1}(\mathbf{w}^1)= v^{1},\cdots, f^{m}(\mathbf{w}^m)=v^{m} }} 
\mathrm{Pr}(Y|\mathbf{W}=\mathbf{w})\mathrm{Pr}(\boldsymbol{\xi}^{\setminus i} |\boldsymbol{V}^{\setminus i})  }
{\sum_{\substack{\forall(w_1^i,\cdots,w_L^i):\\ f^i(w_1^i,\cdots,w_L^i)=0}} \cdot \sum_{\forall\boldsymbol{V}^{\setminus i}\in\mathbb{F}_{\mathbf{q}_{\setminus i}} }\cdot 
\sum_{\substack{\forall(\mathbf{w}_{1}^{\setminus i},\cdots,\mathbf{w}_{L}^{\setminus i}):\\ f^{1}(\mathbf{w}^1)= v^{1},\cdots, f^{m}(\mathbf{w}^m)=v^{m}  }} 
\mathrm{Pr}(Y|\mathbf{W}=\mathbf{w})\mathrm{Pr}(\boldsymbol{\xi}^{\setminus i} |\boldsymbol{V}^{\setminus i})}  	\right) \label{equ:posteriori.6} \\
&= a_k^i + e_k^i \notag
\end{align}

\section{Mutual information for linear combinations} \label{APD:2}
The mutual information between the received signal and the decoded linear combination at the $i^{\mathrm{th}}$ layer is:
\begin{align}
I(Y,H;V^i) &= \boldsymbol E_{(Y,V^i,H)}\left[\log_2\frac{P(Y|V^i,H)}{P(Y)}  \right] \notag \\
&=\sum_{v^i}\mathrm{Pr}(V^i=v_i)\int_{\mathbb{C}}P(H)\int_{\mathbb{C}}P(Y|V^i,H)\log_2\frac{P(Y|V^i,H)}{P(Y)}\mathrm{d}Y\mathrm{d}H
\end{align}

The probability density function $P(Y|V^i,H)$ conditioned on $V^i=v^i$ should be calculated by:
\begin{align}
P(Y|V^i=v^i,H) = \frac{1}{\mathrm{Pr}(V^i=v^i)}\sum_{\forall \boldsymbol{v}^{\setminus i}}\sum_{\substack{\forall(\mathbf{w}_1,\cdots,\mathbf{w}_L):\\ f^{1}(\mathbf{w}^1)= v^{1},\cdots, f^{m}(\mathbf{w}^m)=v^{m} }} P(Y|\mathbf{W=w},h)P(\mathbf{W=w}) 
\end{align}

The conditional mutual information $I(Y;V^i|V^1\cdots V^{i-1})$ gives the maximum achievable rate at the $i^{\mathrm{th}}$ layer when the linear combinations of the preceding stages are perfectly known, which can be calculated by:
\begin{align}
&I(Y;V^i|V^1\cdots V^{i-1})  \notag \\
=& \boldsymbol E_{(Y,V^1,\cdots,V^i,H)}\left[\log_2\frac{P(Y|V^1\cdots V^i,H)}{P(Y|V^1\cdots V^{i-1},H)}  \right] \notag \\
=&\sum_{v^1\cdots v^{i-1}}\mathrm{Pr}(V^1=v^1,\cdots,V^i=v_{i-1})
\sum_{v^i}\int_{\mathbb{C}}P(H)\int_{\mathbb{C}}P(Y,V^i|V^1\cdots V^{i-1},H)\log_2\frac{P(Y|V^1\cdots V^i,H)}{P(Y|V^1\cdots V^{i-1},H)}\mathrm{d}Y\mathrm{d}H
\end{align}
where the conditional probability density function $P(Y,V^i|V^1\cdots V^{i-1},H)$ should be calculated in terms of the random variables of the messages, which is given by:
\begin{align}
&P(Y,V^i|V^1=v^1,\cdots,V^{i-1}=v^{i-1},H) \notag \\
 =& \frac{1}{\mathrm{Pr}(v^1,\cdots,v^{i-1})}\sum_{\forall\mathbf{v}\notin(v^1\cdots v^{i})} \sum_{\substack{\forall(\mathbf{w}_1,\cdots,\mathbf{w}_L):\\ f^{1}(\mathbf{w}^1)= v^{1},\cdots, f^{m}(\mathbf{w}^m)=v^{m} }}P(Y|\mathbf{W=w},h)\mathrm{Pr}(\mathbf{W=w}) 
\end{align} 
where $P(Y|\mathbf{W=w},h)$ is given in (\ref{equ:APD:Gaussin.Dist}). Note that $V^i$, $i=1,2,\cdots,m$ is a random variable defined by the linear function of the $i^{\mathrm{th}}$ layer over $\mathbb{F}_{\tilde{p}_i}$. Every $V^i$ operates over different finite filed or chain ring. 

\section{LIF Quantizer}\label{Appendix.3}\label{APD:3}
We show here a LIF quantizer $\mathcal{Q}_{\mathrm{LIF}}^{(i)}$ implemented via a modified Viterbi decoder. The quantization problem for the $i^{\mathrm{th}}$ layer can be mathematically expressed as:
\begin{align}
&\arg\min_{\mathbf{c}_i}  ||\alpha^i\mathbf{y} - (\tilde{\sigma}^{-1}(\mathbf{c}^i) +\lambda_i^{\prime} ) ||^2 \\
=&\arg\min_{\mathbf{c}_i}|| (\alpha^i\mathbf{y} -\tilde{\sigma}^{-1}(\mathbf{c}^i))-\mathcal{Q}_{\Lambda_i^{\prime}}((\alpha^i\mathbf{y} -\tilde{\sigma}^{-1}(\mathbf{c}^i)) ||^2 \\
&\mathrm{subject~ to:} ~~~~~~~~ \mathbf{c}^i\in\mathcal{C}^i, ~~~~ \lambda_i^{\prime}\in\Lambda_i^{\prime},  \\
&~~~~~~~~~~~~  \tilde{\sigma}(\lambda)\in \mathcal{C}^1\oplus\cdots\oplus (\mathcal{C}^i=\mathbf{c}^i)\oplus\cdots\mathcal{C}^m
\end{align}
where $\mathcal{Q}_{\Lambda_i^{\prime}}(\mathbf{x})$ is the coarse lattice quantizer for the $i^{\mathrm{th}}$ layer and can be expressed as a modulo operation $\mathbf{x}~~\mathrm{mod}~~ \Lambda_i^{\prime}$ (as defined in Theorem \ref{Theorem.NewLattice.Decoding}). $\tilde{\sigma}^{-1}(\cdot)$ is the inverse operation of $\tilde{\sigma}$ which produces a set of lattice points $\lambda$.

We can construct trellis for the non-binary convolutional code $\mathcal{C}^i$. Assume that the states of the $k^{\mathrm{th}}$ and $(k+1)^{\mathrm{th}}$ time slots are $s_k$ and $s_{k+1}$, respectively. The codeword of the branch that exists from $s_k$ and arrives at $s_{k+1}$ is denoted as $c^i_{s_k\rightarrow s_{k+1}}$.   
The metric for each branch is given by 
\begin{align}
 ||(\alpha^i\mathbf{y} - \sigma^{-1}(c^i_{s_k\rightarrow s_{k+1}}))-\mathcal{Q}_{\Lambda_i^{\prime}}((\alpha^i\mathbf{y} - \sigma^{-1}(c^i_{s_k\rightarrow s_{k+1}}))||^2
\end{align}
where $\sigma^{-1}(\cdot)$ is the inverse operation of $\sigma(\cdot)$ defined in section \ref{sec:EDC}. We employ Viterbi algorithm to estimate the best possible outcome $\mathbf{c}^i$. This implements the LIF quantizer $\mathcal{Q}_{\mathrm{LIF}}^{(i)}$ for EDC-based MLNC.

\section{Proof of Propositions \ref{prop:Coding.Gain.A.1} and \ref{prop:Coding.Gain.A.2} } \label{APD:4}
The codeword of the $i^{\mathrm{th}}$ layer is $\mathbf{c}^i = \left(c_1^i + \langle\varpi\rangle, \cdots,c_n^i + \langle\varpi\rangle  \right) \in \mathcal{C}^i $. $\mathcal{C}^i$ is a linear code over $\delta_iS/\langle	\varpi\rangle$ which is generated by  $\tilde{\sigma}_i([\mathbf{I}_{k_i}~~ \mathbf{B}^i_{k_i \times (n-k_i)}])$ where $[\mathbf{I}_{k_i}~~ \mathbf{B}^i_{k_i \times (n-k_i)}]$ is a $k_i\times n$ matrix over $\delta_iS$. These are defined in (\ref{equ:EDC.Primary.Lattice.Definition}) and (\ref{equ:EDC.Primary.Map}) . The minimum-norm coset leader in the $i^{\mathrm{th}} $ layer primary sublattice system is given by:
\begin{align}
\tilde{\sigma}_{i,\bigtriangleup}(\mathbf{c}^i) &= \left(c_1^i - \mathcal{Q}_i(c_1^i/p_i\delta_i)p_i\delta_i,\cdots,c_n^i - \mathcal{Q}_i(c_n^i/p_i\delta_i)p_i\delta_i\right) \notag \\
& = \left(c_1^i - \varpi\mathcal{Q}_i(c_1^i/\varpi),\cdots,c_n^i - \varpi\mathcal{Q}_i(c_n^i/\varpi) \right) \label{equ:minimum.norm.A}
\end{align}
where $\mathcal{Q}(z)$ is a quantizer which sends $z\in\mathbb{C}$ to the closest point in $S$. We denote $d^2(\Lambda_{p_i}/\Lambda^{\prime})$ as the length of the squared shortest vectors in the set $\Lambda_{p_i}\setminus \Lambda^{\prime}$, then
\begin{align}
d^2(\Lambda_{p_i}/\Lambda^{\prime}) = \min_{\mathbf{c}^i\neq\mathbf{0}, \mathbf{c}^i\in\mathcal{C}^i}||\tilde{\sigma}_{i,\bigtriangleup}(\mathbf{c}^i)||^2 = \omega_{\mathrm{min}}^{(i)}(\mathcal{C}^i)\label{equ:dmin.primary.A}
\end{align}
The volume of the Voronoi region of $\Lambda^{\prime}$ is $V(\mathcal{V}(\Lambda^{\prime})) = \vartheta^n|\varpi|^{2n}$, where $\vartheta$ is a scaling factor depending on which PID is used, e,g., $\vartheta = \sqrt{3}/2$ when $S$ is Eisenstein integer. The nominal coding gain for the $i^{\mathrm{th}}$ layer primary sublattices is:
\begin{align}
\varrho(\Lambda_{p_i}/\Lambda^{\prime}) &= \frac{\omega_{\mathrm{min}}^{(i)}(\mathcal{C}^i)}{\left(V(\mathcal{V}(\Lambda_{p_i}))\right)^{\frac{1}{n}}} \notag \\
&= \frac{\omega_{\mathrm{min}}^{(i)}(\mathcal{C}^i)}{(\vartheta^n |p_i|^{2(n-k_i)}|\delta_i|^{2n})^{\frac{1}{n}} } \notag \\
&= \frac{\omega_{\mathrm{min}}^{(i)}(\mathcal{C}^i)}{\vartheta |p_i|^{2(1-\frac{k_i}{n})}|\delta_i|^{2} } \label{NomGain.A.Primary}
\end{align}

We now prove the kissing number for the $i^{\mathrm{th}}$ layer primary subllatices. Let $N(\omega_{\mathrm{min}}^{(i)}(\mathcal{C}^i))$ be the number of codewords in $\mathcal{C}^i$ with the minimum Euclidean weight $\omega_{\mathrm{min}}^{(i)}(\mathcal{C}^i)$, and $\mathcal{N}_{\mathcal{U}(S)}$ be the number of units in $S$, e.g.  $\mathcal{N}_{\mathcal{U}(\mathbb{Z}([\omega]))} = 6$. When $|p_i|^2-1\leq\mathcal{N}_{\mathcal{U}(S)} $, recall that $\mathcal{C}^i$ is a linear code over $\delta_iS/\langle\varpi\rangle$. The number of non-zero elements of coset leaders in $\delta_iS/\langle\varpi\rangle$ is $|p_i|^2-1$, and these elements must be a subset of  $\delta_i\mathcal{U}(S)$. Hence there are $\frac{\mathcal{N}_{\mathcal{U}(S)}}{|p_i|^2-1}$ elements in the coset leaders which are formed by the same codeword and give the shortest vector. This means the number of the non-zero elements in a codeword is precisely $\frac{\omega_{\mathrm{min}}^{(i)}(\mathcal{C}^i)}{|\delta_i|^2}$. When $|p_i|^2-1>\mathcal{N}_{\mathcal{U}(S)} $, every neighbour point is represented by different codewords, and hence the kissing number of the $i^{\mathrm{th}}$ layer primary sublattices is given by:
\begin{equation}
K(\Lambda_{p_i}/\Lambda^{\prime}) = \left\{
             \begin{array}{ll}
              N(\omega_{\mathrm{min}}^{(i)}(\mathcal{C}^i))\left(\frac{\mathcal{N}_{\mathcal{U}(S)}}{|p_i|^2-1}\right)^{\frac{\omega_{\mathrm{min}}^{(i)}(\mathcal{C}^i)}{|\delta_i|^2}}, & |p_i|^2-1 \leq \mathcal{N}_{\mathcal{U}(S) }\\
              N(\omega_{\mathrm{min}}^{(i)}(\mathcal{C}^i)), &  \mathrm{Otherwise}
             \end{array}  
        \right.
\end{equation}
and now proposition \ref{prop:Coding.Gain.A.1} is proved.

From the proof of Theorem \ref{Theorem.Lattice.Decomposition}, we have $\tilde{\mathbf{c}}=  \mathbf{c}^1 + \mathbf{c}^2 + \cdots + \mathbf{c}^m$, $\tilde{\mathbf{c}}\in \tilde{\mathcal{C}}$ and $\tilde{\mathcal{C}}\in (S/\langle \varpi\rangle)^n$. The minimum-norm coset leader for $\Lambda/\Lambda^{\prime}$ can be represented by codewords used for all layers, thus:
\begin{align}
\tilde{\sigma}_{\bigtriangleup}(\tilde{\mathbf{c}}) &= \left(\tilde{c}_1 - \varpi\mathcal{Q}\left(\frac{\tilde{c}_1}{\varpi} \right),\cdots,\tilde{c}_n - \varpi\mathcal{Q}\left(\frac{\tilde{c}_n}{\varpi} \right)   \right) 
\end{align}
where $\tilde{c}_j = c_j^1 +c_j^2 + \cdots + c_j^m$ and $c_j^i$, $j=1,2,\cdots,n$, $i=1,2,\cdots,m$, denotes the $j^{\mathrm{th}}$ element of the codeword $\mathbf{c}^i$. Then, the squared shortest vectors in the set $\Lambda\setminus \Lambda^{\prime}$ can be represented by the $m$ linear codes used at each layer,  
\begin{align*}
d^2(\Lambda/\Lambda^{\prime}) = \min_{\tilde{\mathbf{c}}\neq\mathbf{0}, \tilde{\mathbf{c}}\in\mathcal{C}}||\tilde{\sigma}_{\bigtriangleup}(\tilde{\mathbf{c}})||^2 = \omega_{\mathrm{min}}(\tilde{\mathcal{C}})
\end{align*}
The nominal coding gain for $\Lambda/\Lambda^{\prime}$ is
\begin{align}
\varrho(\Lambda/\Lambda^{\prime}) &= \frac{\omega_{\mathrm{min}} (\tilde{\mathcal{C}})}{\left(V(\mathcal{V}(\Lambda))\right)^{\frac{1}{n}}} \notag \\
&= \frac{\omega_{\mathrm{min}}(\tilde{\mathcal{C}})\prod_{\ell=2}^m|p_{\ell}|^{\frac{2(k_{\ell} - k_1)}{n}}}{\vartheta |p_1|^{2(1-\frac{k_1}{n})}|\delta_1|^{2}}  
\label{NomGain.A.Fine}
\end{align}
where we assume $k_1\leq k_2\leq\cdots\leq k_m$ in (\ref{NomGain.A.Fine}). Let $N(\omega_{\mathrm{min}}(\tilde{\mathcal{C}}))$ be the number of codewords in $\tilde{\mathcal{C}}$ with the minimum Euclidean weight $\omega_{\mathrm{min}}(\tilde{\mathcal{C}})$, the kissing number of this kind of lattices is:
\begin{equation}
K(\Lambda /\Lambda^{\prime}) =  N(\omega_{\mathrm{min}}(\tilde{\mathcal{C}}))   
\end{equation}
proposition \ref{prop:Coding.Gain.A.2} is thereby proved.

\section{Proof of Propositions \ref{prop:Coding.Gain.D.1} and \ref{prop:Coding.Gain.D.2} } \label{APD:5}
As explained in section \ref{sec:EDC}, the codeword $\mathcal{C}^i$ of the $i^{\mathrm{th}}$ layer operates over the finite chain ring $\mathcal{C}^i\in \delta_iS/\langle \varpi\rangle$, where $\delta_i = \varpi/p_i^{\gamma_i}$. 
Following (\ref{equ:FCR.Generator}), $\tilde{\sigma}_i(\mathbf{w}^i\mathbf{G}_{\mathrm{FCR}}^i) \longmapsto (\delta_iS/\langle\varpi\rangle)^n$, here the message space of the $i^{\mathrm{th}}$ layer is defined as: 
\begin{align*}
\mathbf{W}^i \cong (\delta_iS/p_i^{\gamma_i}\delta_iS)^{k_{i,0}^{\prime}}\oplus\cdots\oplus (\delta_iS/p_i\delta_iS)^{k_{i,\gamma_i-1}^{\prime}}
\end{align*}
Then the minimum-norm coset leader of $\Lambda_{p_i}/\Lambda^{\prime}$ has similar form as (\ref{equ:minimum.norm.A}) with $\mathbf{c}^i\in \mathcal{C}^i$. The nominal coding gain $\varrho(\Lambda_{p_i}/\Lambda^{\prime})$ can be obtained based on the same derivation in (\ref{equ:dmin.primary.A}) and (\ref{NomGain.A.Primary}). We are more interested in constructing the primary sublattices with some linear codes over the finite field. This can be implemented via the complex construction D approach, based on a set of nested linear codes, as proved in section \ref{sec:EDC}. The residue field is now defined as $Q\triangleq\delta_iS/\langle p_i\delta_i\rangle$. Let $\mathcal{C}^{i,0}\subseteq\cdots\subseteq\mathcal{C}^{i,\gamma_i-1}$ be nested linear codes of length $n$ over $Q$, where $\mathcal{C}^{i,t}$ is an $[n,\sum_{\ell=0}^tk_{i,\ell}^{\prime}]$ linear code for the $t^{\mathrm{th}}$ level of the $i^{\mathrm{th}}$ layer, $t=0,1,\cdots,\gamma_i-1$. Note that $\mathcal{C}^{i,t} $ is row spanned by the vector space:
\begin{align}
\mathbf{g}_{\mathcal{C}^{i,t}} = \begin{bmatrix}
\mathbf{g}^i_{k_{i,0}^{\prime}}\\
\mathbf{g}^i_{k_{i,1}^{\prime}}\\
\vdots \\
\mathbf{g}^i_{k_{i,t}^{\prime}}\label{equ:basis.construct.D.}
\end{bmatrix}
\end{align}
where $\mathbf{g}^i_{k_{i,t}^{\prime}}\in Q_{k_{i,t}^{\prime}\times n}$. None of the rows of $\mathbf{g}_{\mathcal{C}^{i,t}}$ are linear combination of the other rows. It is obvious that the primary sublattice point $\lambda_{p_i}\in\Lambda_{p_i}\setminus\Lambda^{\prime}$ is given by:
\begin{align}
\lambda_{p_i} = \underbrace{\underbrace{\underbrace{p_i^{\gamma_i}\delta_i\mathbf{s} + p_i^{\gamma_i-1}\mathbf{c}^{i,\gamma_i-1}}_{\Lambda_{p_i}^{\gamma_i-1}} + \cdots + p_i\mathbf{c}^{i,1}}_{{\Lambda_{p_i}^{1}}} + \mathbf{c}^{i,0} }_{\Lambda_{p_i}^0}
\end{align}
where $\mathbf{c}^{i,t}$ must not be zero for all possible $t$ values.  The outer lattice $\Lambda_{p_i}^{\gamma_i-1} = \{p_i^{\gamma_i-1}(p_i\delta_i\mathbf{s} + \mathbf{c}^{i,\gamma_i-1}) = p_i^{\gamma_i-1} \Lambda_{i,\gamma_i-1}^{\perp}: \mathbf{s}\in S^n, \mathbf{c}^{i,\gamma_i-1}\in Q^n \setminus \mathbf{0}\}$ . Here $\Lambda_{i,\gamma_i-1}^{\perp}$ forms a lattice which has the same structure as the one in scenario 1, with ${\Lambda_{i,\gamma_i-1}^{\perp^{\prime}}} = \{p_i\delta_i\mathbf{s}:\mathbf{s}\in S^n\}$. Thus, $\mathbf{c}^{i,t} = (c_1^{i,t}+\langle p_i\delta_i\rangle, \cdots, c_n^{i,t}+\langle p_i\delta_i\rangle)\in\mathcal{C}^{i,t}$. The minimum-norm coset leader for $\Lambda_{i,\gamma_i-1}^{\perp}/\Lambda_{i,\gamma_i-1}^{\perp^{\prime}}$ and the minimum Euclidean weight $\omega_{\mathrm{min}}^{(i,\gamma_i-1)}(\mathcal{C}^{i,\gamma_i-1})$ can be readily obtained in the same way as (\ref{equ:minimum.norm.A}) and (\ref{equ:dmin.primary.A}). It is obvious that $\lambda_{p_i}^{\gamma_i-1}\in p_i^{\gamma_i-1}\Lambda_{i,\gamma_i-1}^{\perp} \setminus p_i^{\gamma_i-1} \Lambda_{i,\gamma_i-1}^{\perp^{\prime}} $ and we have $\parallel \lambda_{p_i}^{\gamma_i-1}\parallel^2\geq |p_i^{\gamma_i-1}|^2\omega_{\mathrm{min}}^{(i,\gamma_i-1)}(\mathcal{C}^{i,\gamma_i-1})$. The squared shortest vectors of the inner lattice, e.g. $\parallel \lambda_{p_i}^0 \parallel^2$ must be at least larger than $\parallel \lambda_{i,0}^{\perp} \parallel^2$ where $\Lambda_{i,0}^{\perp}\triangleq \{\lambda_{i,0}^{\perp} =	 p_i\delta_i\mathbf{s} + \mathbf{c}^{i,0}: \mathbf{s}\in S^n, \mathbf{c}^{i,0}\in Q^n \setminus \mathbf{0}\}$, and we have $\parallel \lambda_{p_i}^0 \parallel^2 \geq \omega_{\mathrm{min}}^{(i,0)}(\mathcal{C}^{i,0})$. The squared shortest vectors in the set $\Lambda_{p_i}\setminus \Lambda^{\prime}$ is therefore lower bounded by
\begin{align*}
d^2(\Lambda_{p_i}/\Lambda^{\prime}) \geq \min_{0\leq t \leq \gamma_i-1}\{|p_i |^{2t}\omega_{\mathrm{min}}^{(i,t)}(\mathcal{C}^{i,t}) \}
\end{align*}
where $\omega_{\mathrm{min}}^{(i,t)}(\mathcal{C}^{i,t})$ is referred to as the minimum Euclidean weight of non-zero codewords in $\mathcal{C}^{i,t}\in Q^n$ (an $[n,\sum_{\ell=0}^tk_{i,\ell}^{\prime}]$ linear code) for the $t^{\mathrm{th}}$ level of the $i^{\mathrm{th}}$ layer.  
The nominal coding gain for the $i^{\mathrm{th}}$ layer primary sublattices in scenario 2 is lower bounded by:
\begin{align}
\varrho(\Lambda_{p_i}/\Lambda^{\prime}) &\geq \frac{\min_{0\leq t \leq \gamma_i-1}\{|p_i|^{2t}\omega_{\mathrm{min}}^{(i,t)}(\mathcal{C}^{i,t}) \}}{\left(V(\mathcal{V}(\Lambda_{p_i}))\right)^{\frac{1}{n}}} \notag \\
&=  \frac{|p_i|^{2\sum_{t=0}^{\gamma_i-1}(\gamma_i-t)\frac{k_{i,t}^{\prime}}{n} }\min_{0\leq t \leq \gamma_i-1}\{|p_i|^{2t}\omega_{\mathrm{min}}^{(i,t)}(\mathcal{C}^{i,t}) \}} { \vartheta|\varpi|^2 }  \label{NomGain.D.Primary}
\end{align}

Let $N_{t}(\omega_{\mathrm{min}}^{(i,t)}(\mathcal{C}^{i,t}))$ be the number of codewords in $\mathcal{C}^{i,t}$ with the minimum Euclidean weight $\omega_{\mathrm{min}}^{(i,t)}(\mathcal{C}^{i,t})$ for the $t^{\mathrm{th}}$ level and the  $i^{\mathrm{th}}$ layer. The kissing number of the $i^{\mathrm{th}}$ layer primary sublattice is upper bounded by
\begin{equation}
K(\Lambda_{p_i}/\Lambda^{\prime}) \leq \left\{
             \begin{array}{ll}
              \sum_{t=0}^{\gamma_i-1} N_t(\omega_{\mathrm{min}}^{(i,t)}(\mathcal{C}^{i,t}))\left(\frac{\mathcal{N}_{\mathcal{U}(S)}}{|p_i|^2-1}\right)^{\frac{\omega_{\mathrm{min}}^{(i,t)}(\mathcal{C}^{i,t})}{|\delta_i|^2}}, & |p_i|^2-1 \leq \mathcal{N}_{\mathcal{U}(S) }\\
              \sum_{t=0}^{\gamma_i-1} N(\omega_{\mathrm{min}}^{(i,t)}(\mathcal{C}^{i,t})), &  \mathrm{Otherwise}
             \end{array}  
        \right.
\end{equation}
This completes the proof of Proposition \ref{prop:Coding.Gain.D.1}.

We now define the code $\mathcal{C}^i$ such that $\mathbf{c}^i = \mathbf{c}^{i,0}+p_i\mathbf{c}^{i,1}+\cdots+p_i^{\gamma_i-1}\mathbf{c}^{i,\gamma_i-1}$, and hence $\mathcal{C}^i\in\delta_iS/\langle\varpi\rangle$. From the theorems developed in sections \ref{sec:Multilevel.Lattice.Network.Coding} and \ref{sec:EDC}, we are able to generate a new code $\tilde{\mathcal{C}}$ such that $\tilde{\mathbf{c}} = \mathbf{c}^1 + \cdots + \mathbf{c}^m$ which makes $\tilde{\mathcal{C}}\in S/\langle\varpi\rangle$. Thus, the codeword of $\tilde{\mathcal{C}}$ is generated by the nested linear codes $\mathcal{C}^{i,t}$ of all layers. The minimum-norm coset leader for $\Lambda/\Lambda^{\prime}$ can be represented by:
\begin{align}
\tilde{\sigma}_{\bigtriangleup}(\tilde{\mathbf{c}}) &= \left(\tilde{c}_1 - \varpi\mathcal{Q}\left(\frac{\tilde{c}_1}{\varpi} \right),\cdots,\tilde{c}_n - \varpi\mathcal{Q}\left(\frac{\tilde{c}_n}{\varpi} \right)   \right) 
\end{align}
where $\tilde{c}_j = c_j^1 +c_j^2 + \cdots + c_j^m$ and $c_j^i$, $j=1,2,\cdots,n$, $i=1,2,\cdots,m$, denotes the $j^{\mathrm{th}}$ element of the codeword $\mathbf{c}^i$. Then, the squared shortest vectors in the set $\Lambda\setminus \Lambda^{\prime}$ can be represented by the $m$ linear codes used at each layer,  
\begin{align*}
d^2(\Lambda/\Lambda^{\prime}) = \min_{\tilde{\mathbf{c}}\neq\mathbf{0}, \tilde{\mathbf{c}}\in\mathcal{C}}||\tilde{\sigma}_{\bigtriangleup}(\tilde{\mathbf{c}})||^2 = \omega_{\mathrm{min}}(\tilde{\mathcal{C}})
\end{align*}
The nominal coding gain for $\Lambda/\Lambda^{\prime}$ is     
\begin{align}
\varrho(\Lambda/\Lambda^{\prime}) &= \frac{\omega_{\mathrm{min}} (\tilde{\mathcal{C}})}{\left(V(\mathcal{V}(\Lambda))\right)^{\frac{1}{n}}} \notag \\
&= \frac{\omega_{\mathrm{min}}(\tilde{\mathcal{C}})\prod_{\ell=1}^m|p_{\ell}|^{2\sum_{t=0}^{\gamma_{\ell}-1}(\gamma_{\ell}-t)\frac{k_{i,t}^{\prime}}{n} }}{\vartheta|\varpi|^2}  
\label{NomGain.D.Fine}
\end{align}
The nominal coding gain of the EDC lattice in scenario 2 is related to the minimum Euclidean weight of the composite code $\tilde{\mathcal{C}}$ and the code rates of all nested linear codes at each layer.   
\end{document}